\documentclass[12pt,preprint]{aastex}

\usepackage{graphicx}
\usepackage{epstopdf}

\newcommand{\fuvmag}{\ifmmode{FUV}\else{\it FUV~}\fi}
\newcommand{\nuvmag}{\ifmmode{NUV}\else{\it NUV~}\fi}
\newcommand{\degree}{\ifmmode{^\circ}\else{$^\circ$~}\fi}
\newcommand{\mh}{\ifmmode{{\rm M}_{\rm H}}\else{M$_{\rm H}$~}\fi}
\newcommand{\irx}{\ifmmode{\langle IRX \rangle}\else{$\langle IRX \rangle$}\fi}
\newcommand{\ssfr}{\ifmmode{\langle SSFR \rangle}\else{$\langle SSFR \rangle$}\fi}
\newcommand{\mass}{\ifmmode{\mathcal{M}}\else{$\mathcal{M}$}\fi}

\slugcomment{Submitted for publication in the Special GALEX Ap.J.Suppl. Issue}

\shorttitle{Co-evolution of Star Formation and Extinction}
\shortauthors{Martin et al.}

\begin{document}

\title{The Star Formation and Extinction Co-Evolution of UV-Selected Galaxies over $0.05<z<1.2$}

\author{
D. Christopher Martin\altaffilmark{1},
Todd Small\altaffilmark{1},
David Schiminovich\altaffilmark{13},
Ted K. Wyder\altaffilmark{1},
Pablo G. P{\'e}rez-Gonz{\'a}lez\altaffilmark{12},
Benjamin Johnson\altaffilmark{13},
Christian Wolf\altaffilmark{10},
Tom A. Barlow\altaffilmark{1},
Karl Forster\altaffilmark{1},
Peter G. Friedman\altaffilmark{1},
Patrick Morrissey\altaffilmark{1},
Susan G. Neff\altaffilmark{8},
Mark Seibert\altaffilmark{1},
Barry Y. Welsh\altaffilmark{6}, 
Luciana Bianchi\altaffilmark{2},
Jose Donas\altaffilmark{4},
Timothy M. Heckman\altaffilmark{5},
Young-Wook Lee\altaffilmark{3},
Barry F. Madore\altaffilmark{7},
Bruno Milliard\altaffilmark{4},
R. Michael Rich\altaffilmark{9},
Alex S. Szalay\altaffilmark{5},
Sukyoung K. Yi\altaffilmark{3},
Klaus Meisenheimer\altaffilmark{11},
George Rieke\altaffilmark{12}.
}

\altaffiltext{1}{California Institute of Technology, MC 405-47, 1200 East
California Boulevard, Pasadena, CA 91125}

\altaffiltext{2}{Center for Astrophysical Sciences, The Johns Hopkins
University, 3400 N. Charles St., Baltimore, MD 21218}

\altaffiltext{3}{Center for Space Astrophysics, Yonsei University, Seoul
120-749, Korea}

\altaffiltext{4}{Laboratoire d'Astrophysique de Marseille, BP 8, Traverse
du Siphon, 13376 Marseille Cedex 12, France}

\altaffiltext{5}{Department of Physics and Astronomy, The Johns Hopkins
University, Homewood Campus, Baltimore, MD 21218}

\altaffiltext{6}{Space Sciences Laboratory, University of California at
Berkeley, 601 Campbell Hall, Berkeley, CA 94720}

\altaffiltext{7}{Observatories of the Carnegie Institution of Washington,
813 Santa Barbara St., Pasadena, CA 91101}

\altaffiltext{8}{Laboratory for Astronomy and Solar Physics, NASA Goddard
Space Flight Center, Greenbelt, MD 20771}

\altaffiltext{9}{Department of Physics and Astronomy, University of
California, Los Angeles, CA 90095}

\altaffiltext{10}{Department of Physics, University of Oxford, Keble Road, Oxfordm OX1 3RHU, U.K.}

\altaffiltext{11}{Max-Planck-Institut fur Astronomie, Konigstuhl 17, D-69117 Heidelberg, Germany.}

\altaffiltext{12}{Steward Observatory, University of Arizona, 933 North Cherry Avenue, Tucson, AZ 85721.}

\altaffiltext{13}{Department of Astronomy, Columbia University, 528 W. 120th St., New York, NY 10027.}


\begin{abstract}

We use a new stacking technique to obtain mean mid IR and far IR
to far UV flux ratios over the rest near-UV/near-IR  color-magnitude diagram. 
We employ COMBO-17 redshifts and COMBO-17 optical, GALEX far and near UV, Spitzer
IRAC and MIPS Mid IR photometry.  This technique permits us to
probe infrared excess (IRX), the ratio of far IR to far UV luminosity, 
and specific star formation rate (SSFR)
and their co-evolution over two orders of magnitude of stellar
mass and redshift $0.1<z<1.2$. We find that the SSFR and the characteristic
mass (M$_0$) above which the SSFR drops increase with redshift
(downsizing).  At any given epoch, IRX is
an increasing function of mass up to M$_0$. Above this
mass IRX falls, suggesting gas exhaustion.  
In a given mass bin below M$_0$ IRX increases with time
in a fashion consistent with enrichment. 
We interpret these trends using a simple model with a Schmidt-Kennicutt
law and extinction that tracks gas density
and enrichment. We find that the average
IRX and SSFR follows a galaxy age parameter $\xi$ which is determined mainly
by the galaxy mass and time since formation. 
We conclude that blue sequence
galaxies have properties which show simple, systematic trends
with mass and time such as the steady build-up of heavy elements in the interstellar media of evolving galaxies
and the exhaustion of gas in galaxies that are evolving off the blue sequence.
The IRX represents a tool for selecting  galaxies
at various stages of evolution. 

\end{abstract}



\section{Introduction}

It has long been recognized that the present day properties of most galaxies can be represented by
relatively simple star formation histories \citep{tinsley68,searle72}. While the physical basis for
exponential star formation histories is almost certainly oversimplified, the resulting spectral energy
distributions generally do an excellent job in representing galaxy spectra and broadband colors. 
If exponential models have a basis in the physics of star formation history, in particular
in the conversion of gas into stars, then they make basic predictions that relate the
specific star formation rate to the gas fraction in galaxies over time. Evidence
for such a picture is growing (e.g., \cite{bell05a,noeske07}). Coupled with 
a model for chemical evolution, and brushing aside for the moment
the complexities of dust reprocessing, this framework could also provide a description
of the evolution of dust extinction in galaxies and the co-evolution of extinction
and star formation rate. In particular, we could expect a growth in the dust-to-gas
ratio as gas is processed through stars and potentially an increase in extinction
over time (e.g., as seen at z$\sim$2 by \cite{reddy06}). At the same time, as galaxies exhaust their gas supply, by whatever
mechanism, we may detect a corresponding drop in extinction.

In order to discern such effects, we need to segregate galaxies by a parameter
which is likely to be related to the timescale for evolution. There is certainly
theoretical motivation for using stellar mass as the fundamental parameter. 
For example, surface density scales with stellar mass \citep{kauffmann03b}, and star formation rate
scales with gas surface density \citep{kennicutt89}. The observational
case for ``downsizing'' seems secure \citep{cowie96,brinchmann00}.
The mass-metallicity relation \citep{tremonti04} suggests that low metallicity in
lower mass galaxies could be related to higher gas fractions and lower
processing through star formation. Lower mass galaxies
have younger stellar ages (e.g., \cite{kauffmann03b}). There is a well-known relationship
between luminosity as a proxy for stellar mass and extinction \citep{wang96}
which is present even at high redshift \citep{meurer99,adelberger00,reddy06,papovich06}.
We have recently established a tight relationship between metallicity and
infrared excess (IRX), the ratio of far infrared to far ultraviolet luminosity \citep{johnson07},
that suggests that IRX may be used as a tracer of metallicity and its evolution.
Finally, there is growing evidence that the so-called ``blue cloud'' of
star forming galaxies on the color-magnitude diagram is
actually a ``blue sequence'' in stellar mass \citep{wyder06,johnson06a} that is
relatively tight in color space when extinction is corrected. 

In order to distinguish trends with stellar mass, it is critical to
have as large a mass dynamic range as possible. At the same
time, dust extinction is likely to be a complex process
that introduces considerable noise into any overall trends.
Inclination variations alone produce much variance for
an otherwise constant dust geometry and extinction law.
We need to develop an approach which reveals the
average trends with stellar mass in spite of this noise.
A major benefit of a large multiwavelength survey
is the ability to extract such trends by averaging over many galaxies.
We have used Spitzer IRAC data to measure stellar mass and MIPS24
data for dust luminosity. We combine this with GALEX UV and COMBO-17
optical photometry and redshifts. A major difficulty we face when combining
IR and UV survey data is the relatively small overlap in
detected sources, with the bulk of the overlap
occuring at high luminosity and mass. Thus we have developed a new stacking approach
which permits us to study IRX over two orders of magnitude in stellar mass
over the redshift range $0.05<z<1.2$. Using this and a bolometric correction
we obtain an average IRX over the UV, H-band color magnitude diagram.
We use this to find total star formation rate, stellar mass, and specific
star formation rate. Finally, we show that the co-evolution of the average IRX and specific star formation
rate can be modelled using simple exponential star formation histories
and closed box chemical evolution to $z\sim 1$. 

We note that \cite{zheng07} have recently used an independent stacking technique
\citep{zheng06} to derive the SF history vs. stellar mass, also
using COMBO-17 and Spitzer data, and reached many conclusions
that are similar to ours, although with important differences which we
discuss in \S 5.2.

We use a concordance cosmology $\Omega_\Lambda=0.70$, $\Omega_m=0.30$,
and $H_0=70$ km s$^{-1}$ Mpc$^{-1}$.  We use AB magnitudes for all bands. 
We also use the following nomenclature: observed magnitudes are given
by $m_i$, for example the observed NUV and R magnitudes are $m_{NUV}$ and $m_R$.
Rest-frame magnitudes are denoted by FUV, NUV, H, etc. Extinction corrected
rest-frame absolute magnitudes are denoted M$_{FUV,0}$, M$_{NUV,0}$, M$_{H,0}$, etc. 
Finally, we define the infrared excess IRX as the log ratio
of FIR to FUV luminosity ($\nu\mathcal{L}_\nu$), unless specifically called out.

\section{Data and Source Catalogs}

\subsection{Primary Datasets}

\subsubsection{GALEX}

The GALEX observations of CDFS consist of a total of 61 orbital visits
over the period 4 November 2003 to 5 November 2005 for a total exposure time of 49,758 seconds.  
Simultaneous exposures were obtained in the Far UV (FUV, 1344-1786\AA, center 1549\AA)
and Near UV (NUV, 1771-2831\AA, center 2316\AA) bands.
The individual and co-added images were processed using version 5.0 of the GALEX data pipeline,
also used to process GALEX data releases GR2 and GR3.
The 1.25 \degree diameter GALEX images completely circumscribe the other two survey footprints.
The GALEX mission, on-orbit performance, and current status
of calibration and pipeline reductions are summarized in \cite{martin05}, \cite{morrissey05}, and
\cite{morrissey07} respectively. Source photometry errors (systematic) should be less than 0.05 mag
and astrometric errors less than 1 \arcsec.
Images of this exposure level in low background
regions should reach a 5$\sigma$ Poisson limited depth of $m_{NUV}\simeq$25.5 AB magnitudes in both bands,
and roughly 3$\sigma$ at $m_{NUV}\simeq 26.0$. 
However NUV data in particular are confusion limited because of the 5-6 \arcsec~ PSF FWHM.
We therefore used a PSF fitting source extraction procedure that uses the CDFS optical
positions as priors. This is described below. 

\subsubsection{COMBO-17}

The COMBO-17 survey  \citep{wolf03} combines a set of medium and wide photometric
bands to obtain robust photometric redshifts and basic object classification to a depth of $m_R \sim$24. 
A complete description of the survey can be found in \cite{wolf04}. The CDFS field is
0.5 by 0.5 degree$^2$ centered on 
$(\alpha,\delta)_{J2000}= (03^h32^m25^s, -27^\circ48^\prime 50^{\prime\prime})$. Other than redshifts, we
use the COMBO-17 survey for two purposes: in order to generate a k-corrected NUV luminosity
and NUV-H color, and as the basis for the PSF fitting extraction of the FUV and NUV source fluxes.
\cite{wolf04} have used a Monte-Carlo technique to derive the
survey completeness vs. object color, type, and magnitude. We have used
these completeness matrices to derive the volume-corrected distributions, as we describe below.

\subsubsection{Spitzer}

The Spitzer data are described in detail in \cite{perez-gonzalez05}, which we briefly
recap here. The $1^\circ.5 \times 0^\circ.5$ rectangular areas
centered on CDFS, $(\alpha,\delta)_{J2000}=(03^h 32^m 02^s, -27^\circ 37^\prime 24^{\prime\prime})$,
are mapped in MIPS 24 $\mu$m in scan map mode, and also in the
four IRAC channels (3.6, 4.5, 5.8, and 8.0 $\mu$m). The MIPS 24 $\mu$m
reduction was performed using the MIPS Data Analysis Tool \citep{gordon05},
resulting in images with average exposures of $\sim$1400 s.  IRAC images
were reduced with the general Spitzer pipeline and mosaiced, yielding
an average exposure time of 500 s. Source catalogs for IRAC 3.6 $\mu$m detections
are used below in a jointly selected sample. We tested catalogs generated
by a simple one-pass SExtractor \citep{bertin96} extraction, and
by the multiband technique used by \cite{perez-gonzalez05}, with
no significant differences noted in our results. Source catalogs
of MIPS 24 $\mu$m objects were used to clean 24 $\mu$m images
for stacking, as we describe below in \S \ref{section_fir}. Again, a single-pass SExtractor catalog
produced very similar results to the mulit-pass PSF-fitting catalog generated
by \cite{perez-gonzalez05}. 

\subsection{Matched Datasets}

\subsubsection{GALEX/COMBO-17 PSF Fitting Catalog\label{section_psf}}

As noted above, deep GALEX images suffer from source confusion, especially in the NUV.  For fields with complementary deep optical photometry, we can use the positions of sources from the optical catalog to deblend the GALEX images and obtain more reliable flux estimates.  The center of the COMBO-17 field is only slightly offset (3.8 arcmin) from the center of the GALEX images and is much smaller than the GALEX image.  Within the COMBO-17 field, the variation of the GALEX PSF is small, and so we have have used one average PSF for each band.  After correcting for the small (less than 1 arcsec) systematic offsets between the GALEX and COMBO-17 astrometry, the deblending proceeds by dividing the region to be deblended into contiguous 100 x 100 pixel chunks and then simultaneously fitting the amplitudes of the sources at positions taken from the optical catalog and the mean background in each chunk.  We assume that the counts in each pixel are Gaussian-distributed, which is a safe assumption for the NUV (where the background level is ~100 counts), but is questionable for the FUV (where the background level is ~10 counts).  In order to test the reliability of our deblending, we have added approximately 1000 artificial point sources to the COMBO-17/GALEX overlap region and then compared the extracted fluxes with known input fluxes.  In the FUV, the deblended magnitudes are systematically fainter than the input magnitudes by 0.04 mag and have errors that are 20\% larger than expected from counting statistics.  In the NUV, the deblended magnitudes are systematically too faint by only 0.01 mag, but the errors, due to the source crowding, are a factor of 2 larger than expected from counting statistics.  For the 49,758 second GALEX images used here, the 95\% confidence detection limits are 25.55 mag in FUV and 25.10 mag in NUV.

\subsubsection{Merged Catalog \label{section_merge}}

There is a low fraction of sources detected in all three catalogs.
We generated individual source catalogs for the five Spitzer images using Sextractor.
We matched these detections to the merged COMBO-17/GALEX catalog
using a 2\arcsec search radius. The common area of the three surveys
is 0.19 degree$^2$. The matched source statistics are 
summarized in Table \ref{tab_matchstats}. Of the 11778 COMBO-17 $m_R<$24 sources
in the common region, 7498 are detected at $m_{NUV}<$26, 1784 in MIPS24 ($>$0.02 mJy),
4955 in NUV and IRAC1 ($>$0.0002 mJy), and 1171 in NUV, IRAC1 and MIPS24.
As we discuss below, the main explanation for the low overlap fraction is that many UV-selected sources
have moderate to low infrared-to-UV ratios and are not directly detected in 
the mid-infrared. 

Because we are keenly interested in the evolution of the average extinction, infrared excess, and
star formation history of galaxies over cosmic time, we have adopted a
stacking approach. We summarize the complete methodology in the next section.

\section{Analysis}

Our goal in this paper is to determine the evolution of the average IRX and extinction and
relate this to the evolution of the star formation rate, as a function of stellar mass.
We would like to exploit a property of the blue sequence of star forming galaxies that is
rapidly becoming clear, that this sequence has a relatively low dispersion of properties
once the mass is given \citep{noeske07,wyder06,martin07a}. The dispersion of various
properties, such as extinction, stellar age and mass, measured in individual bins of the
NUV-r color magnitude diagram, for example, is low. 
The IRAC1 channel and COMBO-17 R-band give a good estimate
of the rest H-band flux over the redshift range $0<z<1.2$, providing
a good stellar mass tracer with low extinction sensitivity. 
We therefore feel it is reasonable to
stack using bins of the rest-frame M$_H$, NUV-H band color-magnitude diagram.

\subsection{Methodology}

Here is a step-by-step summary of our approach, with cross-references to
more detailed discussion:

\begin{enumerate}

\item Use COMBO-17 positions to generate a joint COMBO-17/GALEX catalog using PSF fitting (\S \ref{section_psf})

\item Match IRAC1 sources detected with Sextractor to joint COMBO-17/GALEX catalog (\S \ref{section_merge})

\item Generate rest-frame NUV, H-band magnitudes using SED interpolation (\S \ref{section_cmd})

\item Construct the volume-corrected color-magnitude diagram (CMD)  $\phi$(\mh,NUV-H)  in several redshift bins (\S \ref{section_cmd}).

\item For each (\mh, NUV-H,  redshift) bin, stack all IRAC2-4 and MIPS24 images at the
R-band COMBO-17 source positions falling in that (\mh, NUV-H,  z)  bin.  Our stacking
technique adds detected source fluxes to a stack of undetected source regions, as we discuss in \S \ref{section_fir}.

\item Use a bolometric correction obtained by fitting a local SWIRE/SDSS/GALEX sample \citep{johnson07}
and the total Far IR luminosity and infrared excess IRX=$\log{L_{FIR}/L_{FUV}}$ or $\log{L_{FIR}/L_{NUV}}$. (\S \ref{section_fir})

\item Use a standard extinction law to convert IRX into NUV and H-band extinction. (\S \ref{section_ssfr})

\item Determine the average extinction correction for galaxies in each (\mh, NUV-H,  z) bin. (\S \ref{section_ssfr})

\item Using $(NUV-H)_0$ and $M_{H,0}$ infer  the specific star formation rate and stellar mass
using a simple prescription.  (\S \ref{section_ssfr})

\item Determine the volume-corrected distributions of specific star formation and stellar mass ($\mathcal{M}$) in each redshift
bin $\phi$($\mathcal{M}$,SSFR,z).  (\S \ref{section_ssfr})

\item Finally, calculate the average IRX and specific star formation rate (\irx ~ and \ssfr) as a function of stellar mass and redshift (\S \ref{sec:avgirx}, \ref{sec:avgssfr}).

\end{enumerate}

\subsection{Evolution of the Color-Magnitude Distribution \label{section_cmd}}

We choose to use rest-frame FUV or NUV to derive star formation
rate and H-band magnitudes to obtain stellar mass. Rest-frame H-band flux was obtained by interpolating between
the COMBO-17 R-magnitude and the IRAC 3.6 $\mu$m flux, exploiting
the fact that the SED is essentially constant over this range for most galaxy templates. Rest-frame NUV flux is
an interpolation of the observed NUV and the COMBO-17 catalog rest-frame u-band flux.
Rest-frame FUV flux is an interpolation between observed FUV and NUV which accounts for the Lyman continuum break.
We also tried SED fitting, which produces very similar results. 

We derived the volume-corrected, rest-frame NUV-H vs \mh distributions as follows.
We used five redshift bins, $z=$0.05-0.2. 0.2-0.4, 0.4-0.6, 0.6-0.8, and 0.8-1.2.
Maximum detection volumes in  were derived for a sample jointly selected in observed NUV, r-band, and
IRAC channel 1 with the following limits: $m_{NUV}<26.0$, $m_r<24.0$, and $f(3.6)> 0.5 \mu$Jy. 
The minimum $V_{max}$ of the three bands was used. For each galaxy a completeness was calculated 
$f_c = f_c(NUV) * f_c(C17)$. We assume that the IRAC channel 1 completeness is high to
the flux limit. With PSF fitting, NUV completeness is estimated to be $\sim$80\% at $m_{NUV}$=25.5
and $\sim$56\% at $m_{NUV}$=26.0, based on comparison between the observed magnitude distribution
and that measured by \cite{gardner00}. Because of the soft rolloff of completeness for this PSF-fit catalog we choose to
use a deeper magnitude cutoff corresponding to a 3$\sigma$ detection threshold. 
The COMBO-17 redshift catalog completeness is a function
of object magnitude, color, and type. We used the completeness matrix derived by \cite{wolf03}
to calculate $f_c(C17)$ for each galaxy. The volume-corrected color-magnitude distribution
is then calculated by summing for each galaxy the term 

\begin{equation}
V_{max,i} = min[V_{max}(NUV_i,z_i),V_{max}(r_i,z_i),V_{max}(f(3.6)_i,z_i)] f_c(NUV_i) f_c(r_i; g_i-r_i)
\end{equation}

\begin{equation}
\phi(NUV-H,M_H,z) = \sum {1 \over {V_{max,i}} }
\end{equation}

\begin{figure}
\plottwo{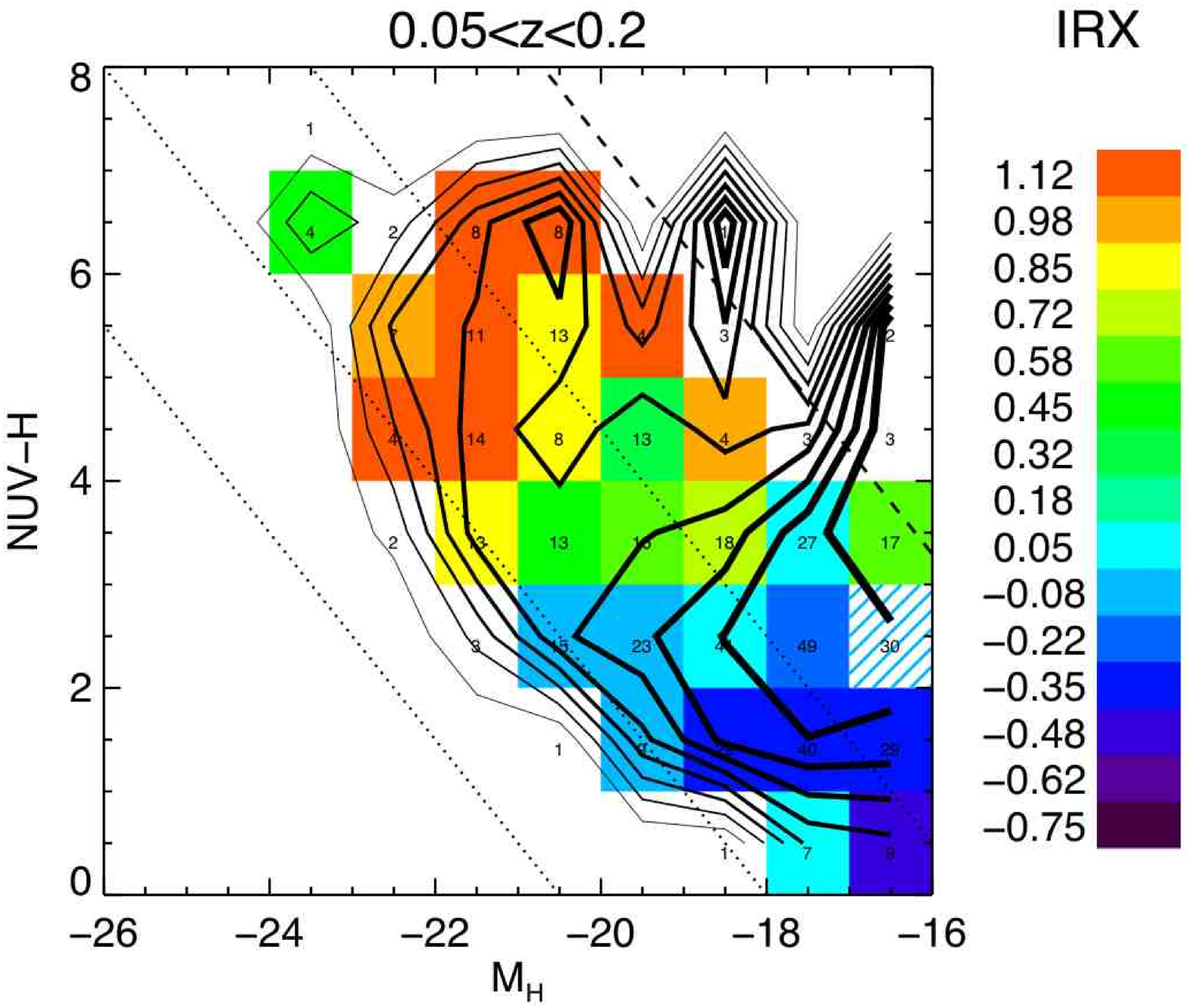}{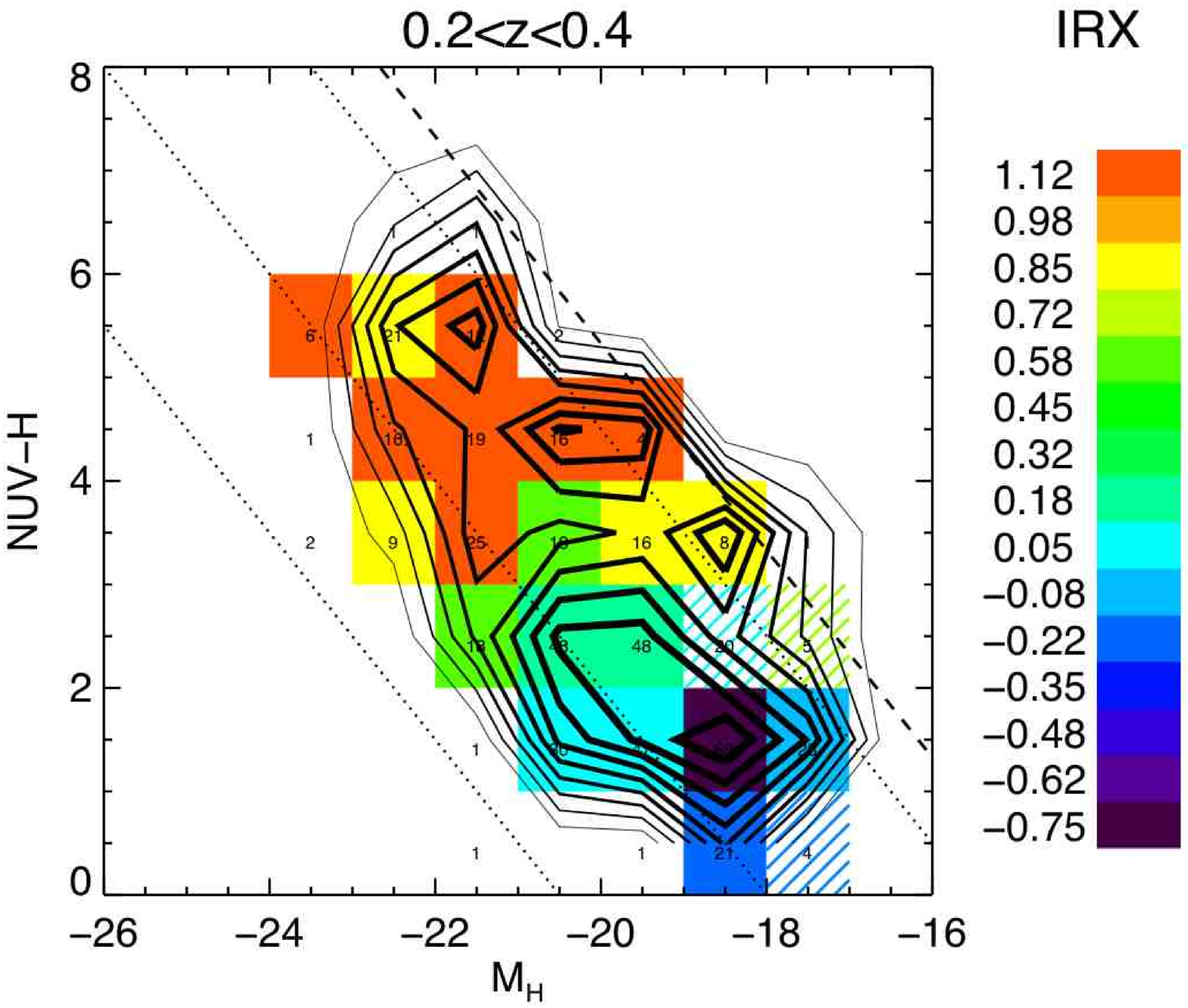}
\plottwo{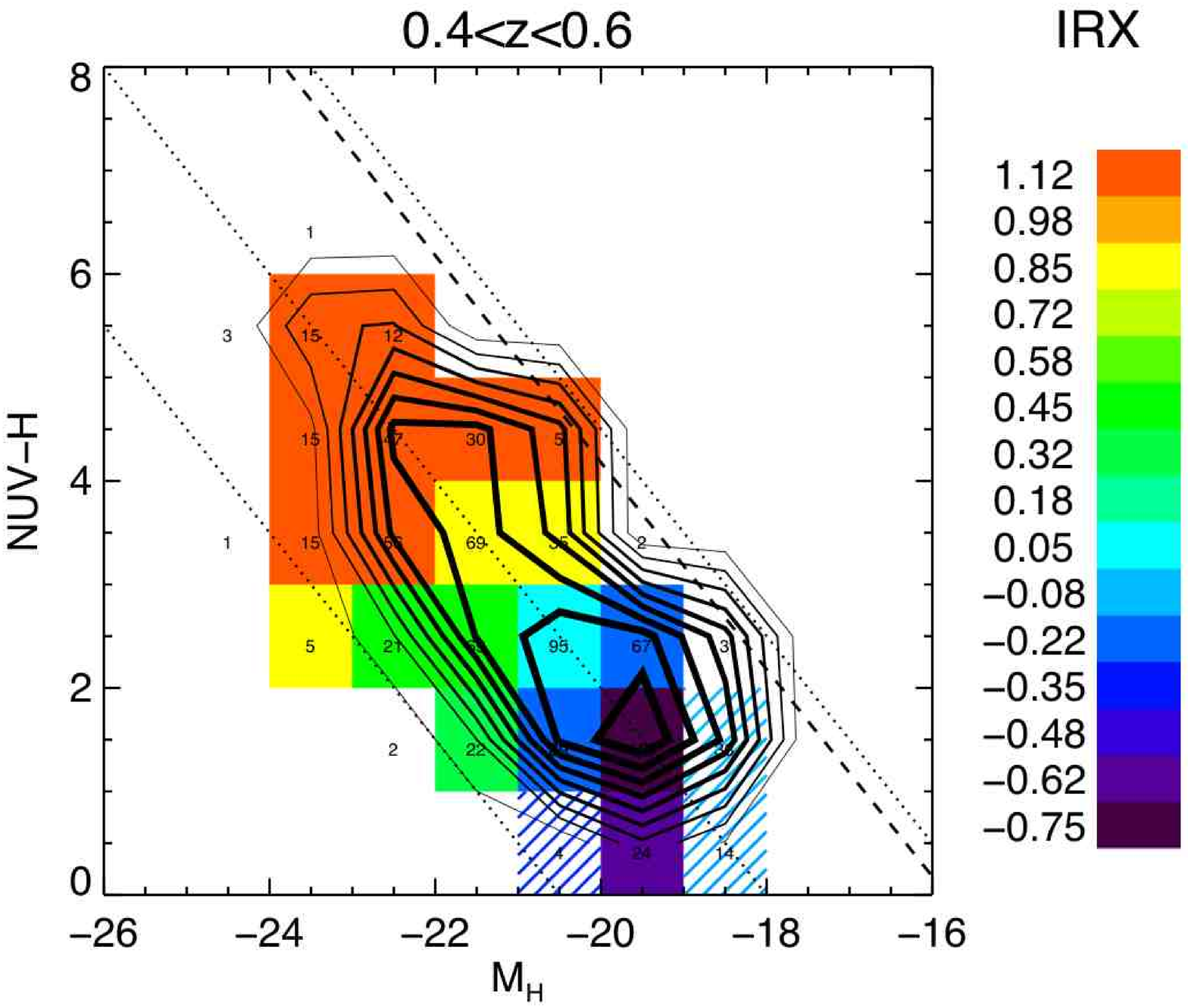}{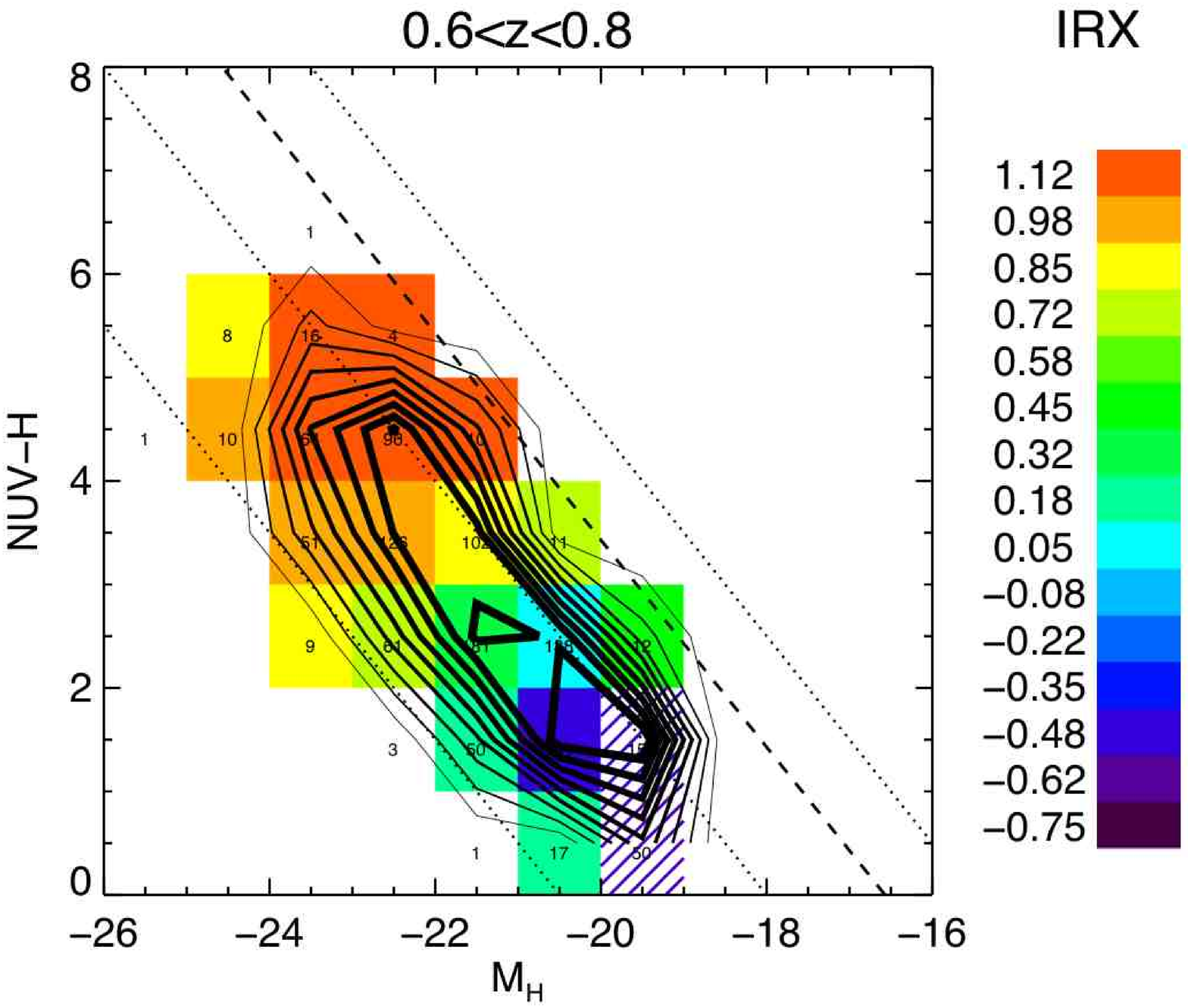}
\plottwo{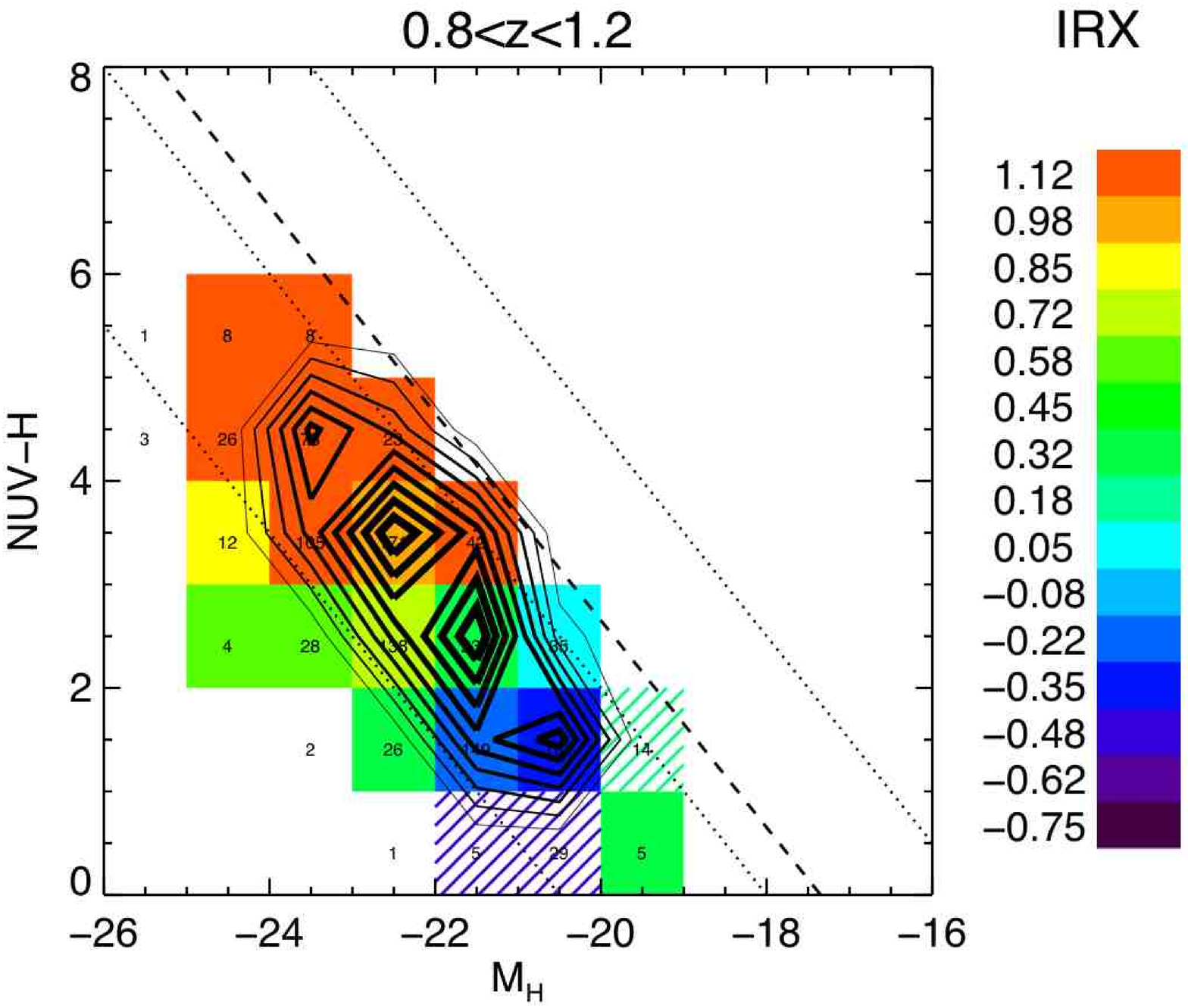}{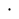}
\caption{Volume corrected bivariate color-magnitude distribution $\phi(M_H,NUV-H)$. Contours are
equally spaced in $\log{\phi}$, with 10 divisions from $-4<\log{\phi}<-2$. Colors give IRX, which is the log of the FUV-to-FIR luminosity ratio.
Hashed color bins are upper limits to the IRX. Dotted lines give loci of constant M$_{NUV}$=-15.5,-18.0,-20.5. 
Dashed diagonal line indicates $m_{NUV}=26$ limit in center of redshift bin.
\label{fig_cmd_observed}}
\end{figure}

The resulting distribution is displayed in contour plots in Figures \ref{fig_cmd_observed}.
The general trend that can be seen is a shift to bluer NUV-H colors and brighter \mh magnitudes.
These plots also show average IRX in each bin, to be discussed in the next section.

\subsection{IR Stacking, Bolometric Correction, Extinction Correction \label{section_fir}}

We have generated an average IRX for each bin in the \mh, NUV-H color magnitude
diagram. As we discussed earlier, we do this because of the small overlap in UV and MIPS
detections. There is considerable evidence that galaxies occupying a single
color magnitude bin have a relatively small dispersion in most properties, including
extinction \citep{martin07a,wyder06,johnson06a}. This approach allows us to estimate the
average IRX over a large range of stellar mass and redshift, offering sensitivity to
quite low IRX. In order to ensure that the technique is not affected by
systematic or random error, we perform a set of tests below.

The basic stacking technique is as follows.  We first generate a catalog of detected
sources in the MIPS 24 $\mu$m band using Sextractor. Using this catalog we
generate a set of cleaned images with detected sources removed.
For each redshift and each color-magnitude bin
(NUV-H, \mh, z) we stack images in each band that do not have detected sources. We then extract either
a source flux or an upper limit, and add this to the detected flux. This results in a flux or upper limit. 

We must then make a bolometric correction to the observed 24 $\mu$m luminosity.
We have used the GALEX/SWIRE catalog generated for
\cite{johnson06a,johnson07} to derive the bolometric correction
of rest frame flux from 12-24 $\mu$m (corresponding to 0$<z<$1).
We use the measured FIR fluxes and fits to \cite{dale02} SEDs
and derive coefficients in the following relationship using log-log fits:
\begin{equation}
\log L_{FIR} = a_\lambda + b_\lambda \log (L[\lambda]/10^{10}L_\odot)
\end{equation}
where $\lambda$ is the observed rest frame luminosity ($L[\lambda]=\nu L_\nu[\lambda]$).
We list in Table 2 the coefficients a$_\lambda$ and b$_\lambda$.
The rms errors in the fits used to derive the coefficients are small, $\sigma \simeq 0.03-0.06$.

Finally, we  correct the NUV-H color and H-band magnitude for internal extinction 
using the IRX and the following prescription based on \cite{calzetti00}:

\begin{equation}
A_i = 2.5 \log_{10} [ {{BC_{dust}} \over {BC_i}} IRX_i + 1 ]
\label{eqn:irx}
\end{equation}
where $i$ corresponds either to FUV or NUV, and the bolometric corrections
are $BC_{dust} = 1.75$, $BC_{FUV} = 1.68$, and $BC_{NUV} = 2.45$.

\begin{figure}
\plottwo{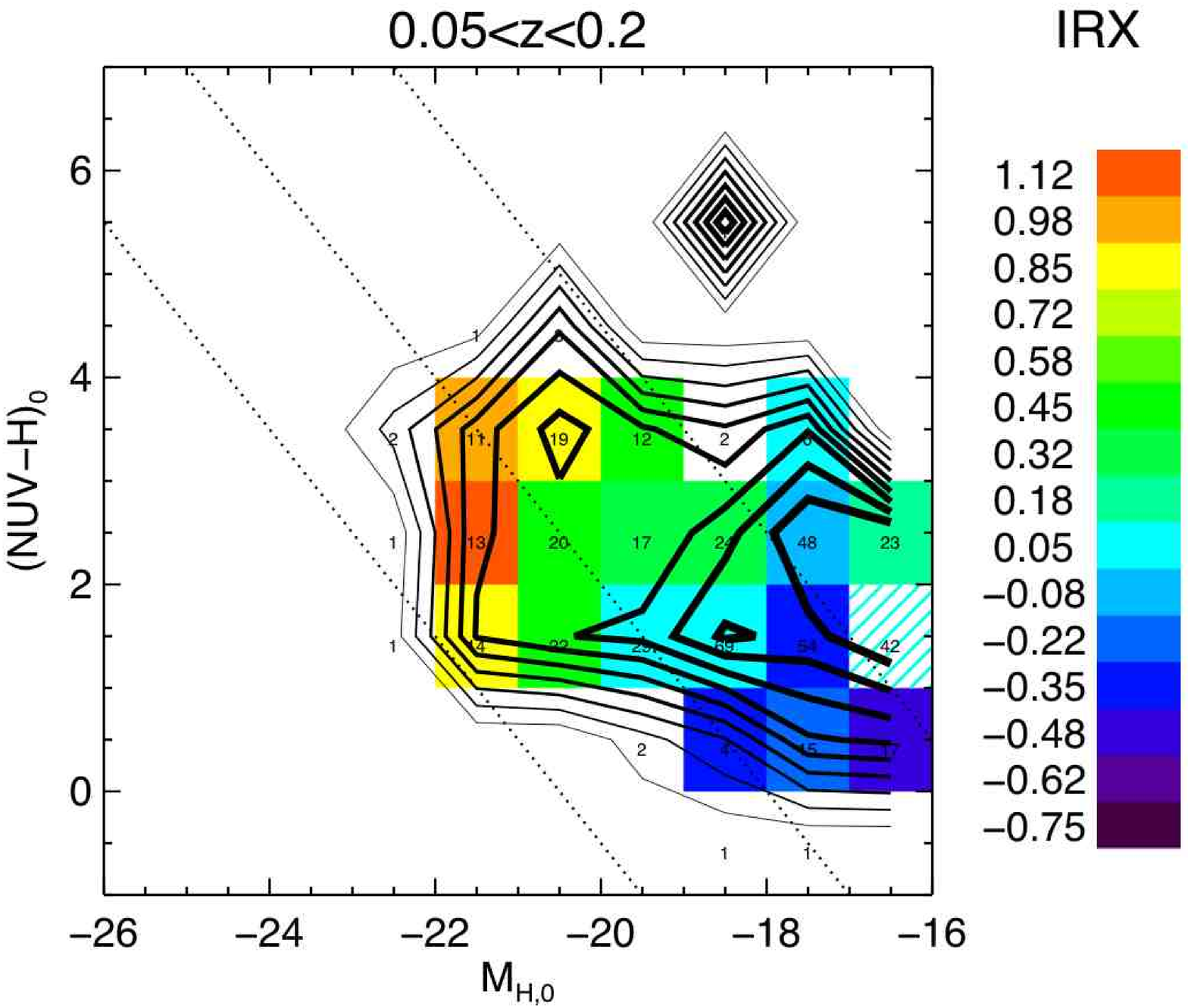}{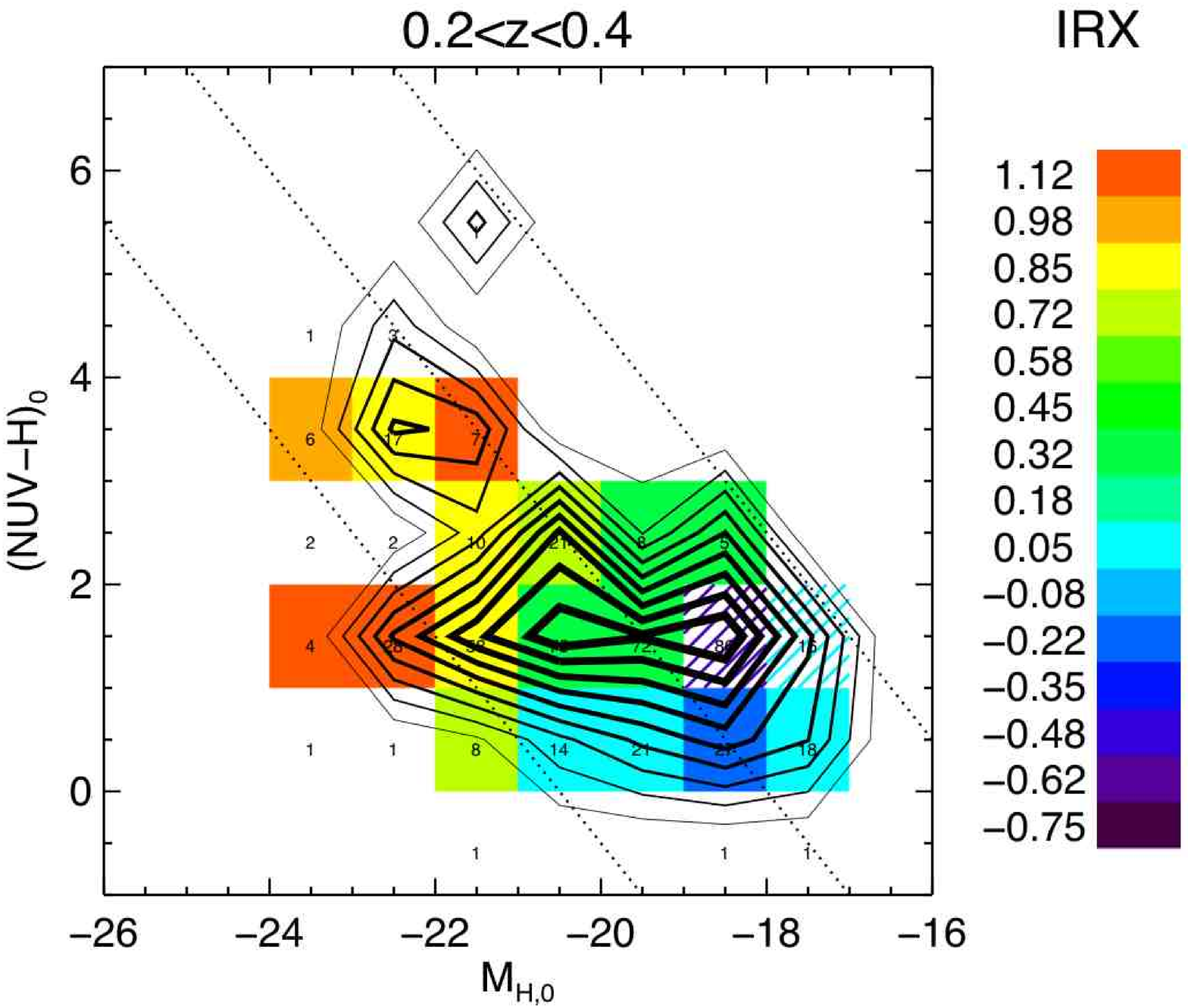}
\plottwo{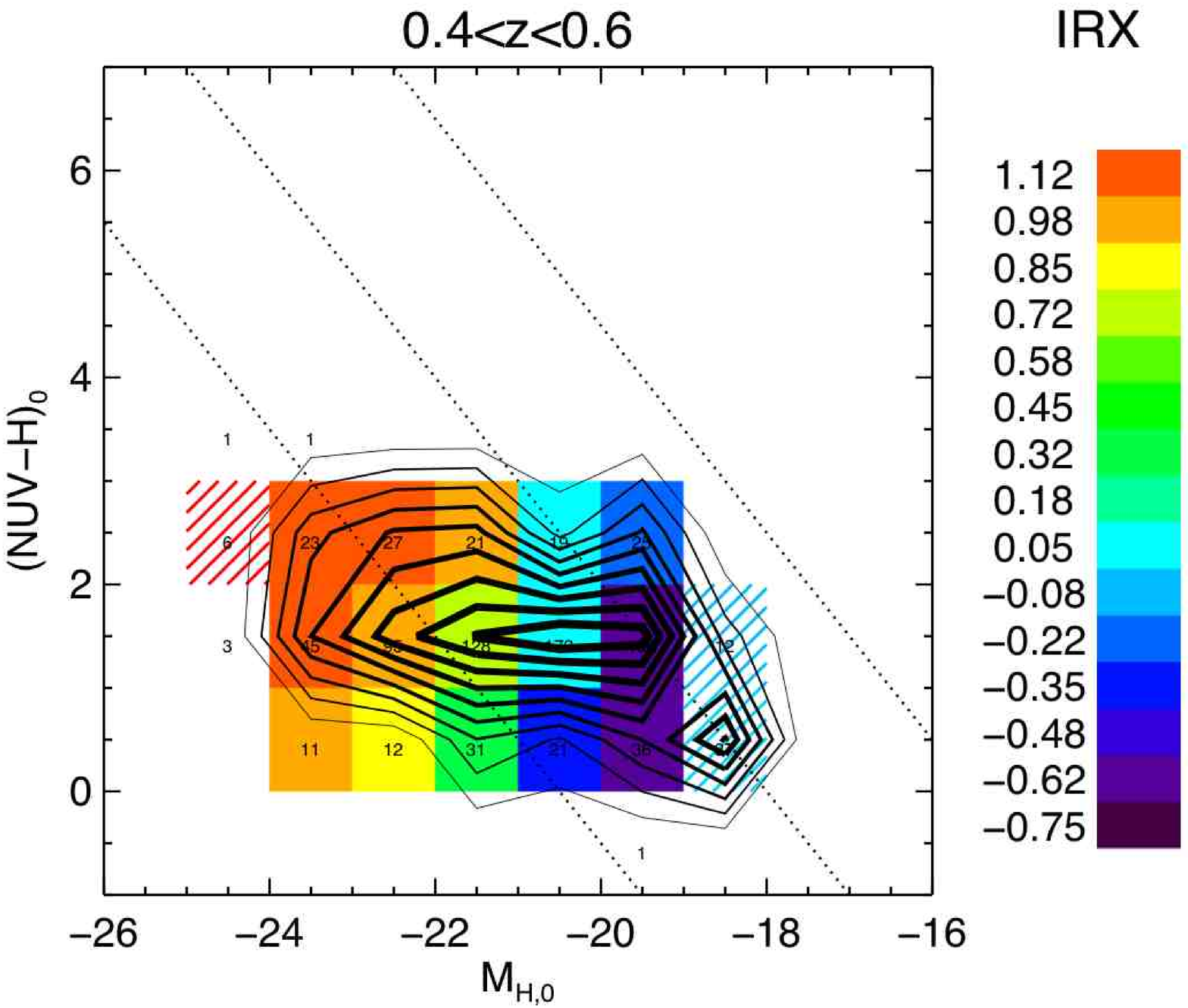}{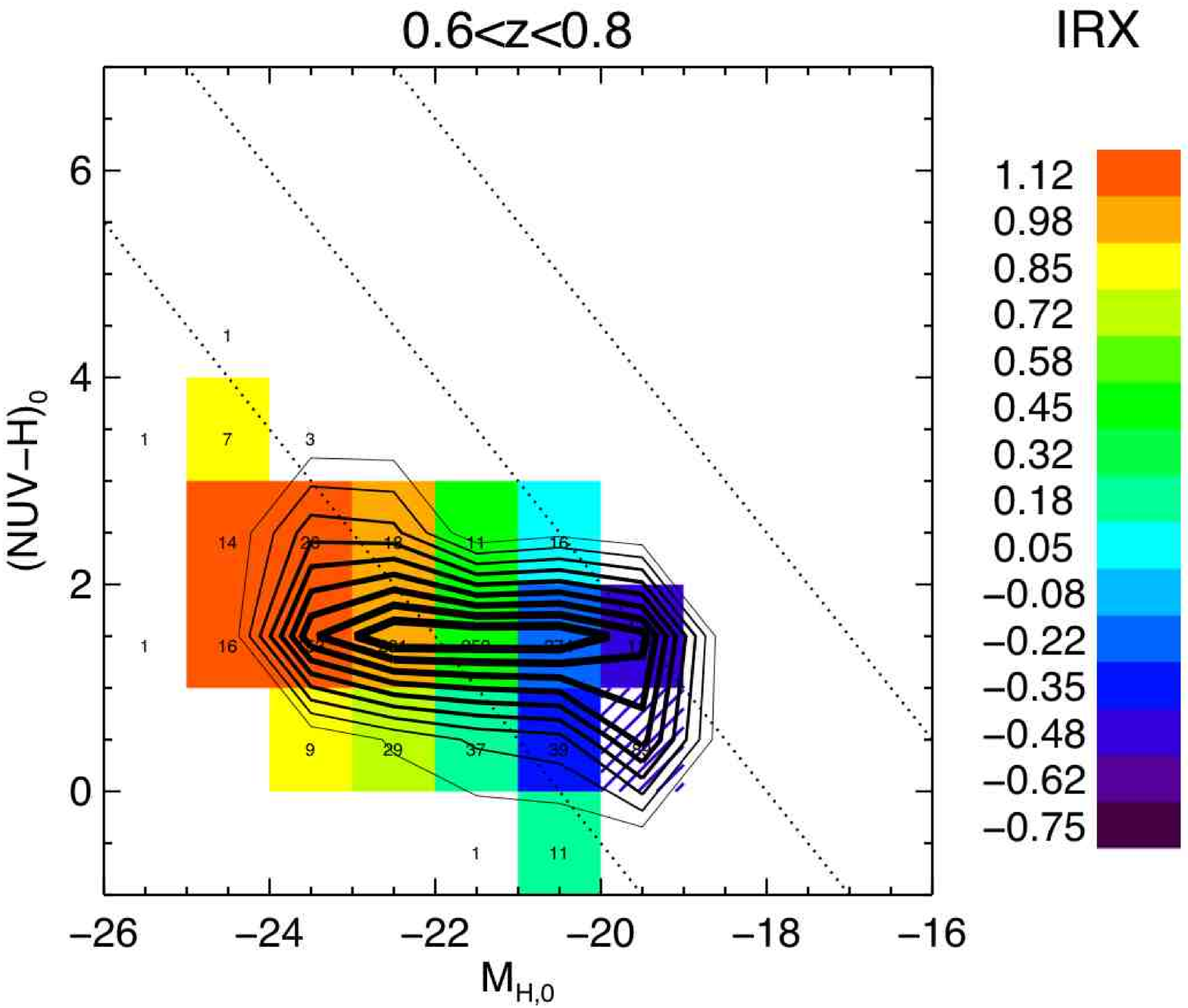}
\plottwo{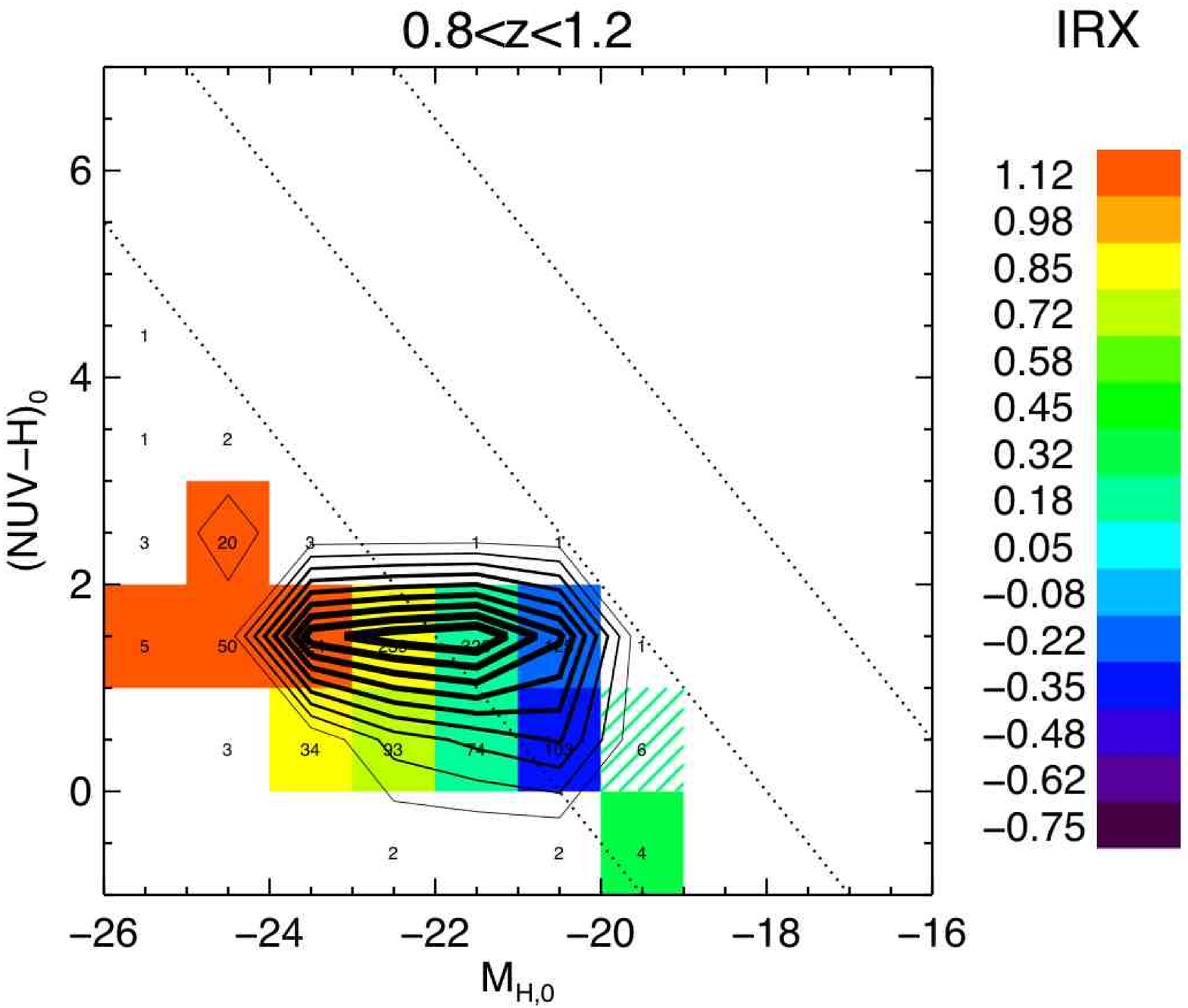}{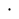}
\caption{Volume corrected bivariate {\it extinction corrected} color-magnitude distribution $\phi(M_{H.0},(NUV-H)_0)$. Contours are
equally spaced in $\log{\phi}$[Mpc$^{-3}$], with 10 divisions from $-4<\log{\phi}<-2$. Colors give IRX, which is the log of the FUV-to-FIR luminosity ratio.
Hashed color bins are upper limits to the IRX. Dotted lines give loci of constant M$_{NUV,0}$=-15.5,-18.0,-20.5. 
\label{fig_cmd_corrected}}
\end{figure}

\begin{figure}
\plottwo{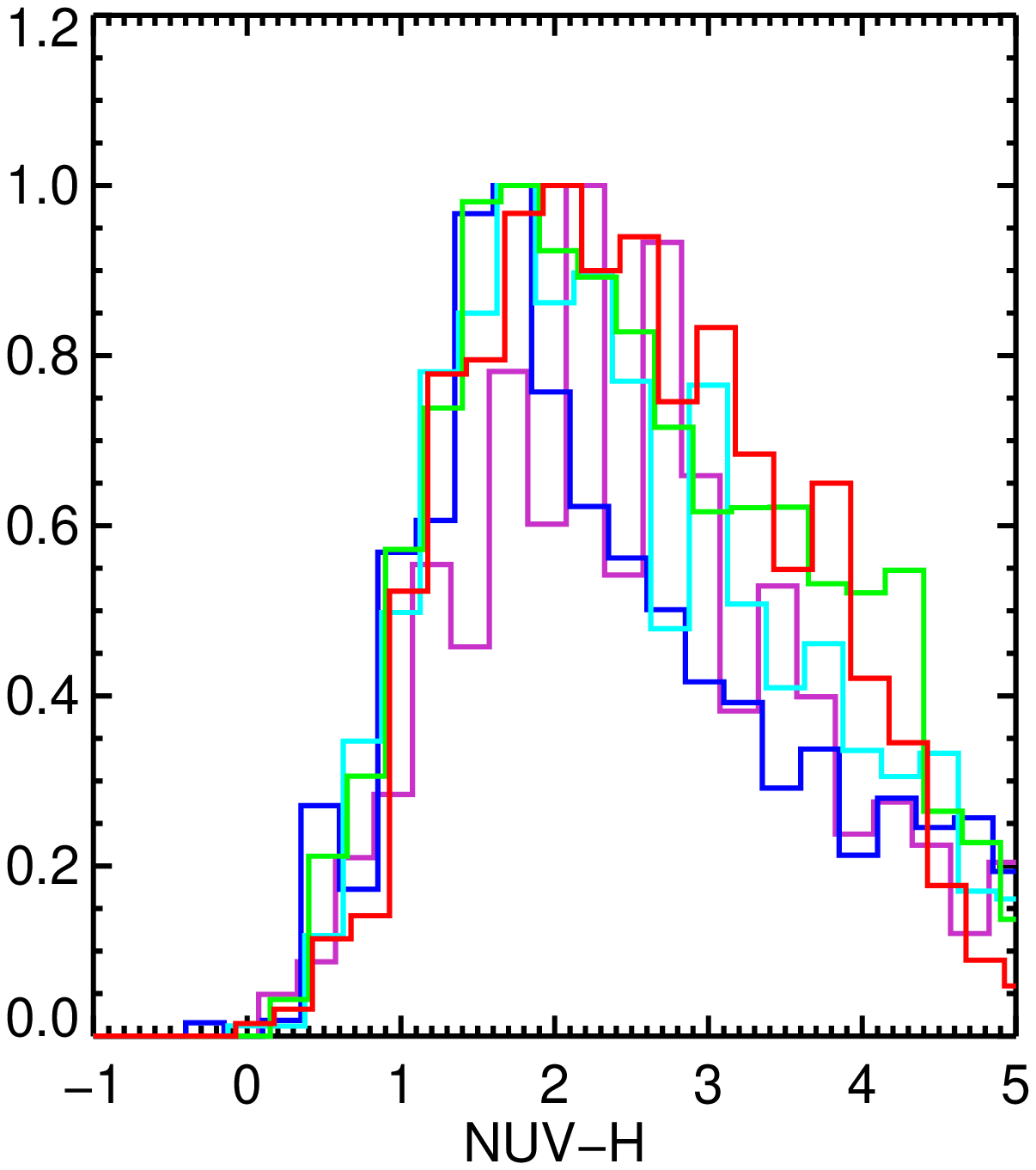}{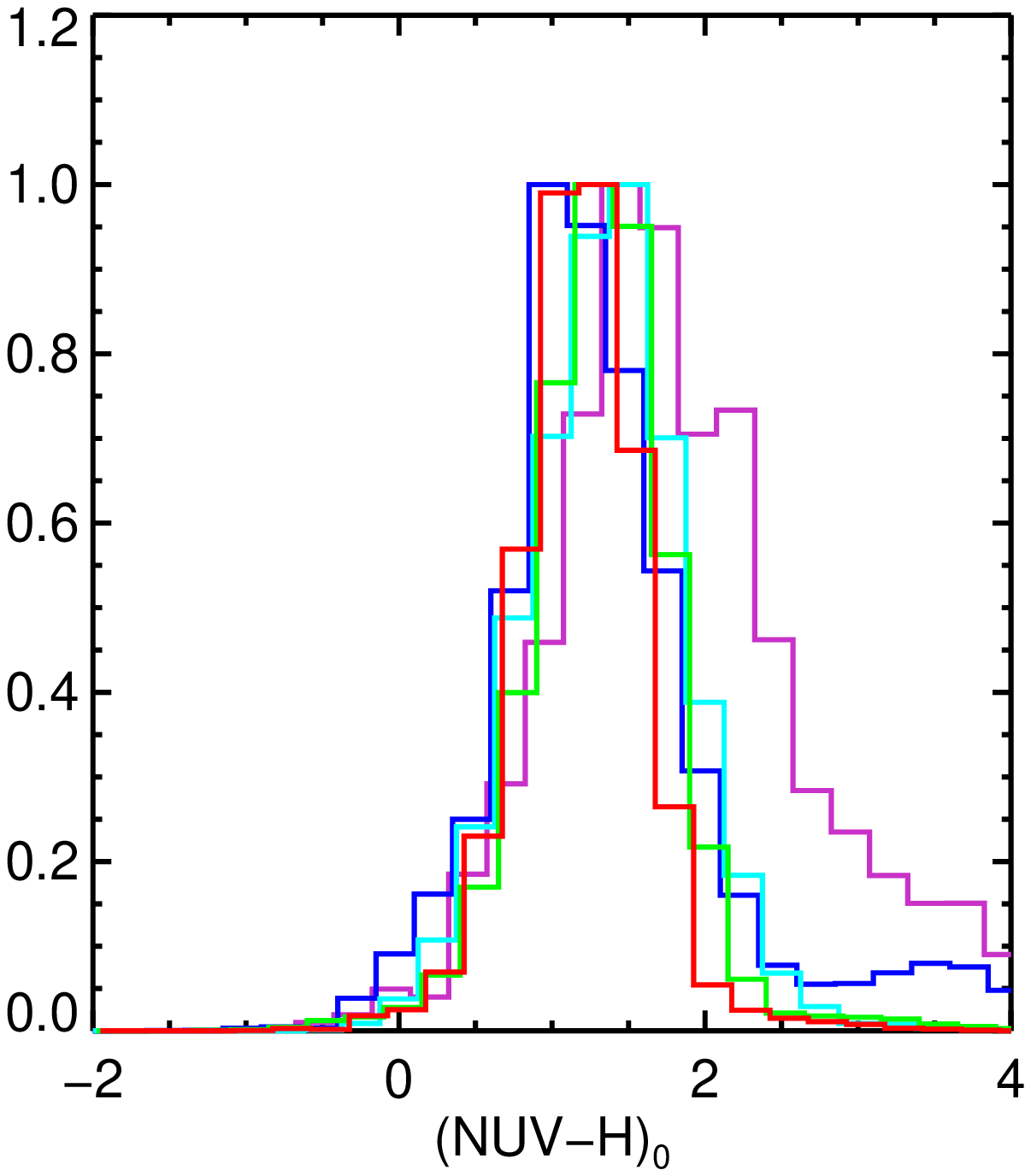}
\caption{Distribution of observed NUV-H (LEFT) and extinction-corrected (NUV-H)$_0$ (RIGHT). Color
gives redshift bin: $0.05<z<0.2$ (purple), $0.2<z<0.4$ (blue), $0.4<z<0.6$ (cyan), $0.6<z<0.8$ (green), and $0.8<z<1.2$ (red).
\label{fig_nuvh_dist}}
\end{figure}

Since the H-band correction is small, the extinction correction is very insensitive
to the extinction law. It enters somewhat if we use L(FUV) to generate IRX$_{FUV}$
and use this to correct (NUV-H). Even if there is evolution in the extinction
law, we have found that using IRX$_{NUV}$,
gives very similar results to those using IRX$_{FUV}$. 
The volume-corrected distribution of extinction-corrected magnitudes $\phi$(\mh$_0$,[NUV-H]$_0$) vs. redshift
are given in Figure \ref{fig_cmd_corrected}. We show in Figure \ref{fig_nuvh_dist}
the uncorrected and corrected distribution in NUV-H.

\subsection{Stellar Mass and Specific Star Formation Rate \label{section_ssfr}}

\begin{figure}
\plottwo{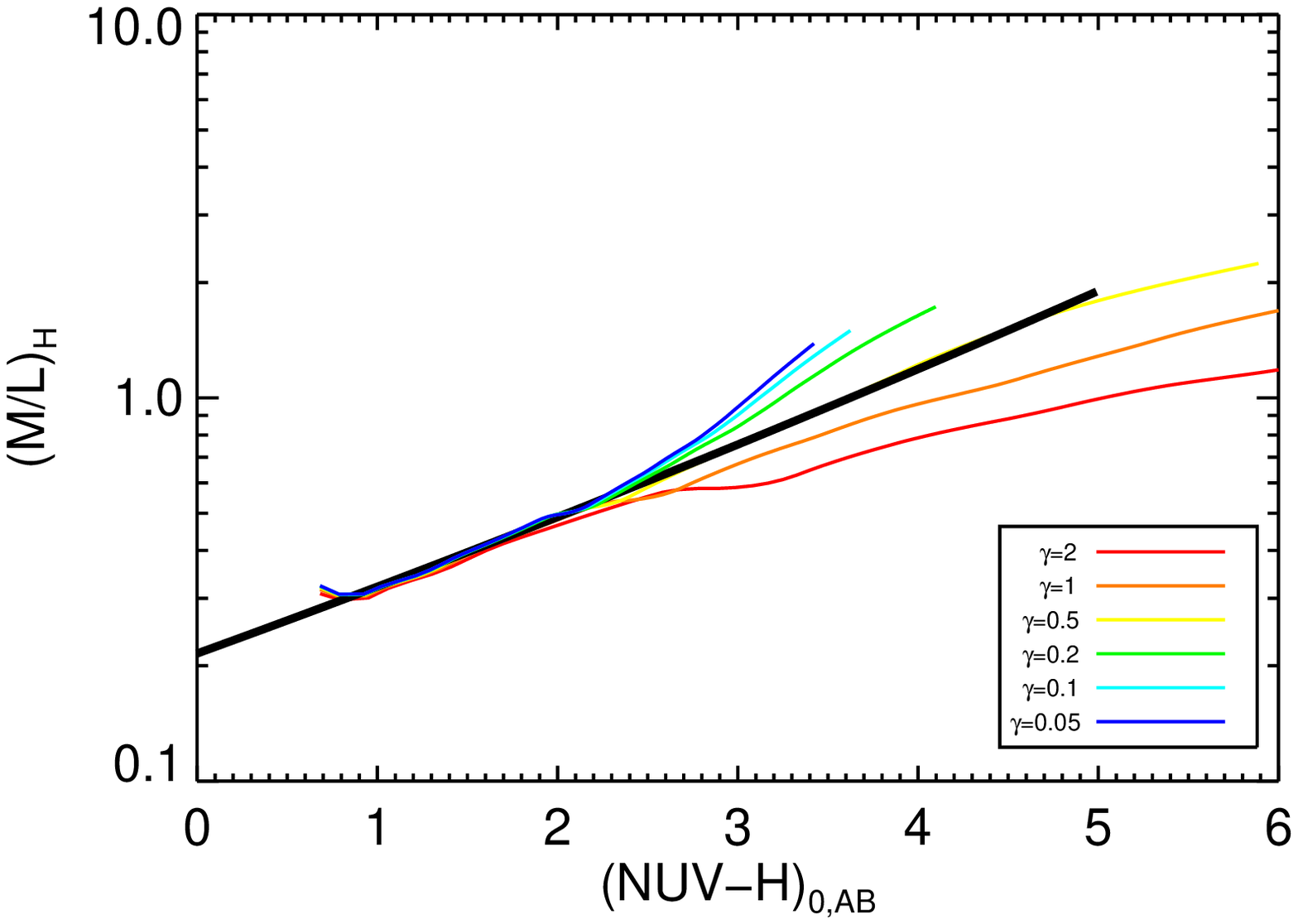}{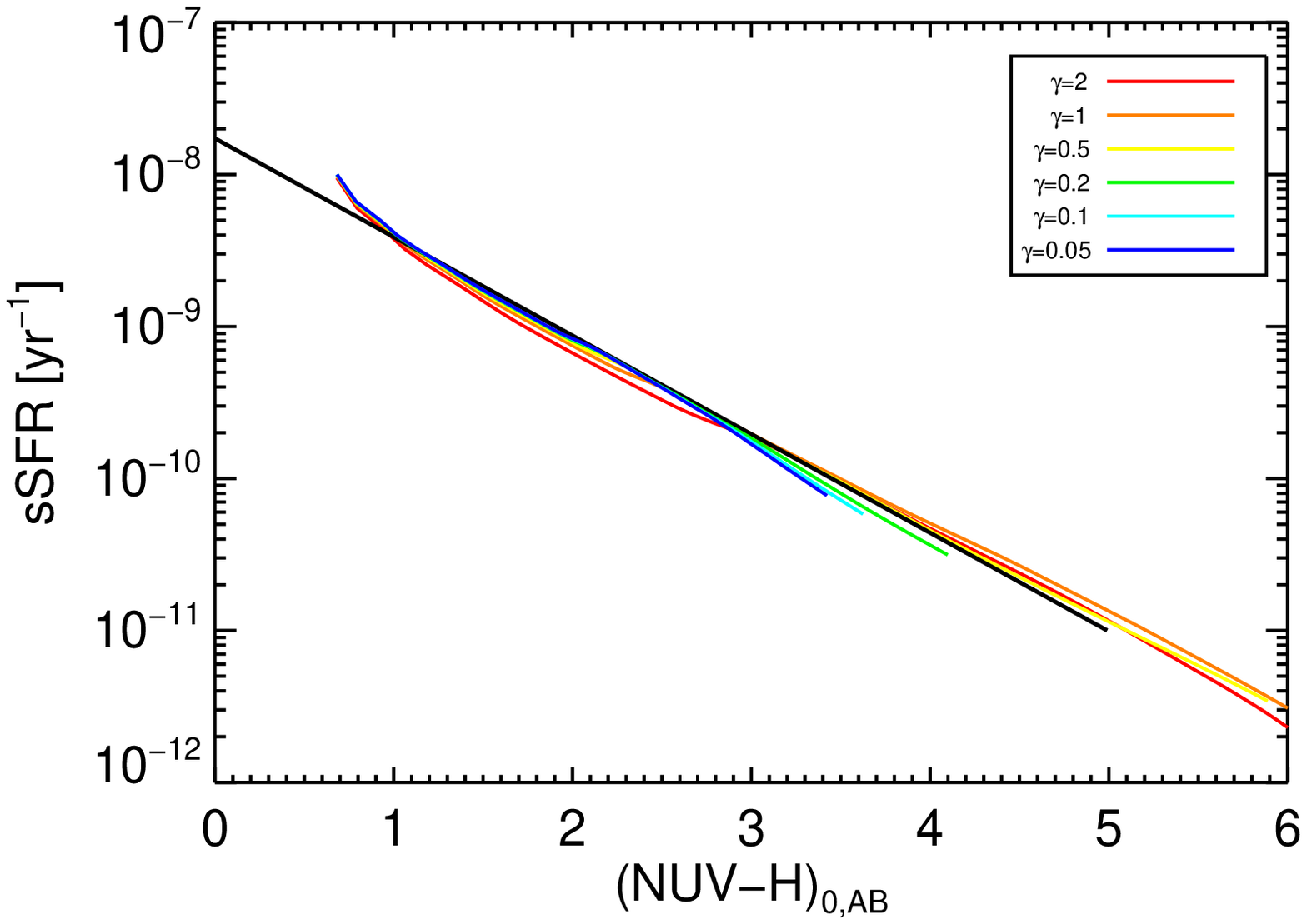}
\caption{LEFT: Mass-to-light ratio at H-band vs (NUV-H) for exponentially declining star formation histories $e^{-\gamma t}$,
obtained using model of \cite{bruzual03}. RIGHT: Specific star formation rate (SSFR) for exponentially declining star formation histories.
\label{fig_sed_model}}
\end{figure}

We derive stellar mass $\mathcal{M}$ using extinction corrected rest-frame H-band absolute
magnitude and NUV-H color, following the basic scheme of \cite{bell03}. 
For smooth star formation histories the stellar mass-to-light ratio
is a single parameter function of a measure of the specific
star formation rate such as the rest frame extinction corrected NUV-H color.
This can be seen in Figure \ref{fig_sed_model}, which shows the predicted $\mathcal{M/L}$
vs. (NUV-H)$_0$ for different values of the exponential SFR decay, based
on solar metallicity models of \cite{bruzual03} and a (standard, non-diet) Salpeter
initial mass function\footnote{Note that all derived stellar masses and star formation rates can
be converted to the ``Diet Salpeter'' IMF of \cite{bell03} by multiplying
by 0.7 Specific star formation rates are unaffected.}.  There is
almost no dependence on the star formation history for (NUV-H)$_0$ for NUV-H$<$2.5,
where the bulk of the extinction-corrected galaxies fall. We use this parabolic fit:

\begin{equation}
\log_{10} (\mathcal{M/L}) = -0.667 + 0.17 (NUV-H)_0 + 0.00373 (NUV-H)_0^2
\end{equation}

We have also tested more complex star formation histories in which starbursts
become significant. These models produce the same general trends
between $\mathcal{M/L}$ and NUV-H color, with some dispersion. 
There is no significant impact on the results described below.

We derive the star formation rate from the extinction-corrected FUV luminosity
using $SFR=1.4 \times 10^{-28} \mathcal{L}_\nu (1500)$ \citep{kennicutt98}.
We obtain the specific star formation rate (SSFR) by dividing by
the stellar mass.

We also note that the specific star formation rate (SSFR) is tightly correlated with (NUV-H)$_0$ and
independent of decay timescale,
as can be seen in Figure \ref{fig_sed_model}.
We can also use the following linear fit to convert (NUV-H)$_0$ to SSFR.

\begin{equation}
\log_{10} (SSFR [yr^{-1}] ) = -7.8 - 0.65 (NUV-H)_0 
\end{equation}
Again, either technique for computing the SSFR produces essentially identical results.

\begin{figure}
\plottwo{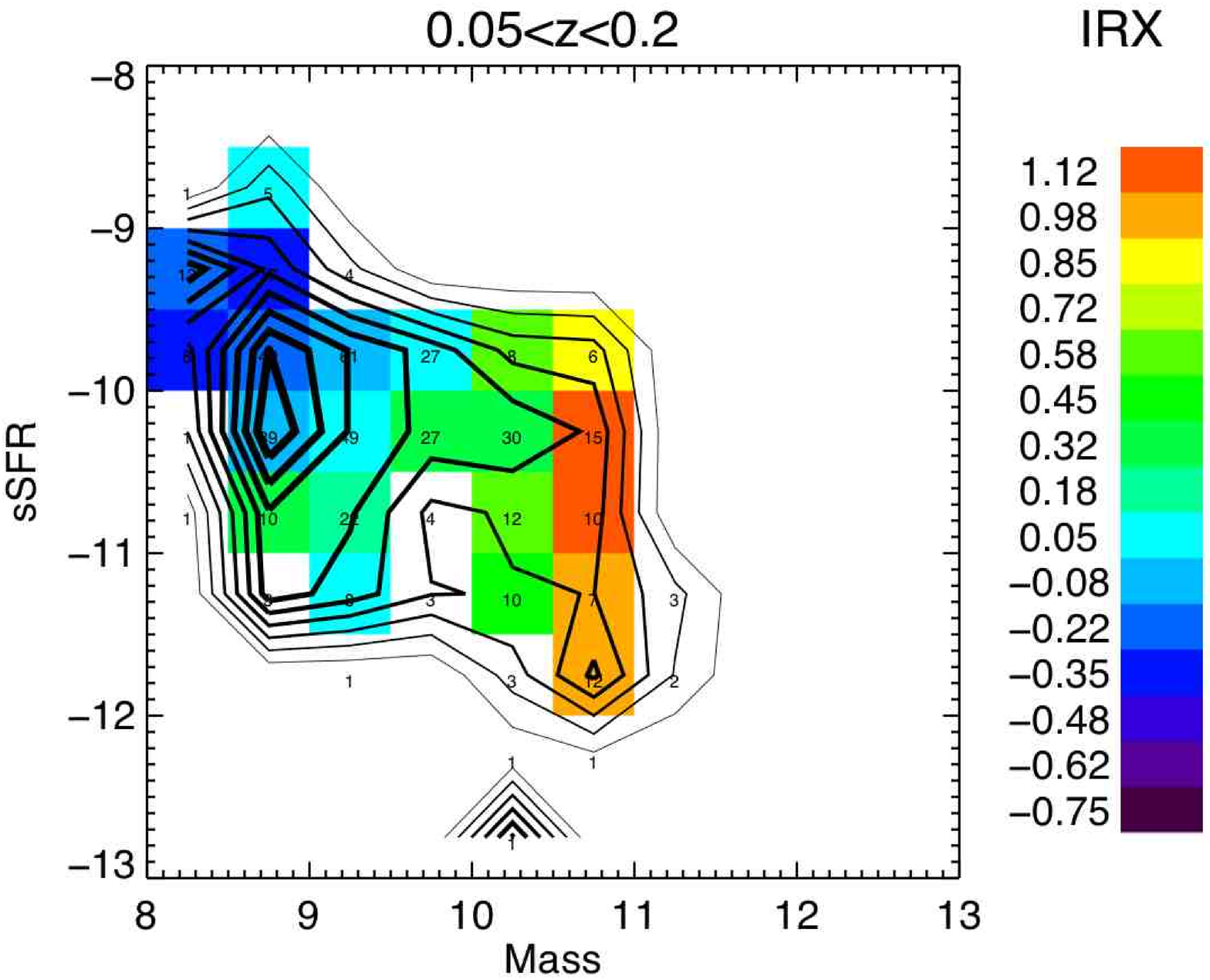}{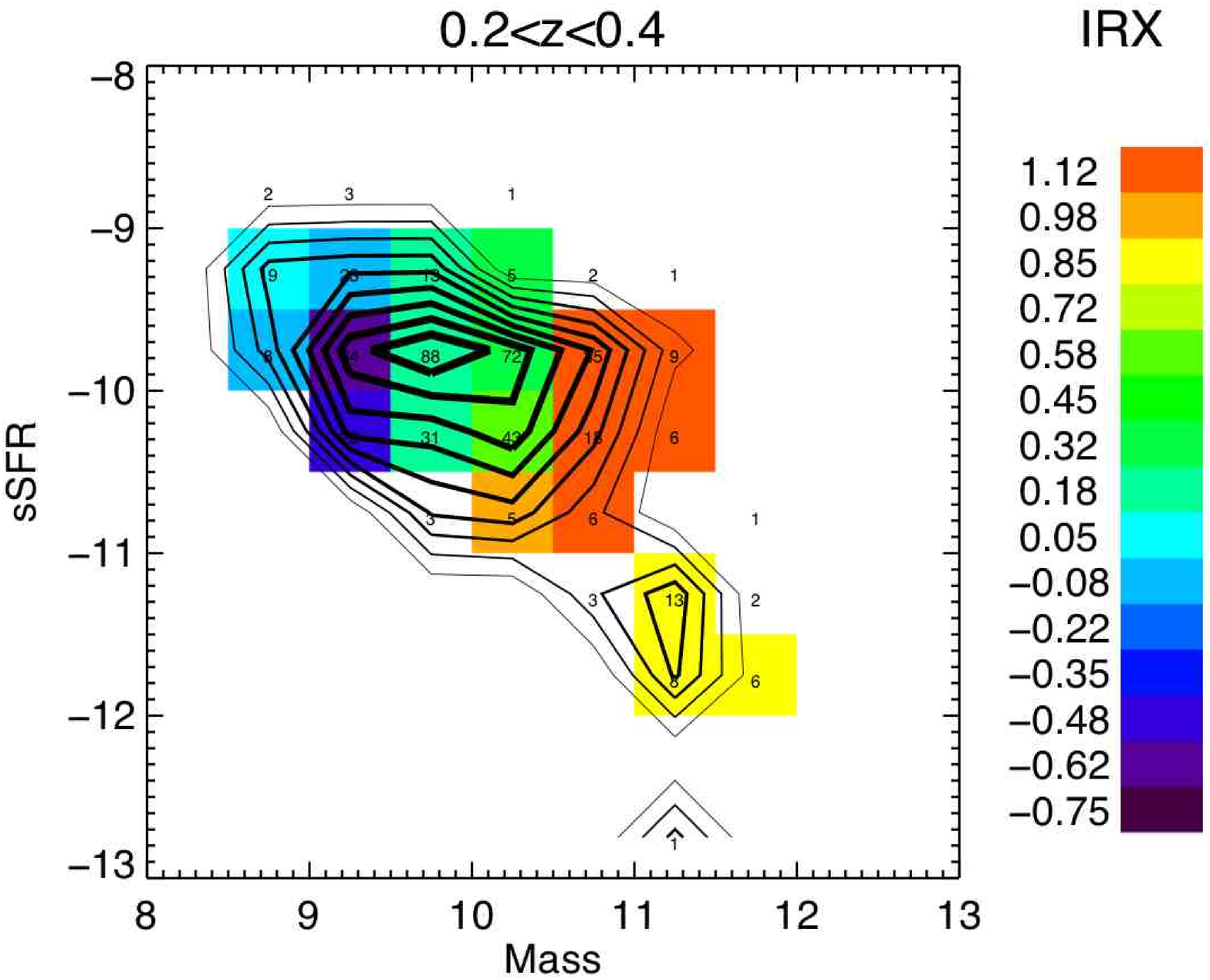}
\plottwo{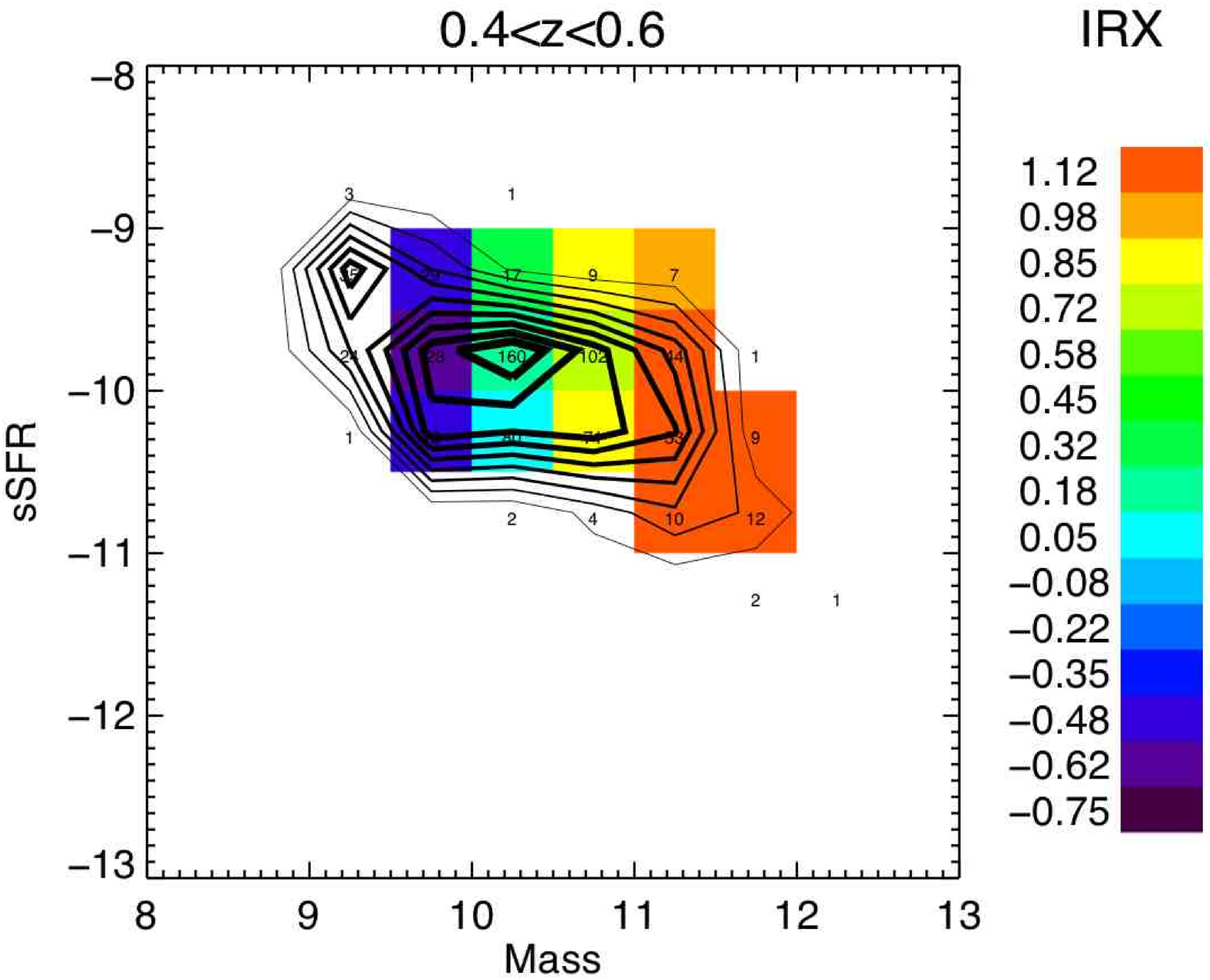}{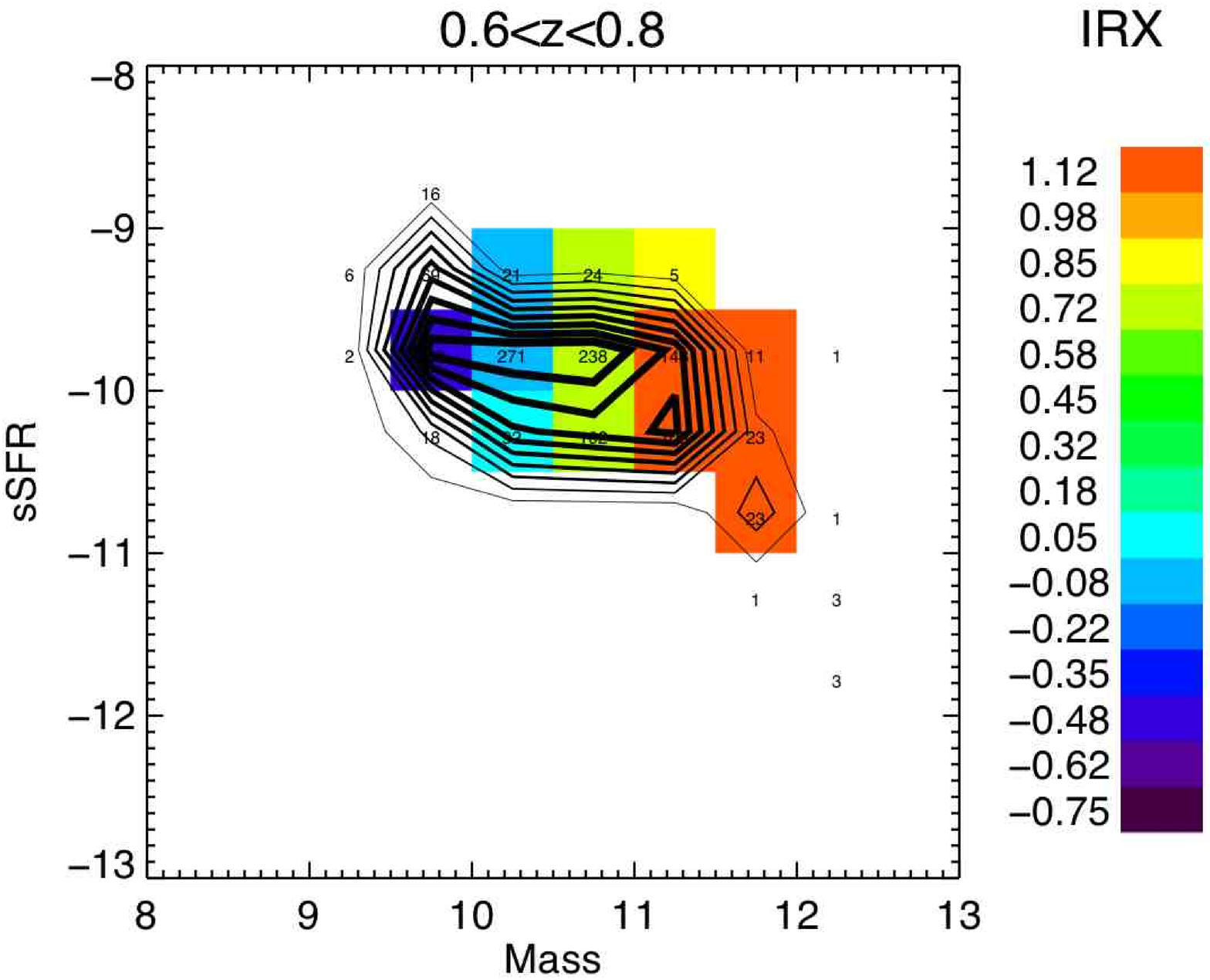}
\plottwo{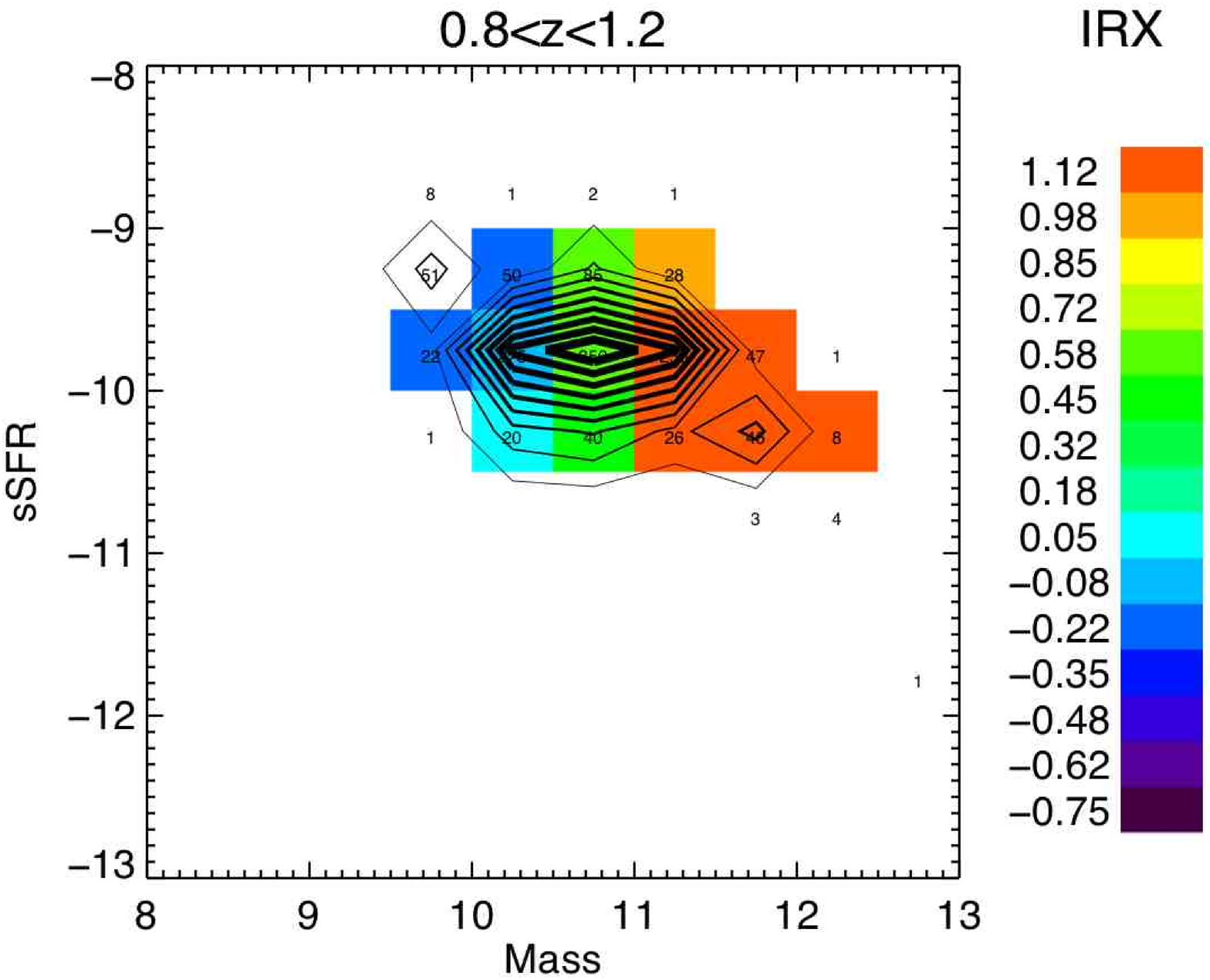}{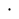}
\caption{Volume corrected bivariate Mass-SSFR distribution $\phi(M, SSFR)$. Contours are
equally spaced in $\log{\phi}$ [Mpc$^{-3}$], with 10 divisions from $-4<\log{\phi}<-1$. Colors give IRX, which is the log of the FUV-to-FIR luminosity ratio.
\label{fig_ssfr_vs_mass}}
\end{figure}

We derive the stellar mass and SSFR for each galaxy by correcting its
rest-frame \mh, M$_{1500}$,  and NUV-H for the (bin-averaged) extinction.
Using these distributions
we derive the volume-corrected  bivariate \mass-SSFR distribution
$\phi$($\mathcal{M}$,SSFR,z) in
the same fashion as the volume-corrected color-magnitude distribution.
We also generate an average IRX in each bin of the \mass-SSFR distribution.
We calculate the mean log infrared excess in each bin,
using the IRX obtained in the previous section. These distributions
are displayed in Figure \ref{fig_ssfr_vs_mass}.

\subsection{Errors}

We use the bootstrap method \citep{bootstrap} to derive errors to all bivariate
distributions discussed above as well as the average
distributions discussed in the next section. 
Specifically,
in each redshift bin we randomly select objects, with replacement,
until we have the same number of objects found in that redshift bin.
We then proceed to determine the color-magnitude distribution,
the mean 24 micron flux in each color-magnitude bin by stacking
this new sample, the corrected CMD, the Mass-SSFR distribution, and the mean
IRX in each Mass-SSFR bin (cf. \S\ref{sec:avgirx}). Of order one hundred trials are used
to generate a standard deviation in each bin of every distribution calculated.
Such errors will not, however, account for cosmic variance due to
large scale structure. (cf. \S\ref{sec:sfhist}).

\section{Results}

\subsection{Infrared Excess vs. Stellar Mass}

We begin by examining the trends in the NUV-H, \mh CMD.
At a fixed \mh redder galaxies have a higher IRX.
In general, the blue sequence shows a significant tilt in the color-magnitude diagram,
much of which appears to be produced by this extinction-luminosity relationship.
Much of the color width of the blue sequence, which leads some authors to
refer to it as the ``blue cloud'', is also
produced by variance in extinction \citep{wyder06}, some of which is
simply due to inclination variations \citep{martin07a}.
Extinction correction produces a much tighter distribution
in the CMD, as we see in Figure \ref{fig_cmd_corrected} and Figure \ref{fig_nuvh_dist}.
The trend of increasing IRX with H-band
luminosity is even more apparent in the extinction-corrected CMD.
Consequently, there is a strong increase in IRX with stellar mass as is expected
from the trend in the CMD. This trend 
can be clearly detected in Figure \ref{fig_ssfr_vs_mass}. This trend persists
in all redshift bins. 

\subsection{Evolution of the Bivariate CMD}

There is clear evolution in the NUV-H, \mh color-magnitude
diagram, in the sense that the density of H-band luminous
galaxies is increasing with redshift.
This is consistent with the increase in characteristic
UV luminosity \citep{schiminovich05}
and B luminosity \citep{bell04}. 
As expected from the previous section,
this is accompanied by an increase in the contribution
from higher IRX galaxies.   The evolutionary trend is even easier to discern
in the extinction corrected CMD, Figure \ref{fig_cmd_corrected}.

\subsection{Infrared Excess vs. Stellar Mass and Redshift}
\label{sec:avgirx}
\begin{figure}
\plottwo{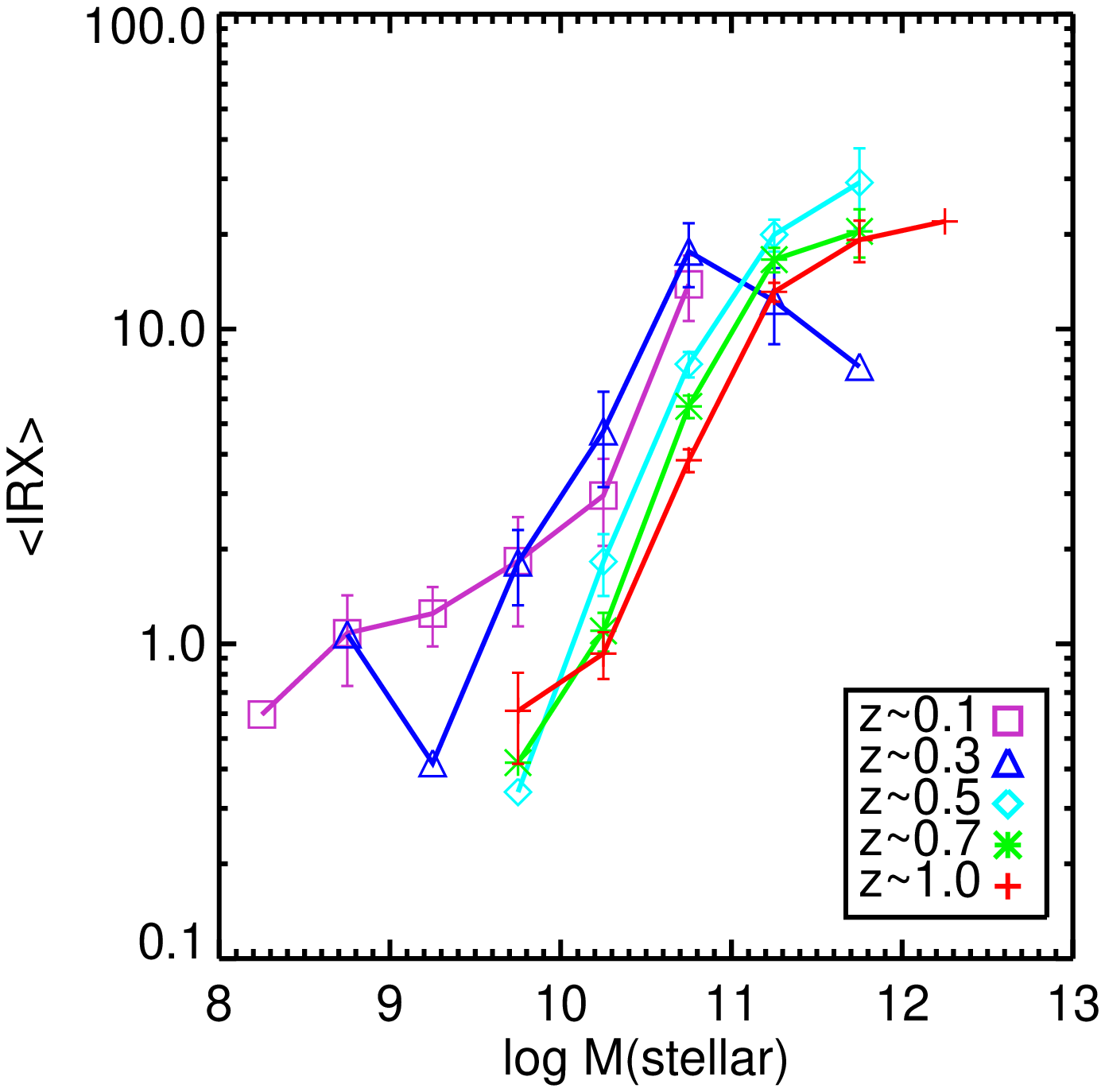}{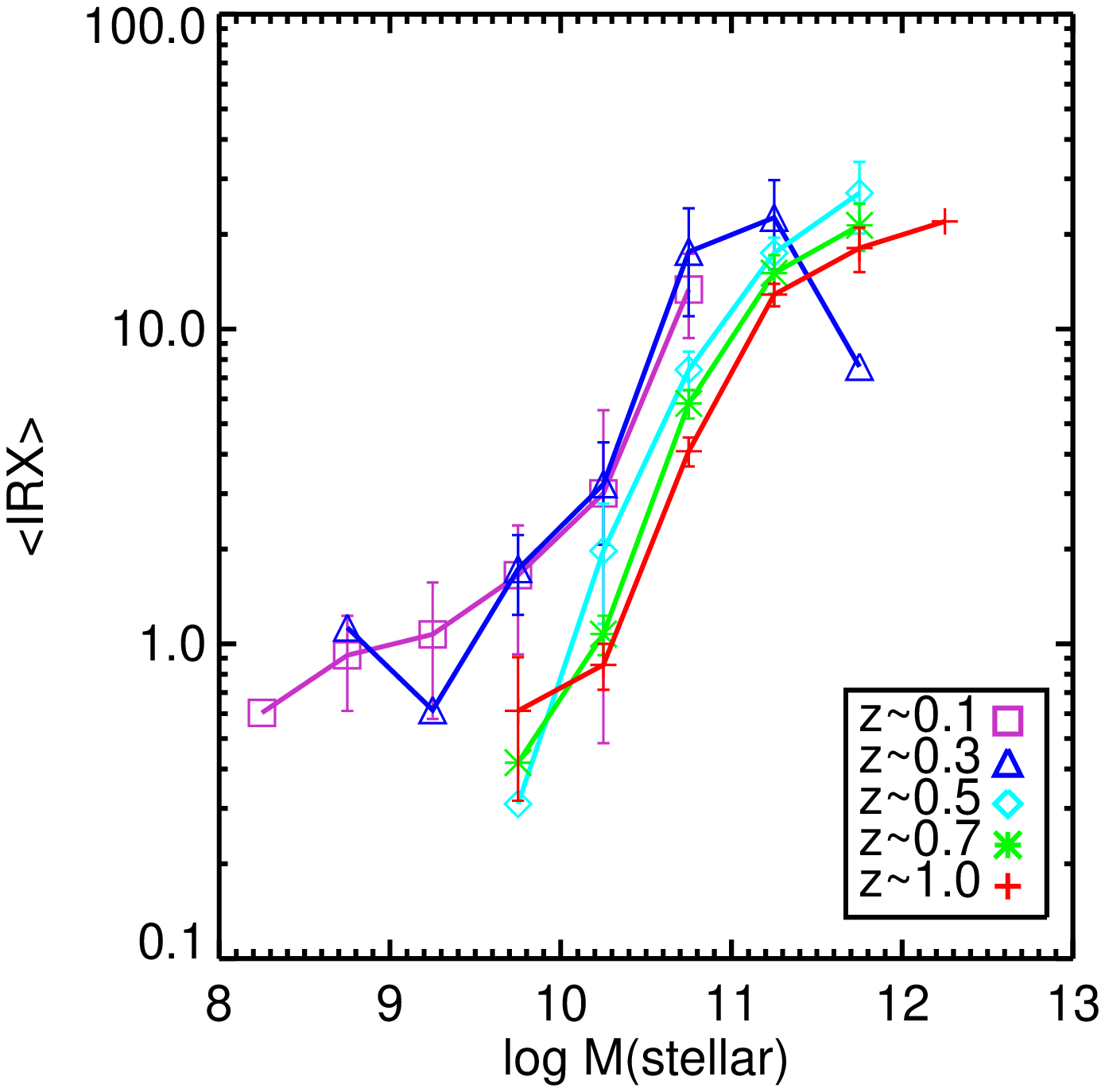}
\caption{Average IRX \irx~ (linear version plotted on a logarithmic scale) vs. mass in each redshift bin. LEFT: numerical average. RIGHT: SFR-weighted average.
Errors are derived from bootstrapping.
\label{fig_irx_mass}}
\end{figure}

In order to further explore IRX-mass relationship and its evolution
we derive the average IRX in each mass and redshift bin.
We have calculated this average using the number density \irx(\mass)
and weighted by the star formation rate \irx$_{SFR}$(\mass).
The mass trend in redshift bins is shown in Figure \ref{fig_irx_mass}.
The average IRX increases sharply with mass up to a critical
mass. The slope in the in the IRX-log mass
relation is greater than one. The critical mass is lower at low
redshift, $\simeq 10.5$ at z$\sim$0.3, but
appears to move to higher mass at higher redshift, with
 $\sim11.5$ at z$\sim$1.

\begin{figure}
\plottwo{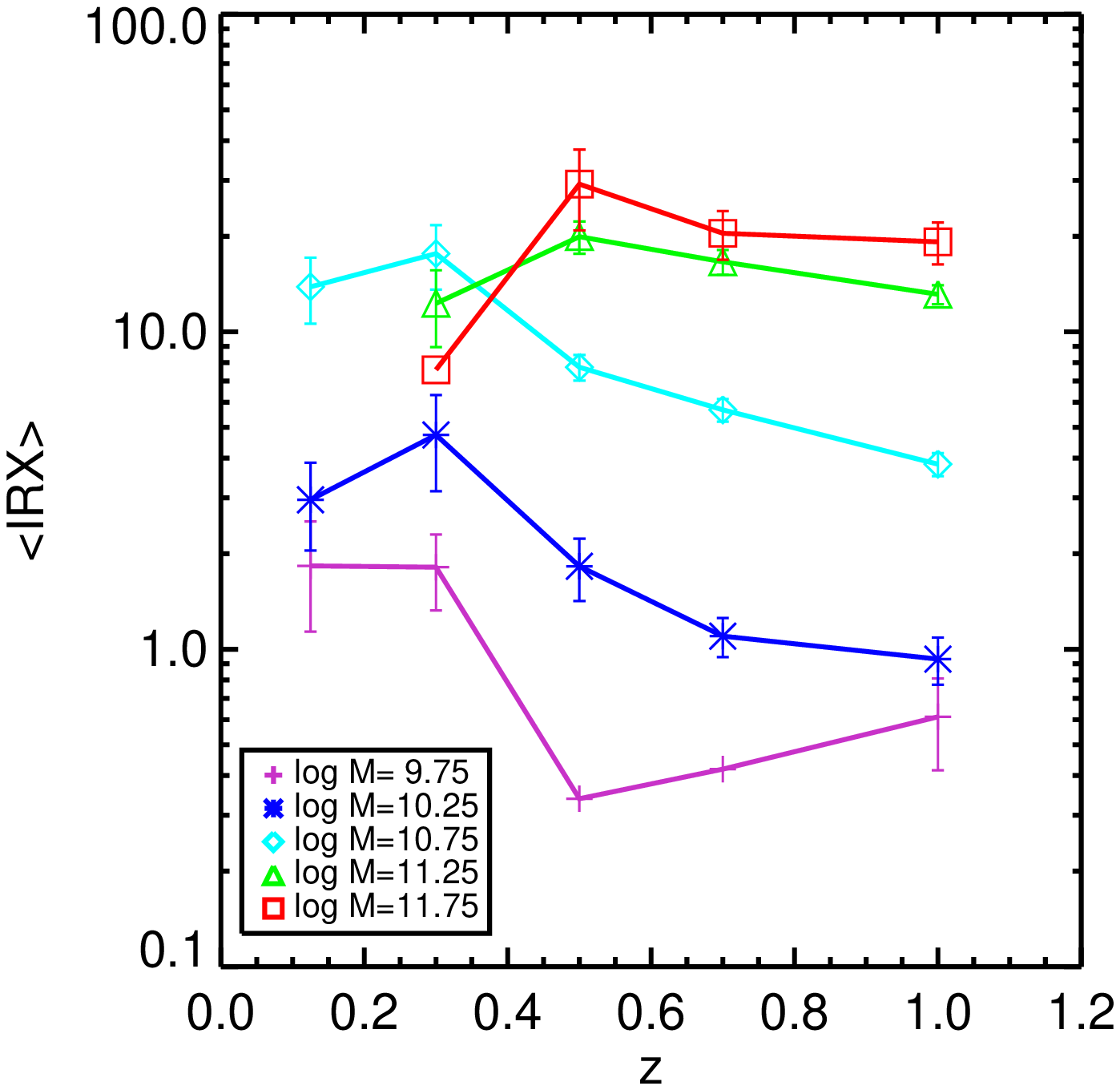}{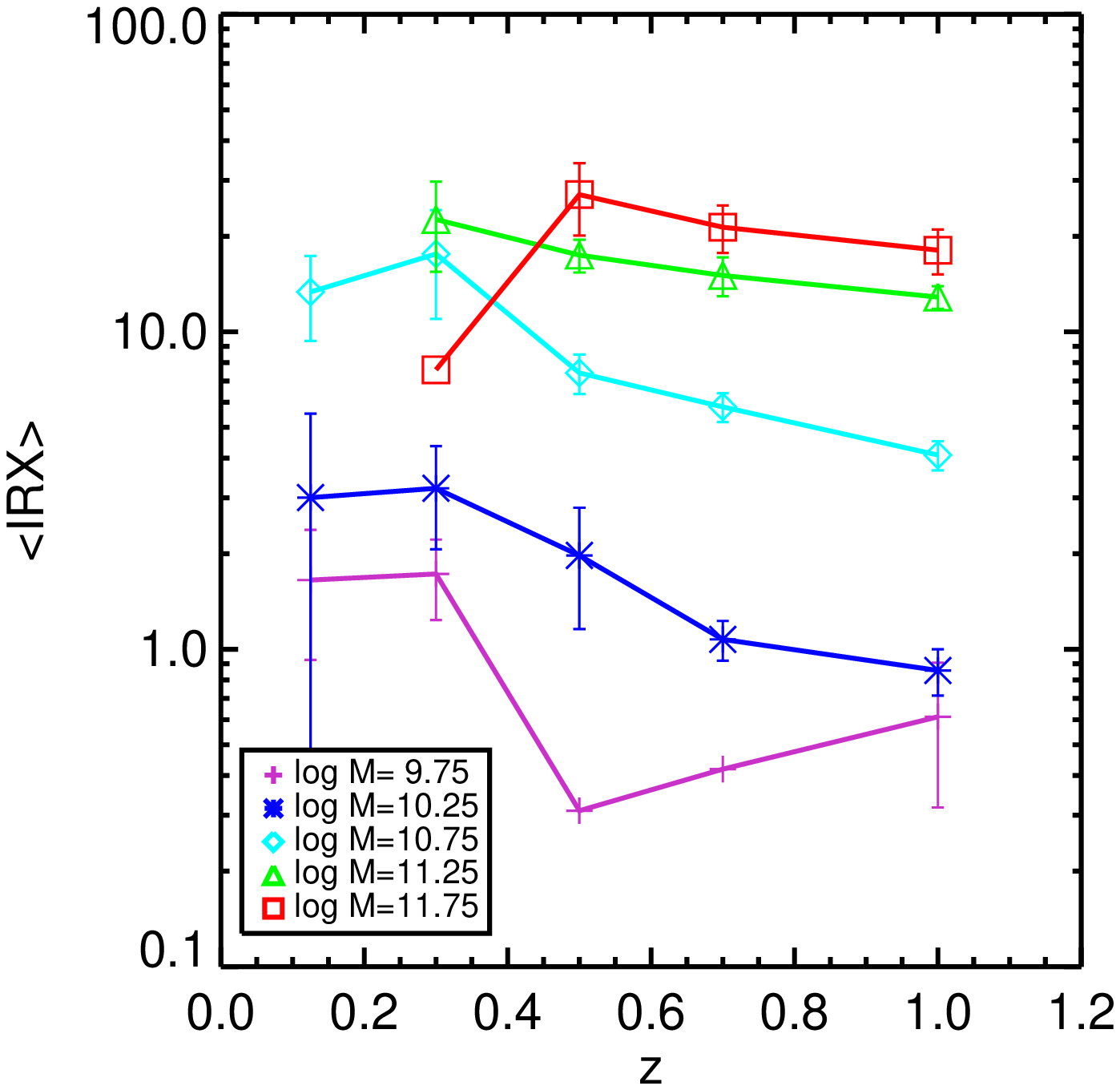}
\caption{Average IRX \irx ~ (linear version plotted on a logarithmic scale) vs. redshift in each mass bin. LEFT: numerical average. RIGHT: SFR-weighted average.
Errors are derived from bootstrapping.
\label{fig_irx_z}}
\end{figure}

The redshift trend in mass bins is shown in Figure \ref{fig_irx_z}.
In the highest mass bin with good redshift coverage ($\log M_{\rm crit}=11.5$), IRX
increases slowly with time then sharply decreases for z$<$0.5.
In the lowest mass bin, IRX appears to increase with time to the lowest
redshift bin. These trends appear in both the number and SFR-weighted
average IRX. For our subsequent analysis we use the number-weighted average.

\subsection{Co-evolution of SFR and IRX}
\label{sec:avgssfr}
\begin{figure}
\plottwo{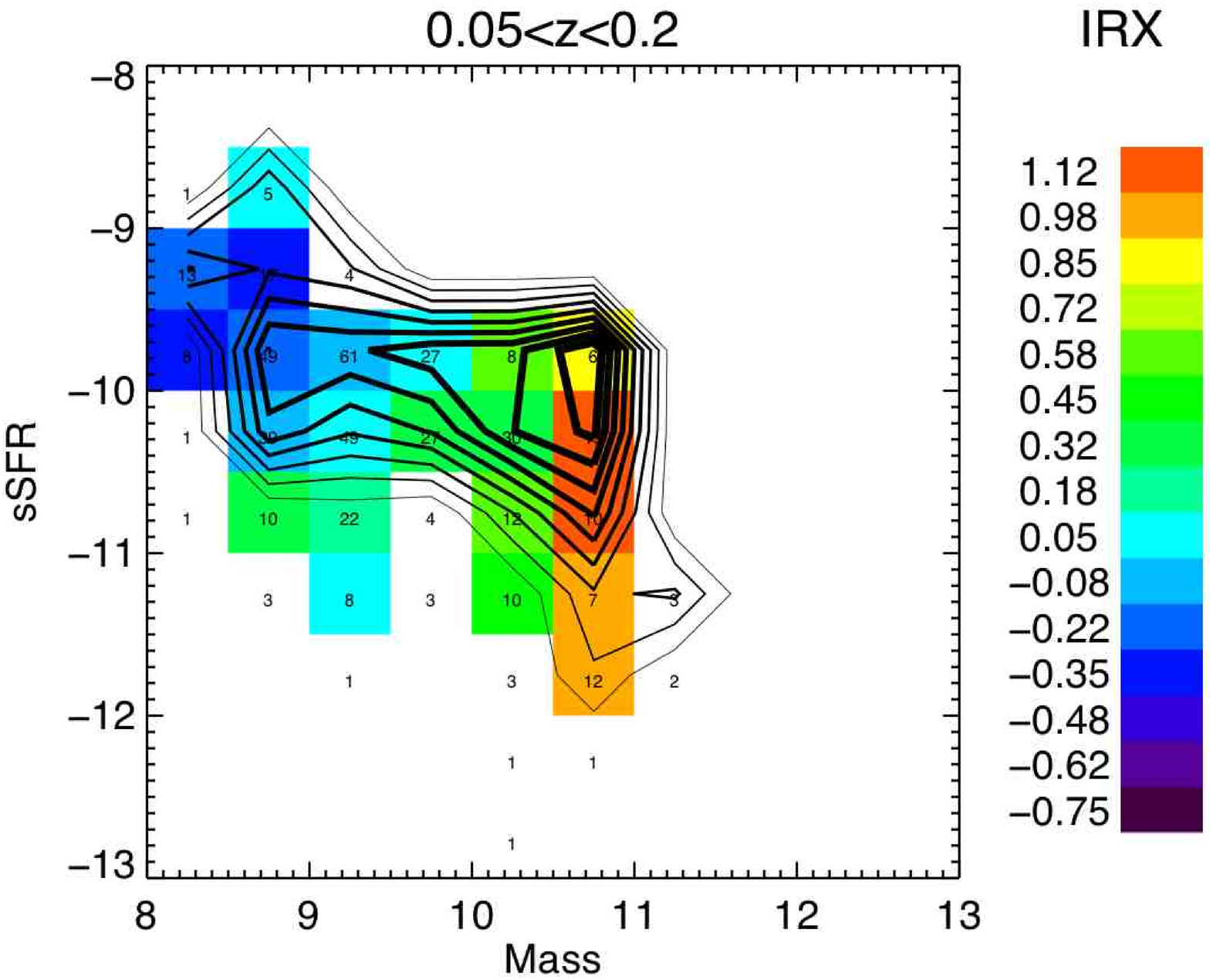}{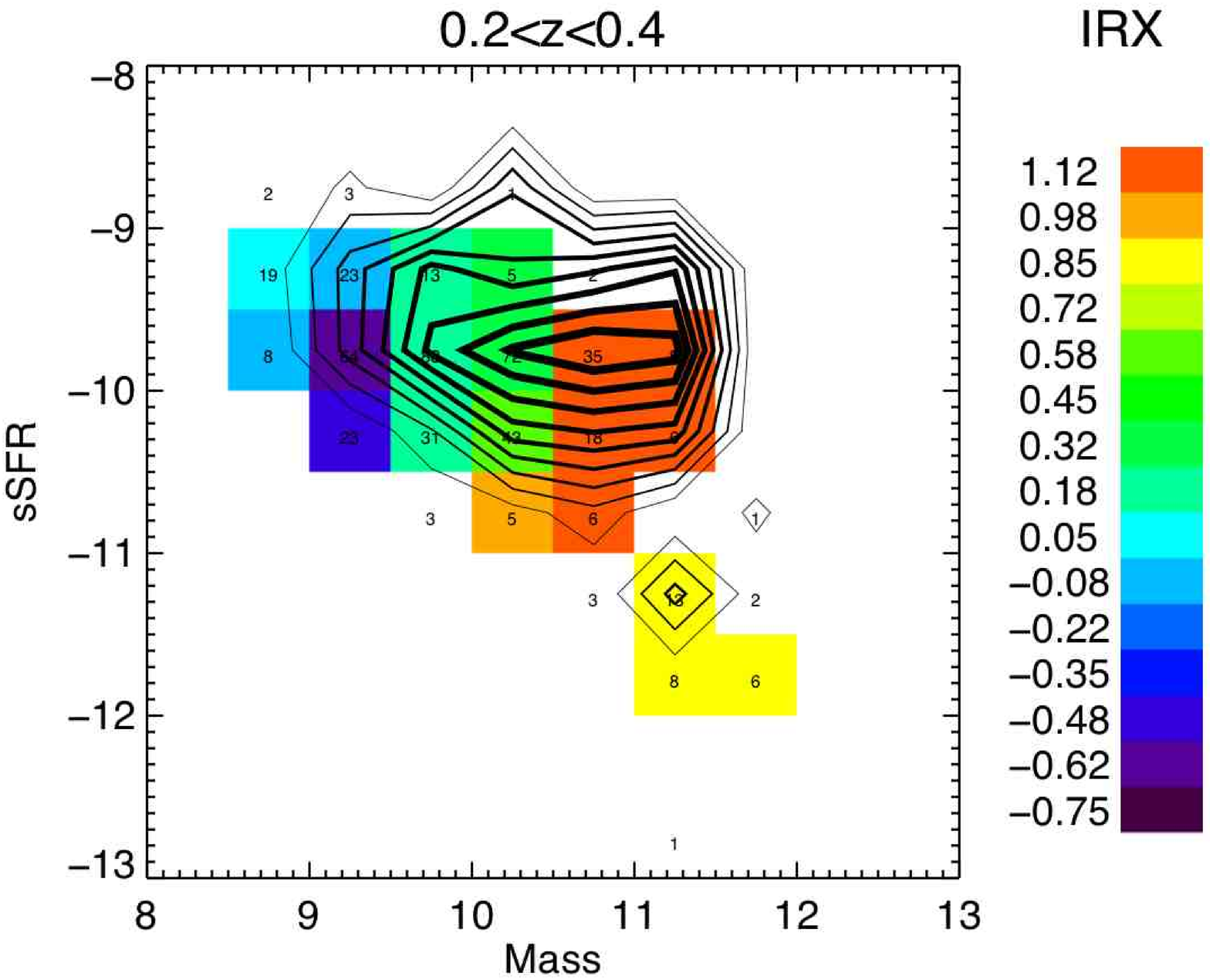}
\plottwo{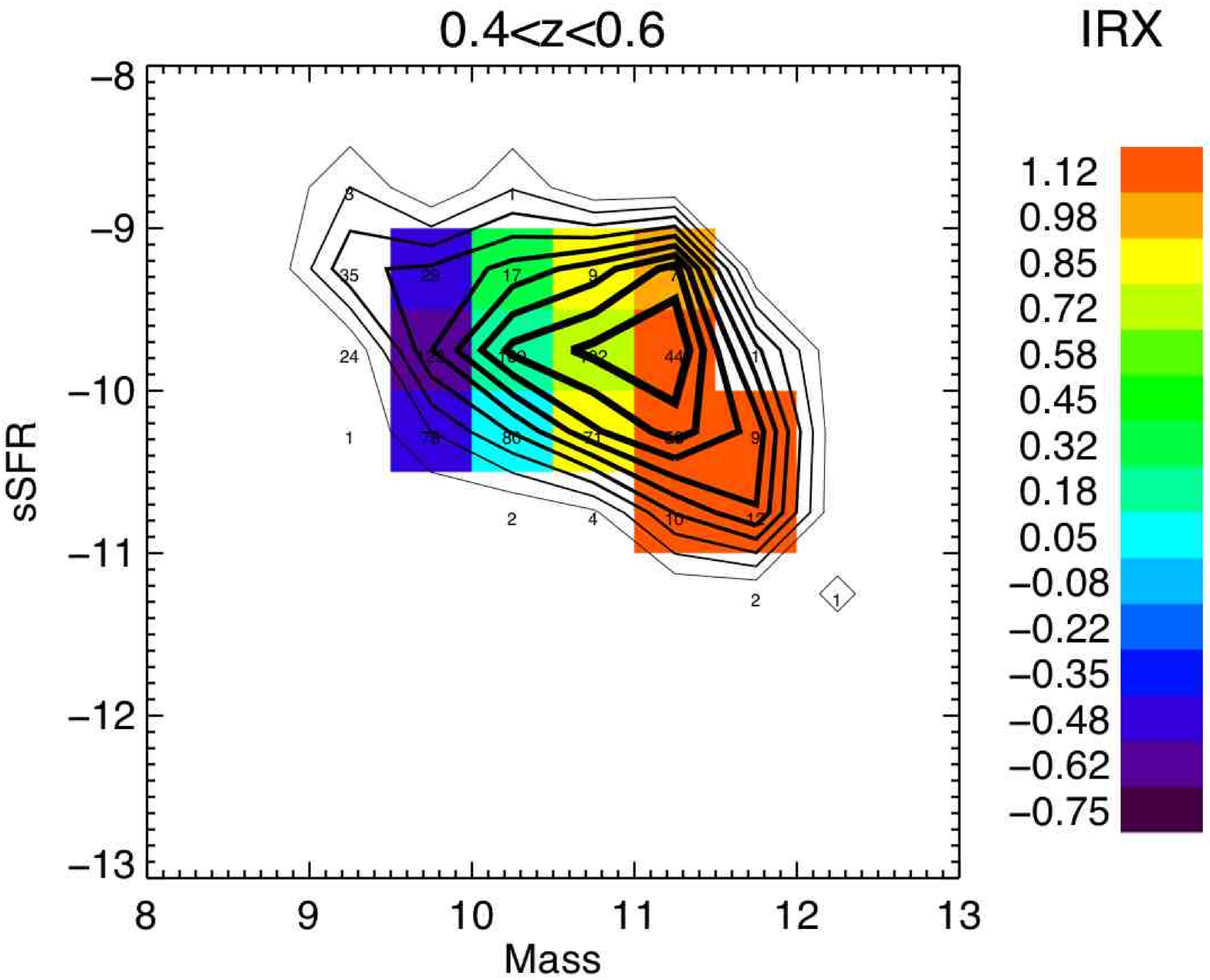}{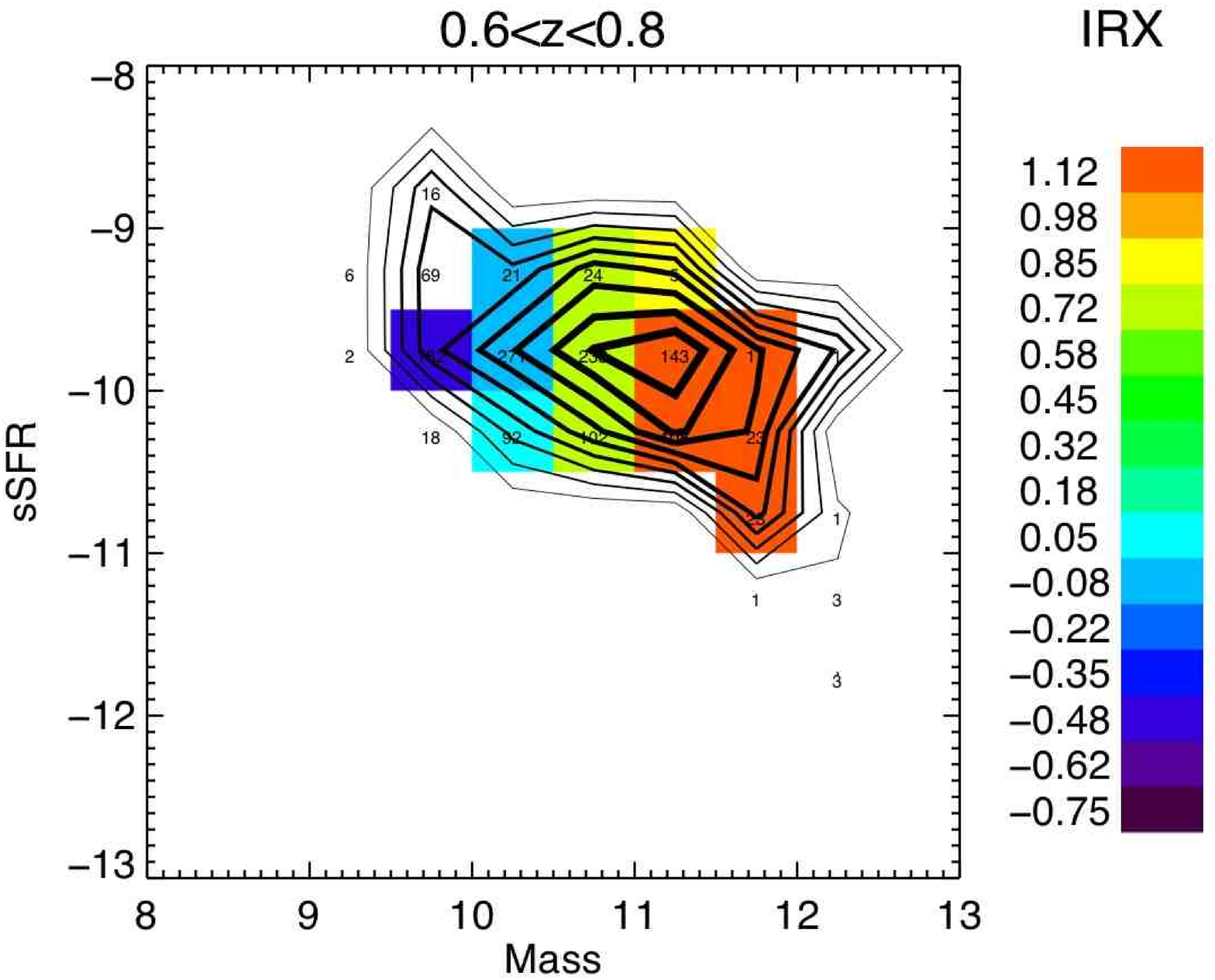}
\plottwo{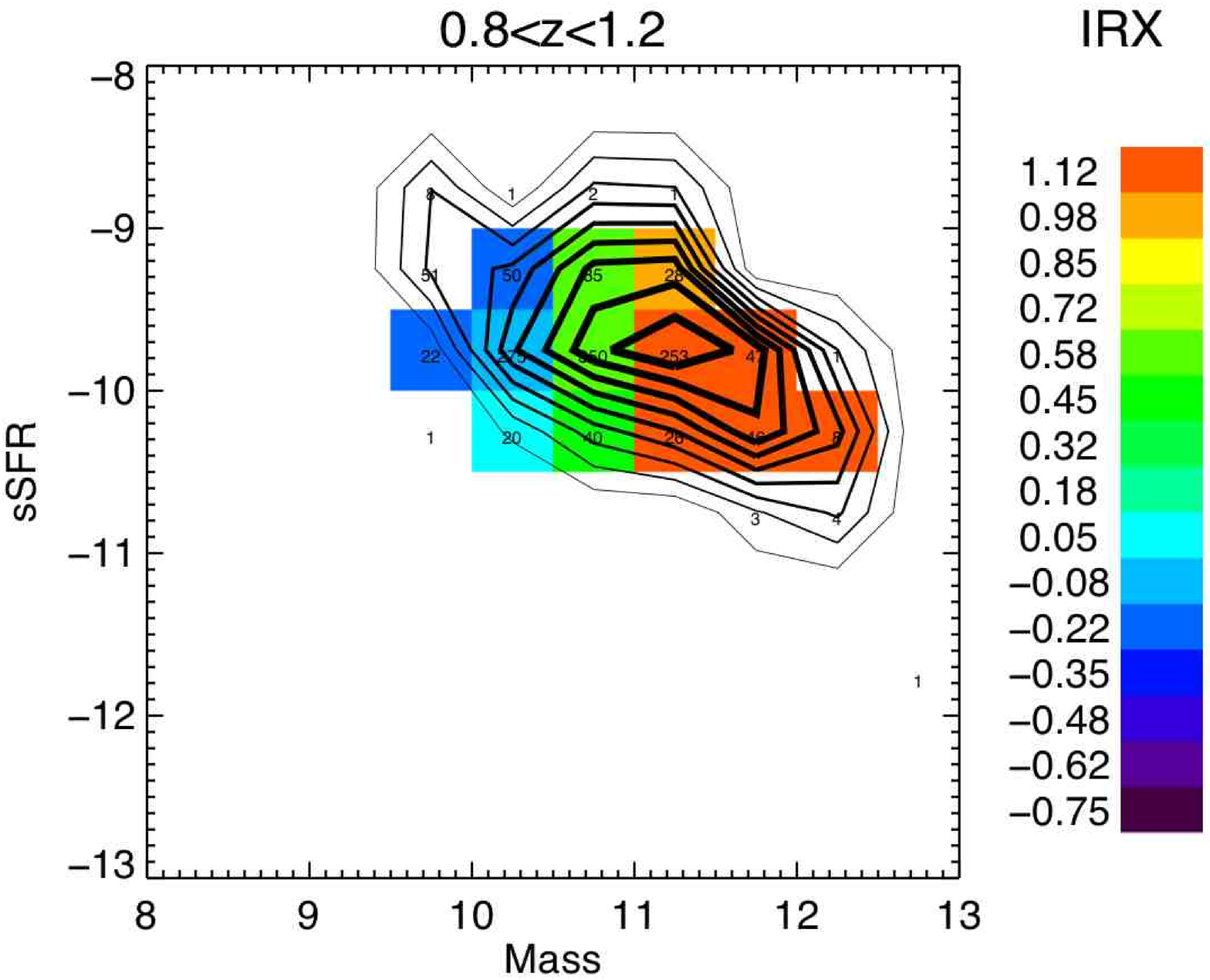}{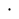}
\caption{Volume corrected bivariate Mass-SSFR distribution weighted by the SFR,  $SFR * \phi(M, SSFR)$. Contours are
equally spaced in $\log{SFR*\phi} $ [M$_\odot$ yr$^{-1}$ Mpc$^{-3}$], with 10 divisions from $-4<\log{SFR*\phi}<-1$. Colors give IRX, which is the log of the FUV-to-FIR luminosity ratio.
\label{fig_ssfr_vs_mass_sfr}}
\end{figure}

\begin{figure}
\plotone{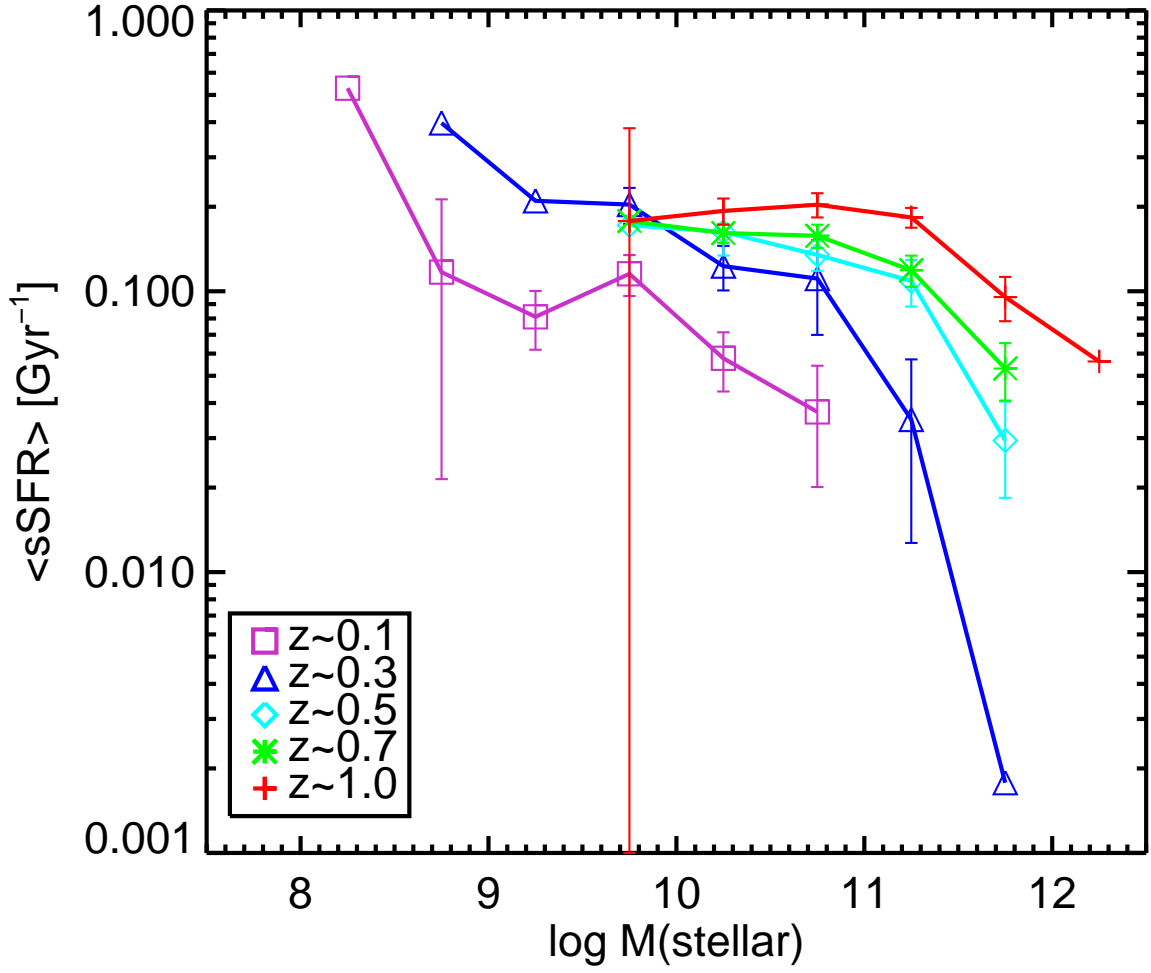}
\caption{Average specific star formation rate (\ssfr) in each mass and redshift bin. Errors are derived from bootstrapping.
\label{fig_ssfr_phi}}
\end{figure}

We have seen that the star formation rate density is moving to
higher masses at higher redshift. This can best be seen in the
SFR-weighted bivariate \mass-SSFR distributions shown
in Figure \ref{fig_ssfr_vs_mass_sfr}. It is interesting to
determine the average SSFR vs. mass and redshift as we
did for IRX. The number-averaged SSFR is given in Figure \ref{fig_ssfr_phi}.
This shows that at lower mass, $\log{\mass}=9.5$, \ssfr~
evolves slowly, while at higher masses the SSFR falls
rapidly with time. 

The behavior in Figures  \ref{fig_irx_mass}-\ref{fig_ssfr_phi}
can be explained by a simple model. We suppose that average
IRX is determined principally by the gas surface
density and by the metallicity.  This naturally produces
a rising then falling IRX as the gas becomes enriched
(in a closed box model) and then exhausted.
Downsizing implies that star formation, enrichment,
and ultimate gas exhaustion move to lower masses with time,
consistent with these resullts. 

We examine this model further in the next section.
But first we ask whether the observed trends could be
an artifact of selection effects or other aspects of
our technique. 

\subsection{Issues and Caveats}

We have performed a number of tests to ensure that the
results presented above are not a product
of the samples or analysis approach.

We could use either FUV (1530\AA) or NUV (2270\AA) flux
to derive star formation rates. Since we bin and stack
sources in the NUV-H CMD, there could be systematic
effects introduced by either the use of an extinction law
to correct FUV given the NUV-derived IRX, or there could be
effects introduced by the different data samples used to
derive FUV and NUV rest luminosities. The former
come completely from GALEX data, while the later
come from interpolating GALEX and COMBO-17 data.
We find however that there is no significant difference
in the results using FUV or NUV to derive IRX and SFR.

We tested stacking the MIPS24 data using detected sources and cleaned
images, and using only the fluxed images (stacking
detected and undetected sources together). This
produced no statistically significant difference.
We also were concerned about the MIPS24 detection
limit and whether the low limit of 0.02 mJy used
to detect and clean the images would include
some spuriously detected sources due to confusion,
yielding an artificially high 24 micron flux in the
stacked result. We checked this by increasing
the detection limit by a factor of two and repeating
the entire analysis. Again, these results showed
only minor quantitative changes. 

A very important question is whether our census
of objects is complete. We could be missing FIR
luminous objects that fall below the UV magnitude
limits of the sample. Moreover, it is likely that
as we move to higher redshift, more of the high IRX
and/or low SSFR sources are lost due to the UV magnitude limit.
This could clearly introduce a spurious
blueing trend as redshift increases, which is exactly
what we detect.

\begin{figure}
\plotone{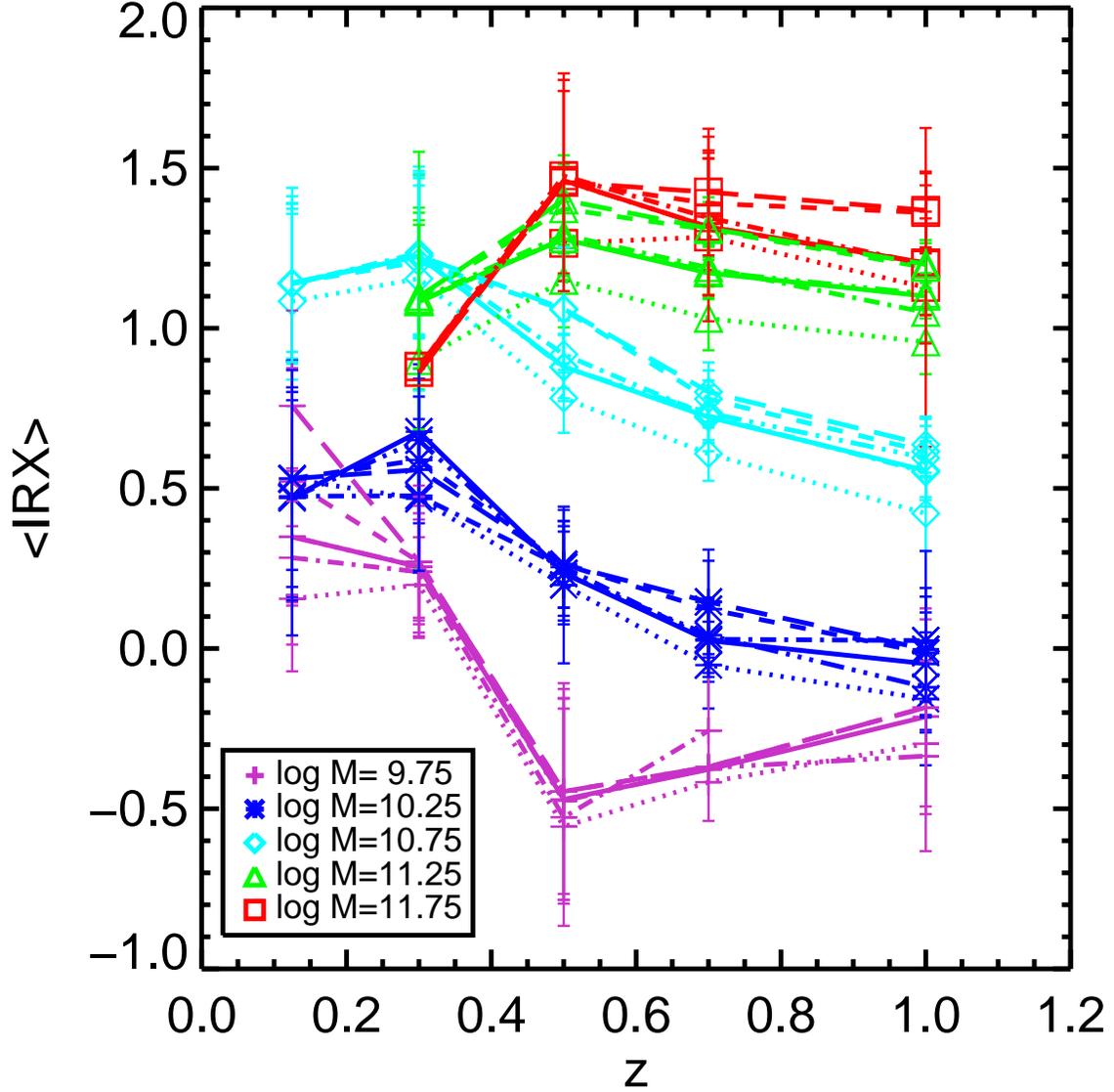}
\caption{Average IRX (\irx) vs. redshift in each mass bin using various techniques as indicated by line styles.
SOLID: 1. Baseline: NUV$<$26.0, r$<$24.0
DOTTED:  Case 2: NUV$<$25.0, r$<$24.0.
DASHED: Case 3: NUV$<$27.0, r$<$24.0.
DASH-DOT: Case 4: NUV$<$26.0, r$<$25.0
DASH-DOT-DOT: Case 5: all r$<$24.0 objects, whether or not detected in NUV.
\label{fig_irx_cases}}
\end{figure}

To test for this effect, we repeated the analysis on the following samples:
1. Baseline: NUV$<$26.0, r$<$24.0;
2. Case 2: NUV$<$25.0, r$<$24.0;
3. Case 3: NUV$<$27.0, r$<$24.0;
4. Case 4: NUV$<$26.0, r$<$25.0;
5. Case 5: all r$<$24.0 objects, whether or not detected in NUV. Those objects below the NUV detection limit
are given an artificial magnitude NUV=27.0. 
The average IRX vs. redshift for these cases is shown in Figure \ref{fig_irx_cases}.
There are no significant changes to \irx~ or in the observed trends with mass and redshift.

\begin{figure}
\plottwo{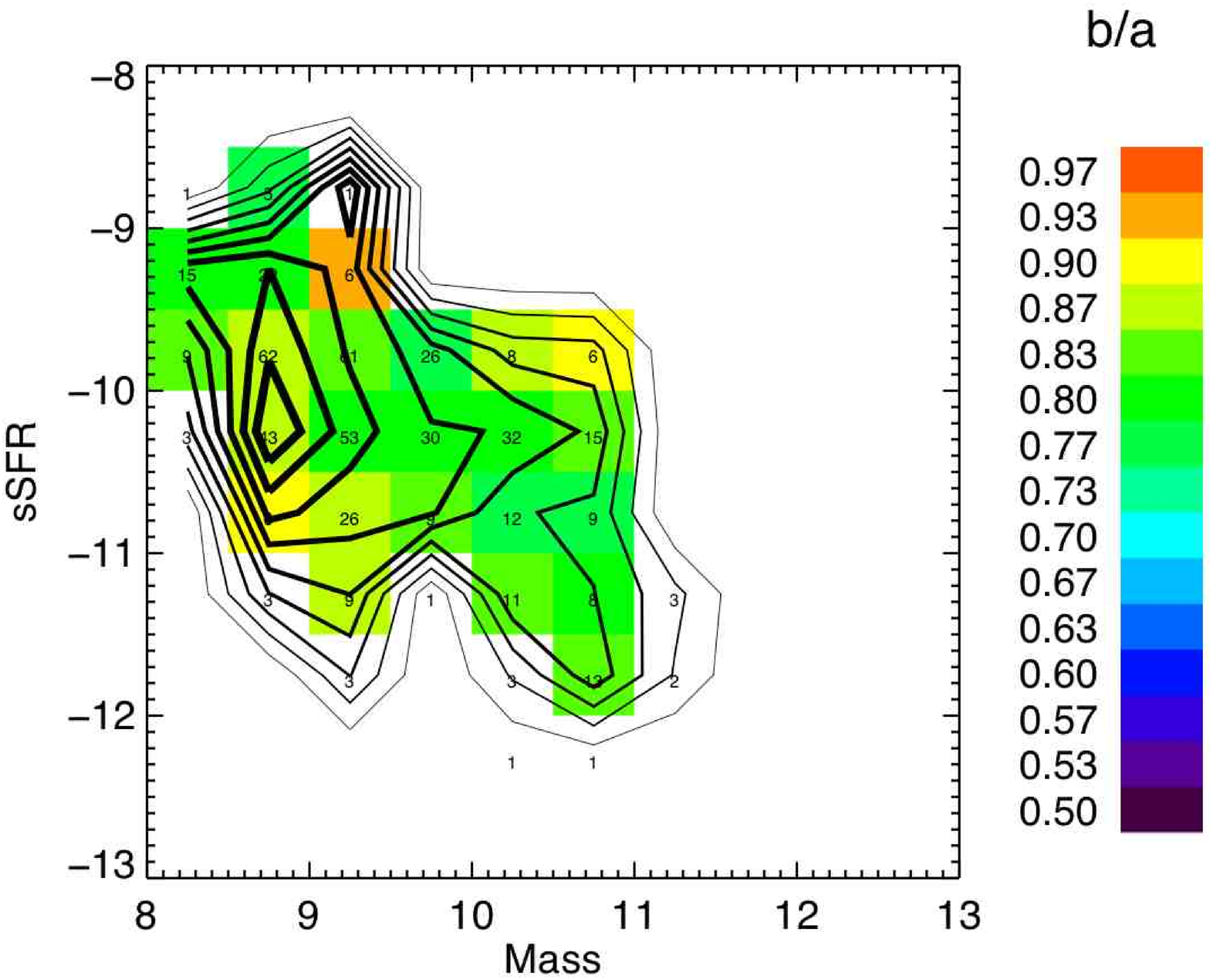}{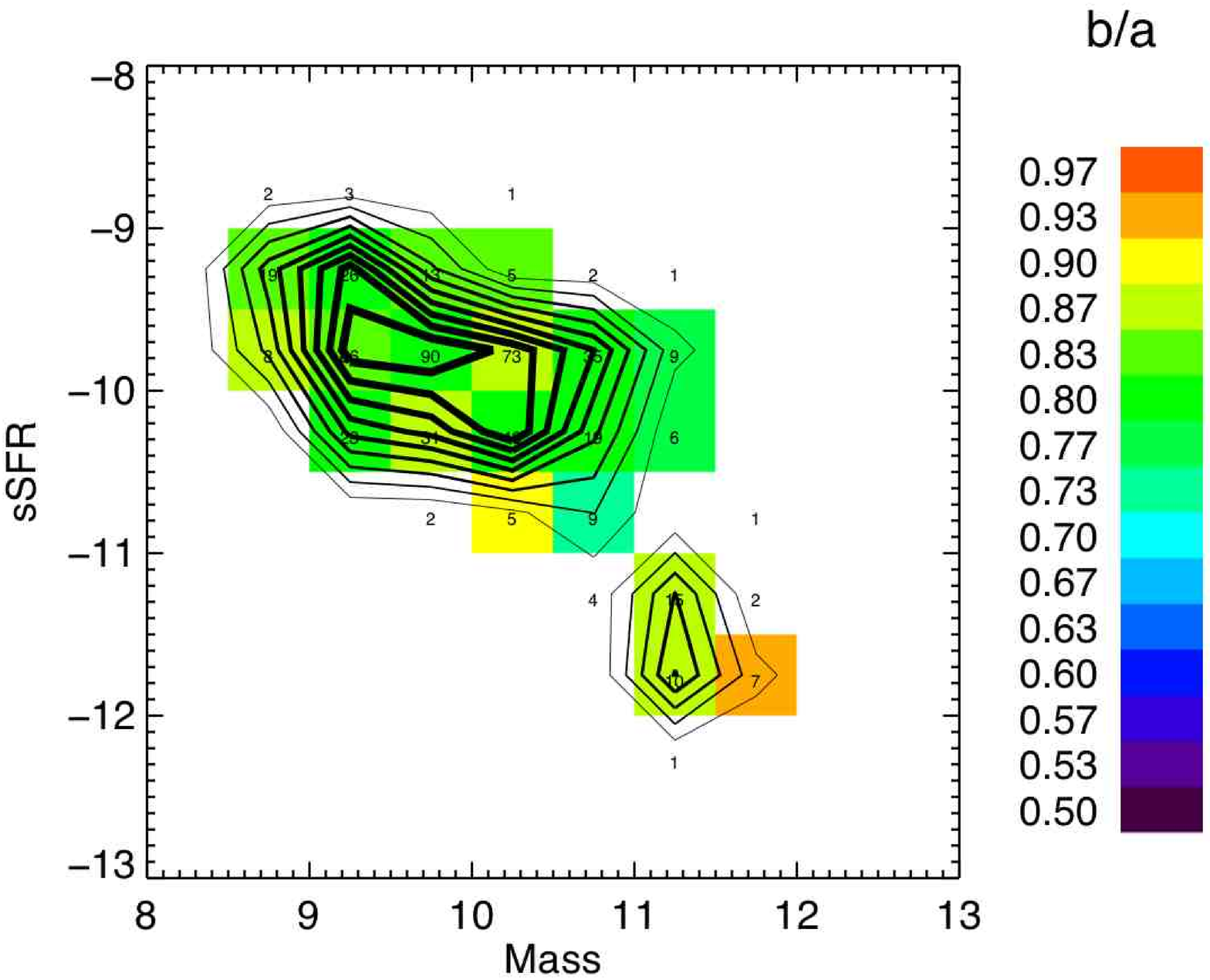}
\plottwo{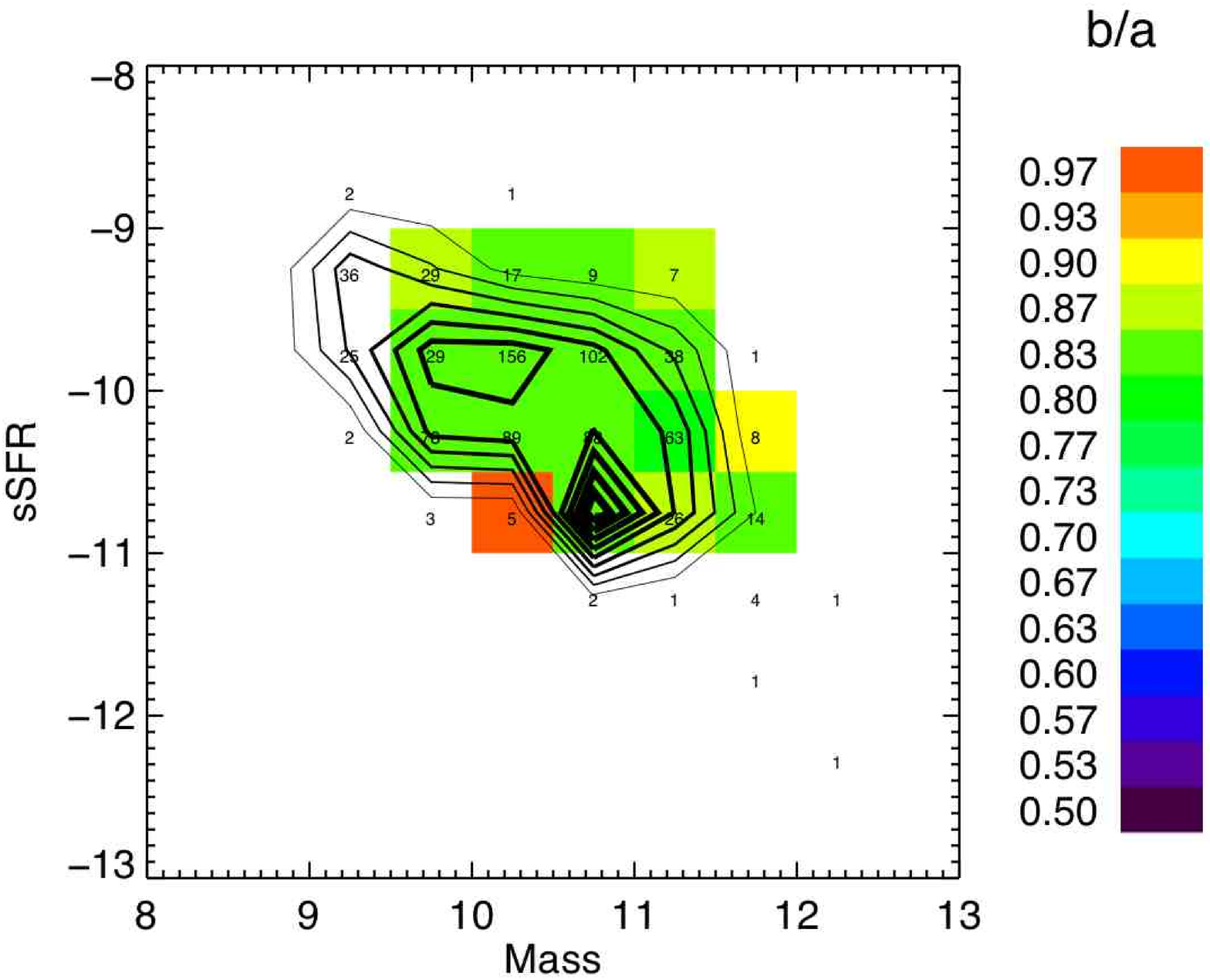}{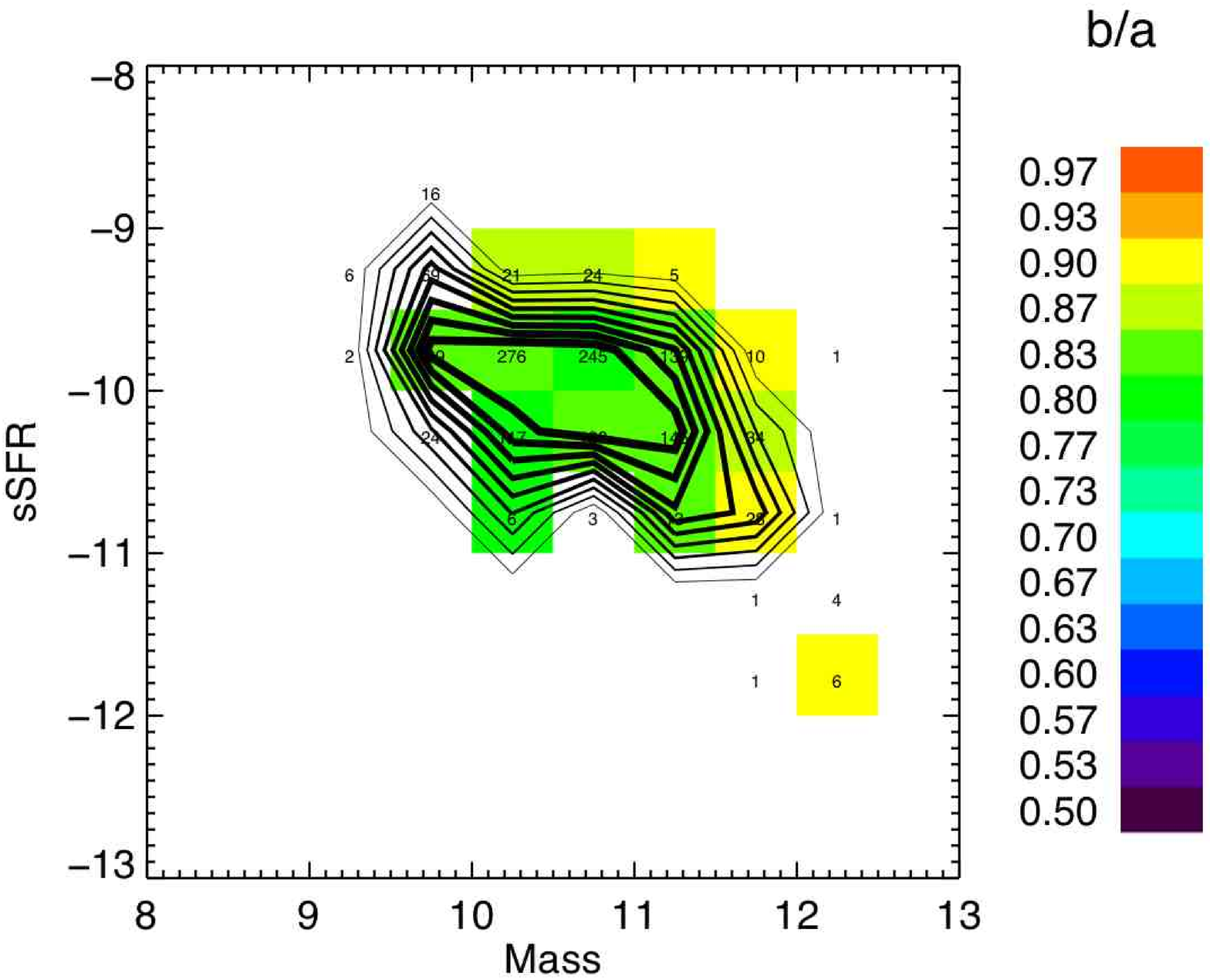}
\plottwo{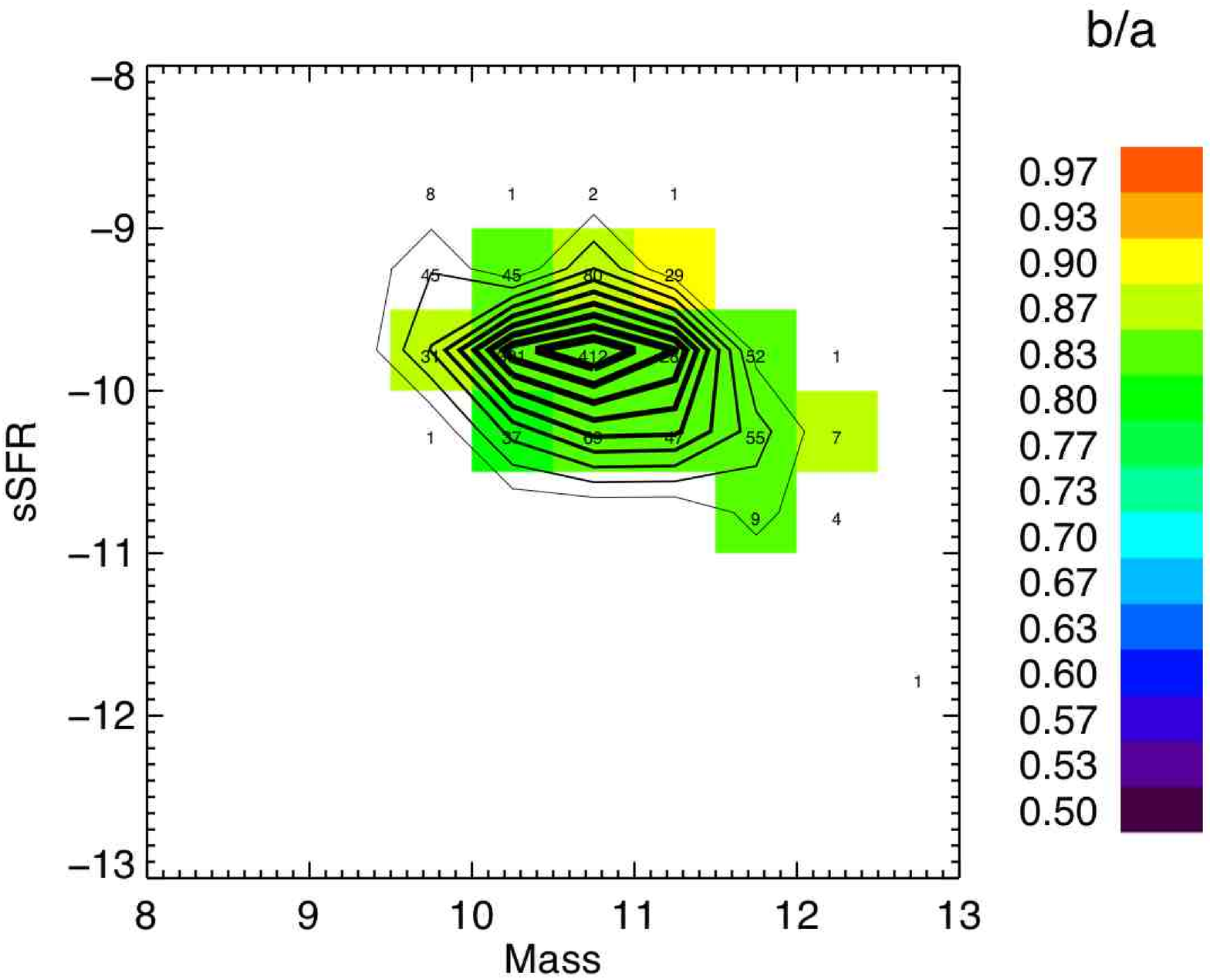}{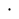}
\caption{Same as Figure 5, with color-coding changed to axis ratio b/a.
\label{fig_aoverb_ssfr}}
\end{figure}

Another test is to consider the inclination bias of the sample.
A sample at high redshift which has not included highly
inclined, more reddened galaxies of otherwise similar
overall dust content will display a higher
average minor to major axis ratio than the low redshift counterpart.
The average axis ratio obtained from the (seeing limited) COMBO-17
data shows no significant trends with redshift. This is true in the \mh, NUV-H
CMD and as we show in Figure \ref{fig_aoverb_ssfr}, the final \mass-SSFR diagram.

Finally, we used a Monte Carlo model to test whether the evolving
IRX-mass relationship could be an artifact of the sample selection. 
The model is semiempirical and we briefly summarize it here. The model predicts the
bivariate luminosity function in the extincted, rest-frame NUV-H, \mh CMD, and
the distribution of \irx~ over the same CMD (\irx[\mh, NUV-H]).
The number distribution
is given by a Schechter function in mass. The SSFR
is log-normal with a constant mean and variance to a certain critical mass, then falls. 
IRX is log-normal and the mean IRX scales with mass. 
Evolution is introduced into the
number density, mean SSFR, SFR cutoff mass, and the
IRX-mass relationship. 
In the latter case, the following relationship is introduced:
\begin{equation}
IRX = IRX_0 + a_m  (\log{M/M_0})  +  a_{m,z} (\log{M/M_0}) \log{(1+z)} + \sigma_{IRX} * \xi
\end{equation}
where $\xi$ is a normally-distributed random variable. 
This assumption allows for IRX dependence on mass, and evolution
of this dependence in a mass-dependent fashion, in other
words the evolutionary trends we appear to detect in the data.

We convert mass, SSFR, and IRX into observed SED's using (in reverse) the
identical transformations that we used for the data. SEDs are redshifted
and run through detection filters with appropriate completeness
functions. We then subject the list of objects and observed FUV, NUV, R-band, IRAC,
and MIPS24 fluxes to the identical analysis steps as the actual
sources, producing the various distributions. (We do not simulate the
actual image formation and detection process). Finally, we compare
the Monte Carlo and data distributions using a chi-square
statistic. For this comparison we combine data and Monte Carlo
errors (data errors calculated from bootstrap, Monte Carlo
errors calculated from Poisson statistics). To calculate $\chi^2$
we use all bins in which either data or Monte Carlo results
are predicted. We simultaneously fit both $\phi(M_{\rm H}, NUV-H)$
and \irx $(M_{\rm H}, NUV-H)$ over all five redshift bins.

Because the model has $\sim 10-14$ parameters, it is difficult to guarantee any
given local minimum is the global minimum. Extensive experimentation
has shown that the $\phi$ distribution and IRX distributions
are mainly influenced by separate variables, so some
minimization can be decoupled. We find best fits with
$a_{m,z}\simeq$1.5 and $a_m\simeq 0.3$, with $\log{M_0}\simeq 11.0$ and with small formal errors ($<0.05$).
The latter are derived in the usual way by fixing the parameter of interest
and marginalizing over all others. 
The key conclusion is that the significant non-zero value of $a_{m,z}$ 
provides additional evidence that the evolving IRX-mass relationship 
is not an artifact of the sample selection.

\section{Discussion\label{sec_discussion}}

\subsection{Simple Extinction, Metallicity, and Star Formation Evolution}

The phenomena displayed in Figures \ref{fig_irx_z} and \ref{fig_irx_mass}
have a very simple interpretation. At low mass, ongoing enrichment by
star formation increases the dust-to-gas ratio and mean extinction
per unit gas, resulting in a steady growth in IRX. Higher mass
galaxies had their periods of peak star formation in the past,
and the exhaustion of their star forming gas supply (by whatever
mechanism) leads to an IRX which falls with time. We reiterate that
our determination of IRX is obtained directly from
the Far UV-to-Far IR ratio (the latter from 12-24 $\mu$m luminosity
and a bolometric correction), and is independent of
the extinction law.

We can model this using the classical exponential SFR models introduced
by \cite{tinsley68} and a very simple extinction model. 
The SFR obeys:
\begin{equation}
SFR \sim e^{-(t-t_0)/\tau} = e^{-\xi}
\end{equation}
where $\xi$ is a scale-free ``age'' parameter. 

For simplicity, we characterize the extinction
as if it occurs in a simple foreground absorbing slab
of dust, and that its strength tracks the amount
of gas responsible for star formation:
\begin{equation}
A_{FUV} \sim  Z \Sigma_{gas}
\end{equation}
where $Z$ is the metallicity (we assume that gas-to-dust ratio scales accordingly), and
$\Sigma_{gas}$ is the gas surface density.
Ignoring inclination induced anisotropies, we have $IRX=\log{{(10^{(0.4A_{FUV})}-1)}}$.
Let us further assume a Schmidt-Kennicutt scaling law \citep{kennicutt89}
$SFR\sim \Sigma_{gas}^\beta$. Here we note
$\beta\simeq 1.5$. Then
\begin{equation}
\Sigma_{gas} \sim e^{-\xi/\beta}
\end{equation}

In a closed-box enrichment model, metallicity grows as 
\begin{equation}
Z = y \ln{\mu_{gas}^{-1}} = y~\xi
\label{eqn:metallicity}
\end{equation}
for an exponentially decaying SFR. 
The gas fraction is $\mu_{gas}$ and $y$ is the average yield \citep{searle72}.
Thus we expect
\begin{equation}
A_{FUV} = C_0 \xi e^{-\xi/\beta}
\label{eqn:afuv}
\end{equation}
Here $C_0$ is a scaling constant appropriate
for an average inclination.

This can be generalized to the leaky-box case \citep{hartwick76}
in which the outflow is proportional to the star formation rate
$\mathcal{M}_{wind}=-c~SFR$. In this case
\begin{equation}
Z = {y \over {1+c}}  \ln{\mu_{gas}^{-1}} = {y \over {1+c}} ~\xi.
\end{equation}
If $c<1$ then an accreting-box case \citep{binney98} with
infall proportional to SFR would obtain.
We do not consider other accreting-box scenarios.

Finally, we need to relate the age parameter to galaxy mass.
\footnote{We label galaxies by their
stellar mass. Over the redshift range we consider, a constant SFR would increase the stellar
mass by 0.5 dex, or one mass bin. We ignore this subtlety in order to make
the arithmetic simple for this very basic model.} 
We make
a simple {\it ansatz} that the age scales as mass to a constant power $\alpha$,
that a single formation time $t_f$ is appropriate, and that the
past age is reduced by the relative elapsed time from formation:
\begin{equation}
\xi (\mass,z) = \left( {\mass \over {\mass_0}} \right)^\alpha \left( {t(z) - t_f} \over {t(0) - t_f} \right)
\label{eqn:xi}
\end{equation}

The specific star formation rate (SSFR) is given for this
model very simply by:
\begin{equation}
SSFR = {1 \over \tau} {1 \over {e^\xi} - 1 }
\label{eqn:ssfr}
\end{equation}
where $\tau$ is a function of mass for a coeval
population: $\tau = (\mass/\mass_0)^{-\alpha} (13.47-t_f)$.

\begin{figure}
\plottwo{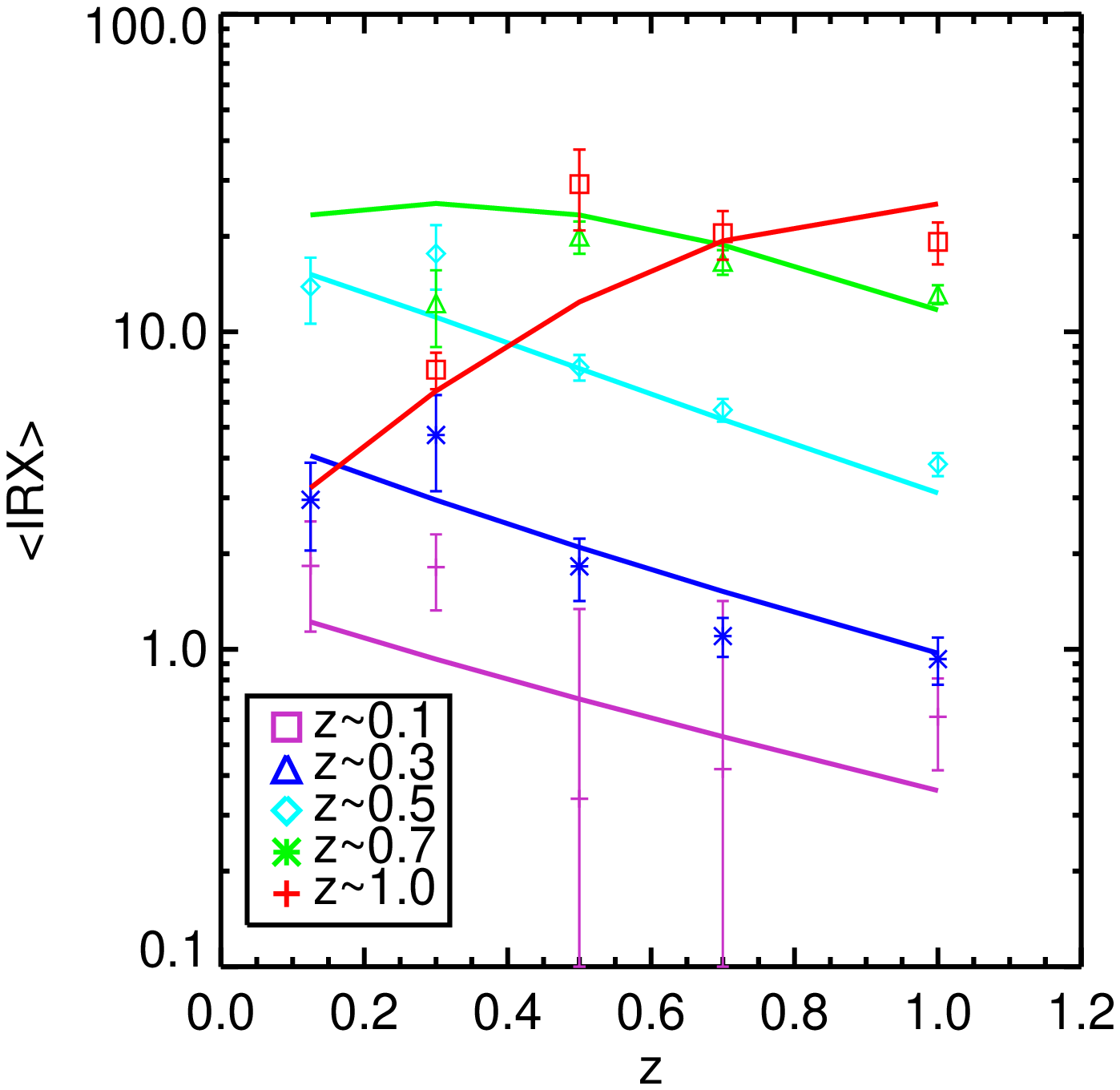}{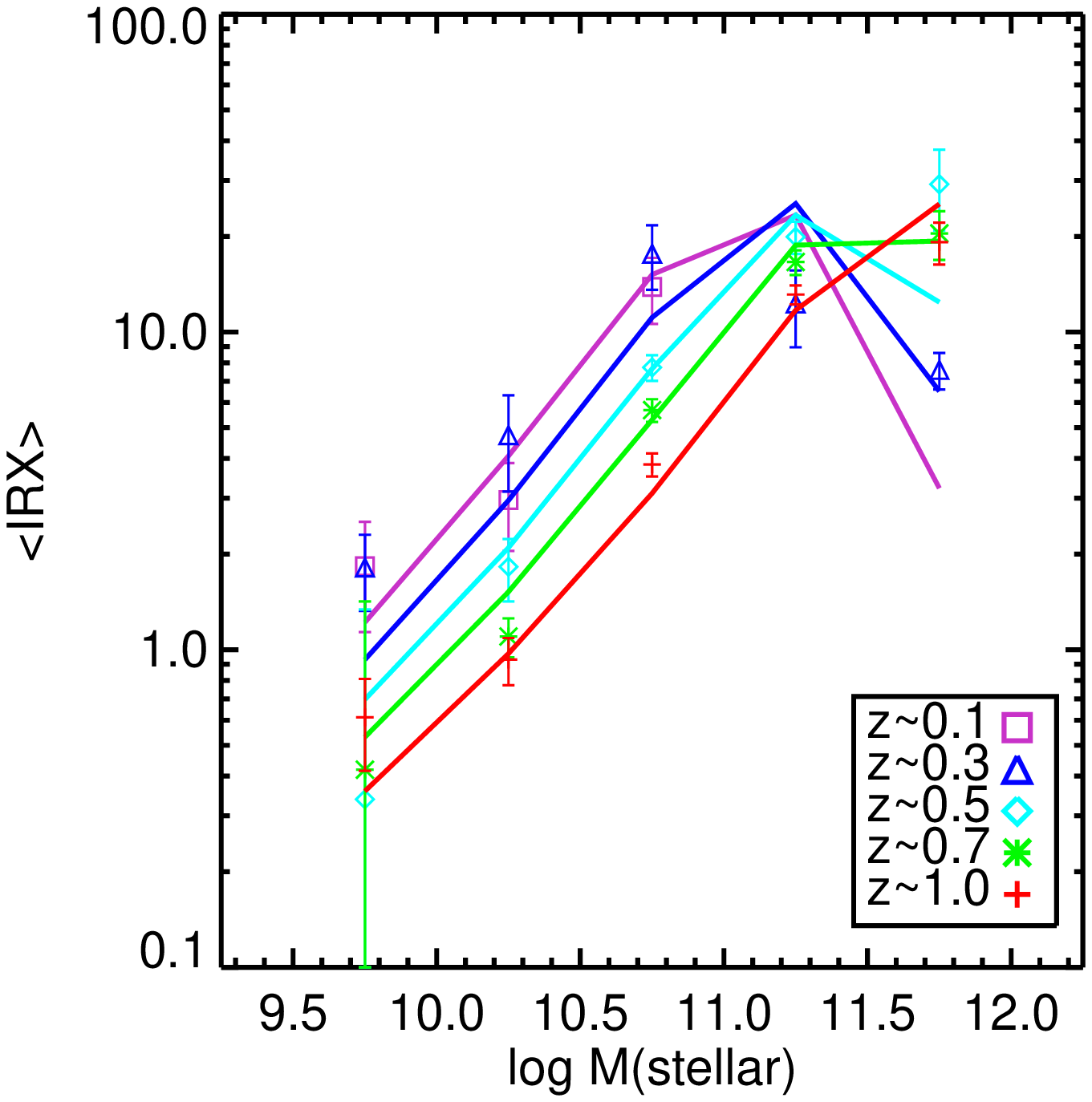}
\caption{Simple SSFR-IRX model fits (lines) to \irx~ (linear version plotted on a logarithmic scale) vs. z (LEFT) and \irx ~vs. mass (RIGHT). 
LEFT: Color gives mass, with $9.5<\log{M}<10.0$ (purple), 
$10.0<\log{M}<10.5$ (blue), 
$10.5<\log{M}<11.0$ (cyan), 
$11.0<\log{M}<11.5$ (green), 
$11.5<\log{M}<12.0$ (red). RIGHT: Color gives
redshift bin: $0.05<z<0.2$ (purple), $0.2<z<0.4$ (blue), $0.4<z<0.6$ (cyan), $0.6<z<0.8$ (green), and $0.8<z<1.2$ (red).}
\label{fig_irx_model}
\end{figure}

\begin{figure}
\plottwo{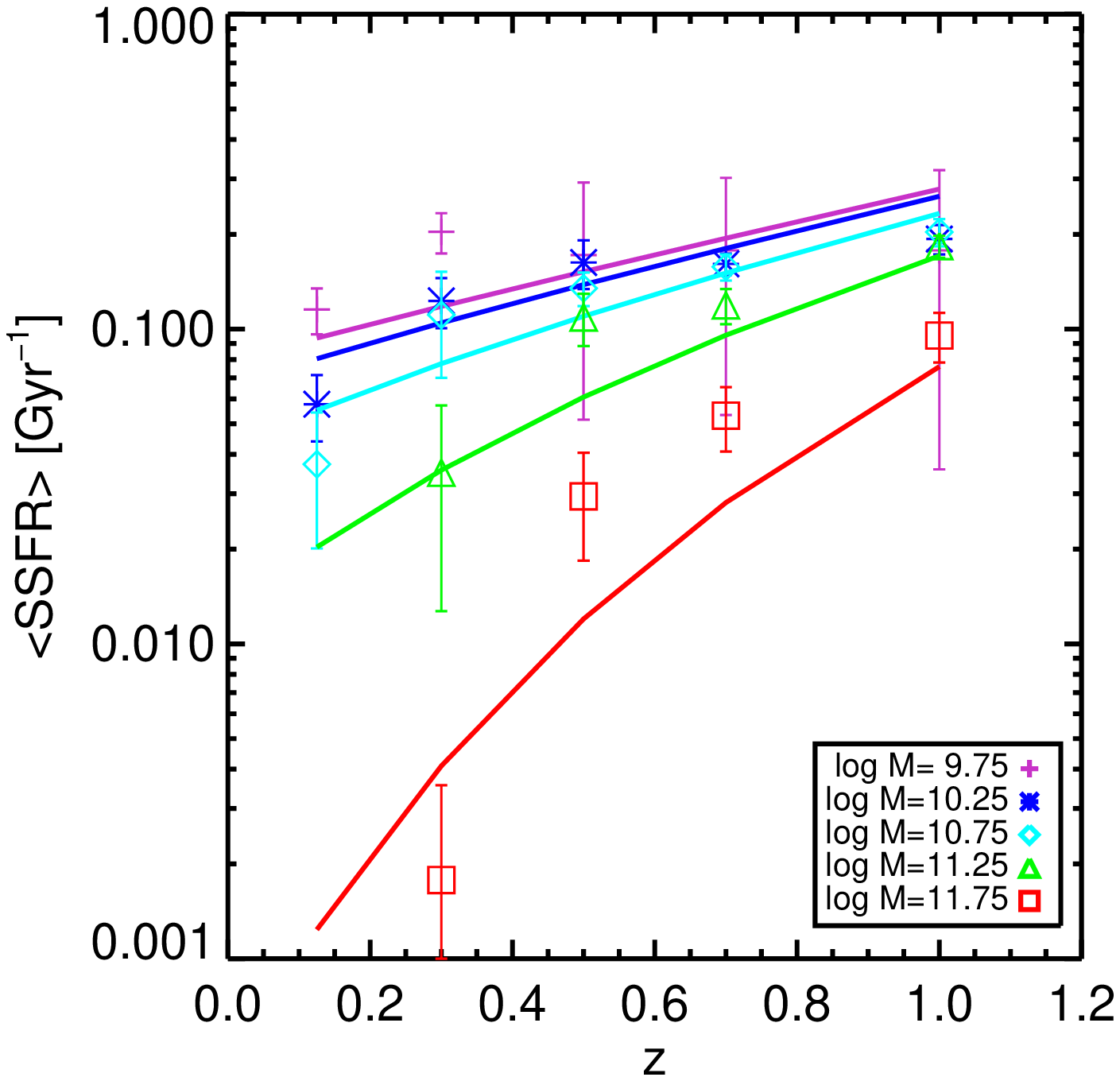}{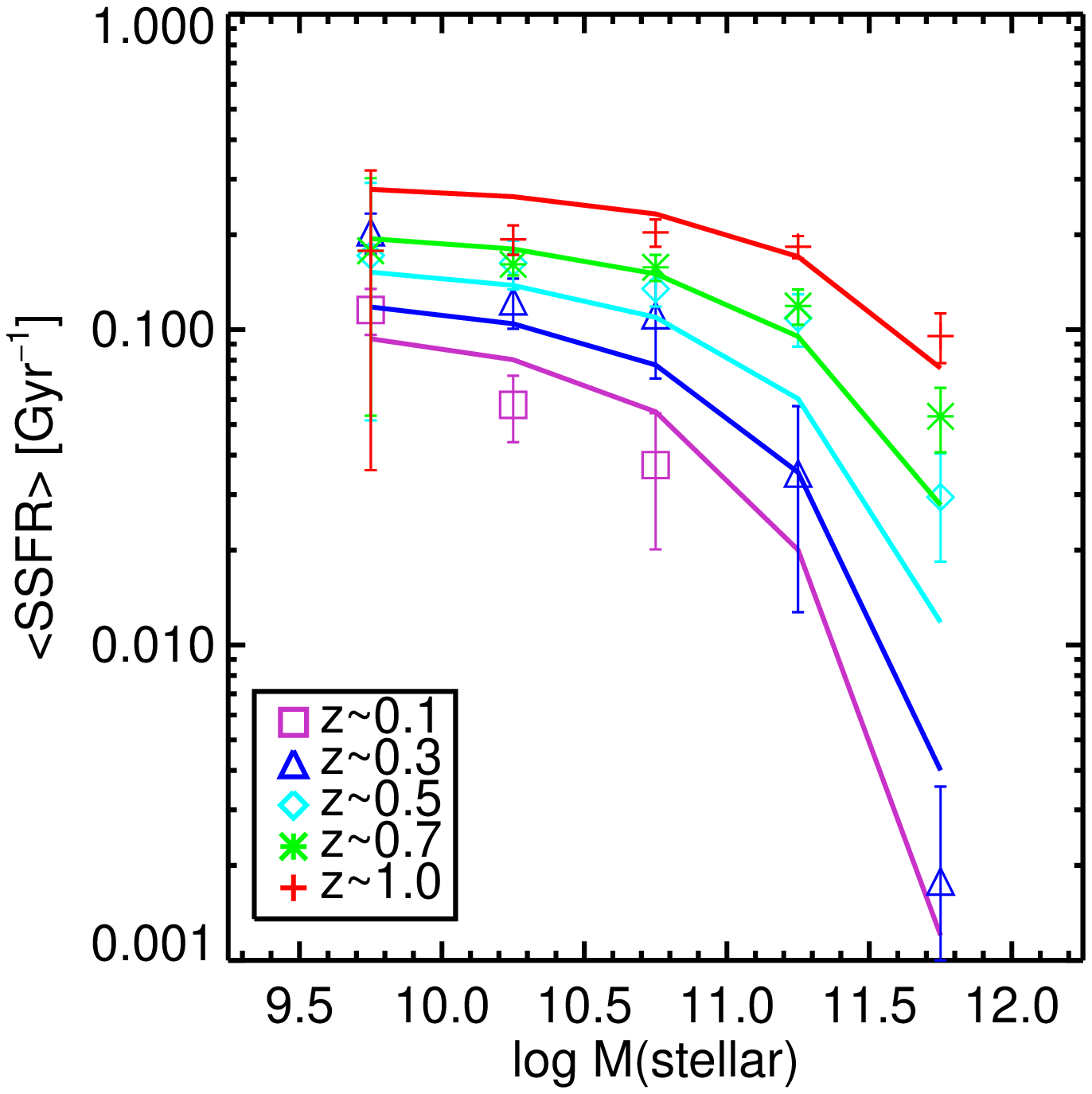}
\caption{Simple SSFR-IRX model fits (lines) to \ssfr ~vs. z (LEFT) and \ssfr~  vs. mass (RIGHT). 
LEFT: Color gives mass, with $9.5<\log{M}<10.0$ (purple), 
$10.0<\log{M}<10.5$ (blue), 
$10.5<\log{M}<11.0$ (cyan), 
$11.0<\log{M}<11.5$ (green), 
$11.5<\log{M}<12.0$ (red). RIGHT: Color gives
redshift bin: $0.05<z<0.2$ (purple), $0.2<z<0.4$ (blue), $0.4<z<0.6$ (cyan), $0.6<z<0.8$ (green), and $0.8<z<1.2$ (red).}
\label{fig_ssfr_model}
\end{figure}

We jointly fit the \irx~ and \ssfr~ vs. mass and redshift with five free parameters:
$C_0$, $t_f$, $\mass_0$, $\alpha$, and $\beta$. We restrict the fits
to the mass range $9.5 \leq \log{\mass} \leq 11.5$,
over which the survey appears reasonably complete at all
redshifts (cf. below).
There are 46 independent
data points and four free parameters. The result is
a reasonable fit, with $\chi^2=31$, with
$C_0 = 4.5$, $t_f = 2.3$ Gyr, $\log{\mass_0}=10.58$, $\alpha=0.72$, and $\beta=2.1$.
Note that for Schmidt-Kennicutt star formation law we expected
$\beta\simeq1.5$.
The fits are displayed in Figure \ref{fig_irx_model} and \ref{fig_ssfr_model}.
Using this we predict that the mass metallicity relation shifts
towards higher masses roughly $\Delta (\log{\mass})\simeq 0.53$ at $z=0.7$
and $0.75$ at $z=1$, not inconsistent with \cite{savaglio05}.
This also predicts a peak equivalent extinction at $\xi_{peak}=\beta=2.2$
of $A_{FUV,peak}\simeq3.0$, or \irx$_{peak}$=1.2.
The only way to distinguish between the closed and open box cases
is to provide an independent measurement or prediction of the proportionality
constant $C_0$, which is beyond the scope of this paper.

\begin{figure}
\plotone{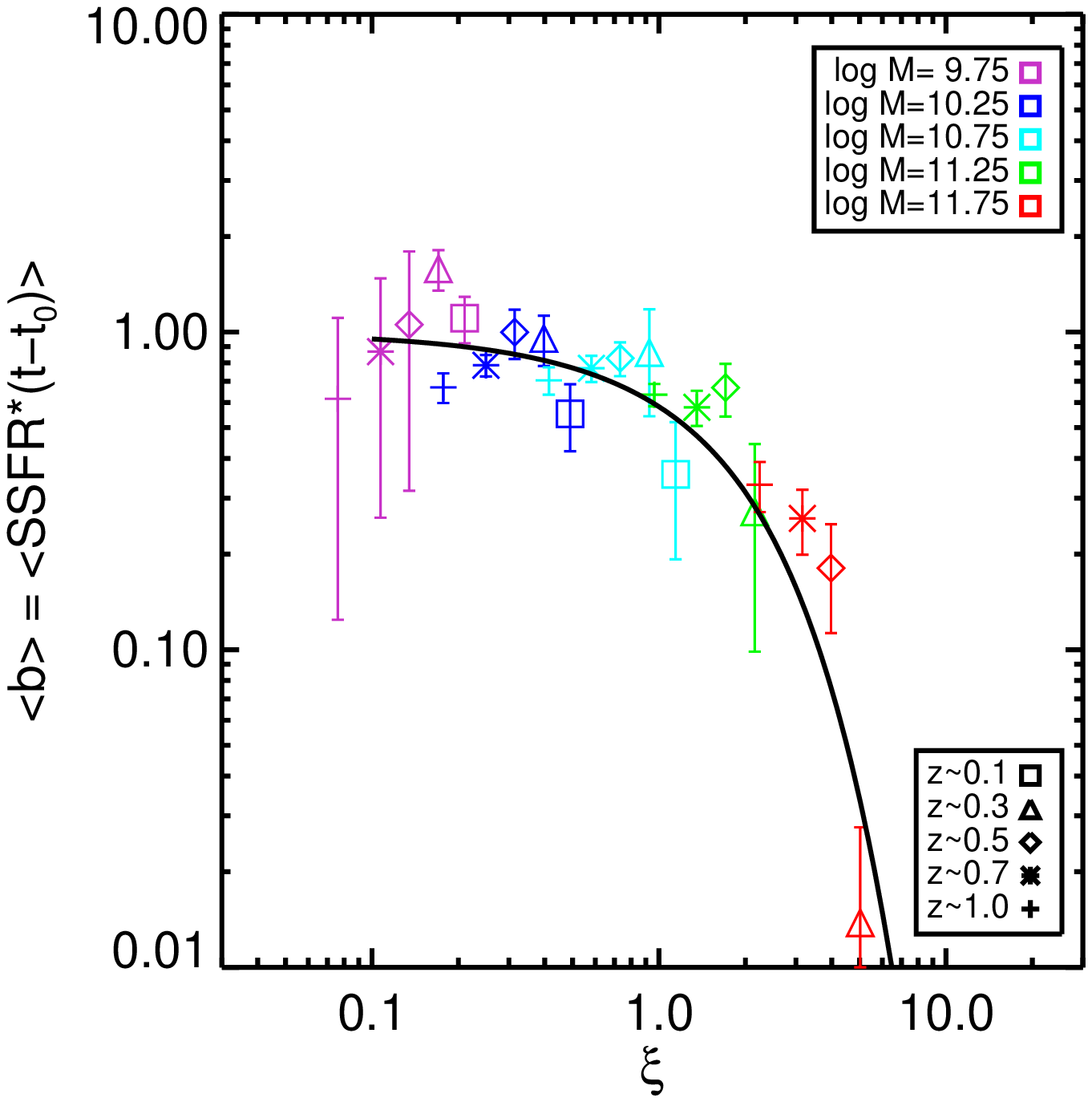}
\caption{Average b-parameter (\ssfr~ times time since formation t$_0$) vs. galaxy age parameter $\xi$ for all masses $9.5<\log{\mass}<12.0$ and all redshifts.
Color gives mass, with $9.5<\log{M}<10.0$ (purple), 
$10.0<\log{M}<10.5$ (blue), 
$10.5<\log{M}<11.0$ (cyan), 
$11.0<\log{M}<11.5$ (green), 
$11.5<\log{M}<12.0$ (red).
Symbol type gives redshift: 
 $0.05<z<0.2$ (plus), $0.2<z<0.4$ (star), $0.4<z<0.6$ (diamond), $0.6<z<0.8$ (triangle), and $0.8<z<1.2$ (square).
Solid line is from equation \ref{eqn:ssfr}.}
\label{fig_ssfr_xi}
\end{figure}

\begin{figure}
\plotone{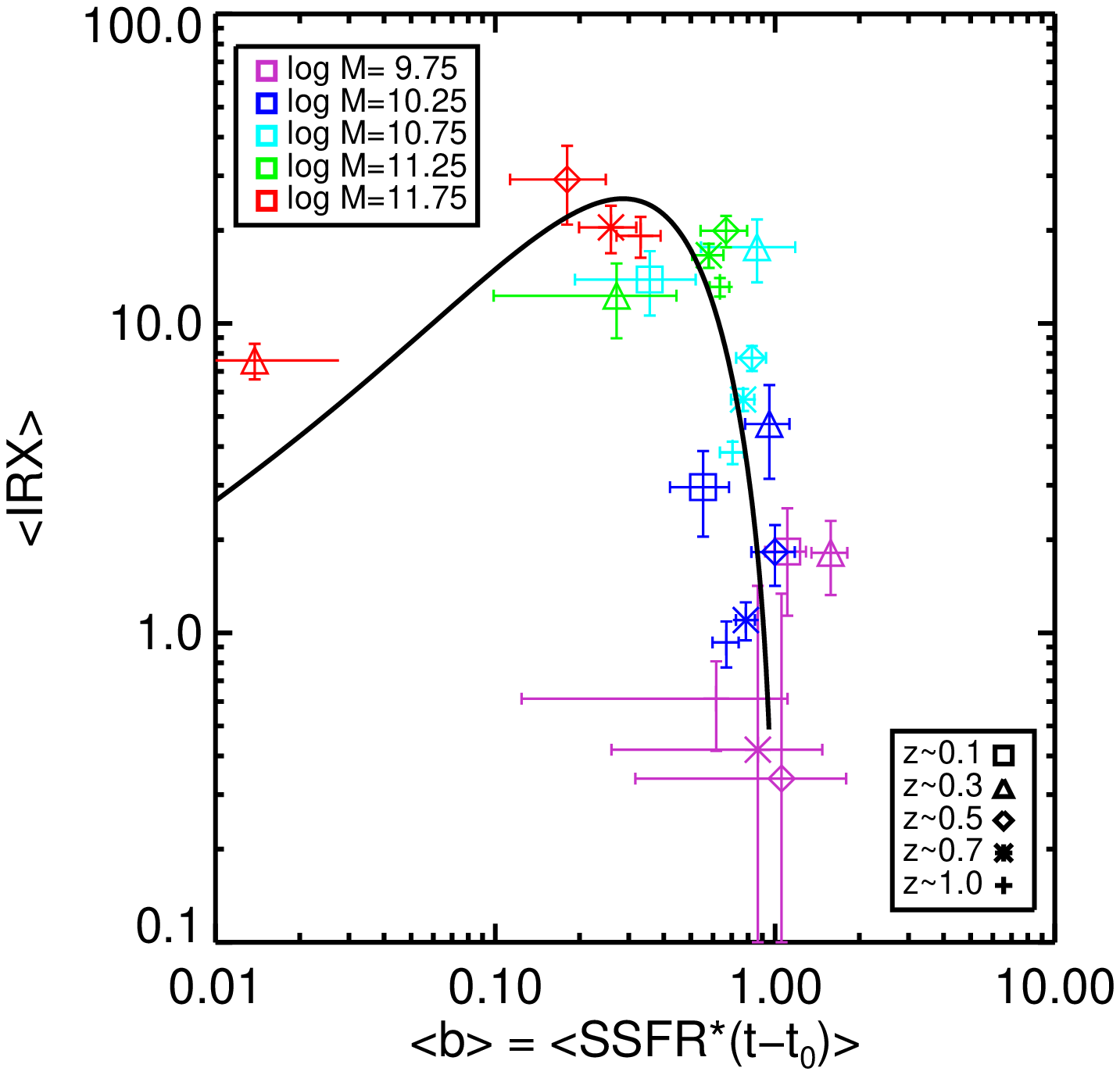}
\caption{IRX (linear version plotted on a logarithmic scale) vs. b-parameter (\ssfr~ times time since formation t$_0$) for all masses $9.5<\log{\mass}<12.0$ and all redshifts.
Color gives mass, with $9.5<\log{M}<10.0$ (purple), 
$10.0<\log{M}<10.5$ (blue), 
$10.5<\log{M}<11.0$ (cyan), 
$11.0<\log{M}<11.5$ (green), 
$11.5<\log{M}<12.0$ (red).
Symbol type gives redshift: 
 $0.05<z<0.2$ (plus), $0.2<z<0.4$ (star), $0.4<z<0.6$ (diamond), $0.6<z<0.8$ (triangle), and $0.8<z<1.2$ (square).
Solid line is from equation \ref{eqn:irx}, \ref{eqn:afuv}, and \ref{eqn:ssfr}.}
\label{fig_irx_ssfr}
\end{figure}

We show the tight relationship between the age parameter $\xi$,
derived from the stellar mass and the best fit parameters:
\begin{equation}
\xi (\mass,z) = \left( {\mass \over {10^{10.5} \mass_\odot}} \right)^{0.70} \left( {t(z) - 2.37} \over {t(0) - 2.37} \right),
\end{equation}
and the mean IRX at  in Figure \ref{fig_irx_xi}.  
There is an equally tight relationship between the ``b-parameter'' which in terms of our
simple parameterization is:
\begin{equation}
b =  <SSFR> [t(z)-t(0)] = {\xi \over {e^\xi -1}}
\end{equation}
This is displayed in Figure \ref{fig_ssfr_xi}. Finally, we
show the co-evolution of \ssfr~ and \irx~ in Figure \ref{fig_irx_ssfr}.

We can also define a ``turnoff'' mass
where $\xi(\mass_t,z)=1.0$ so that
\begin{equation}
\mass_t(z) = 10^{10.5}  \left( {t(z) - 2.37} \over {t(0) - 2.37} \right)^{-1.43},
\end{equation}
which rises from $\log{\mass_t}=10.5$ at z=0 to $\log{\mass_t}=11.24$ at z=1. 
We note that $\log{\mass_t}=10.5$ is exactly the transition mass between
star forming and passively evolving galaxies noted by \cite{kauffmann03b}
in the SDSS spectroscopic sample.

Metallicity and age are simply related in this picture
(cf. equation \ref{eqn:metallicity}).
\cite{johnson07} have shown that
IRX correlates well with metallicity.
The mass-metallicity relationship \citep{tremonti04}
is a result of the lower net astration
in galaxies with a younger effective age $\xi$.
This simple picture does not predict the observed saturation
at high mass. This could be a result of
selection effects, since metallicity can only be measured
using emission lines in galaxies with high SSFR.
It is plausible that transition galaxies with low emission line equivalent widths
display a higher metallicity than the actively star forming sample at high mass.
It could also indicate a second quenching mechanism
in addition to simple gas exhaustion, or
a more complex enrichment picture
than simple closed box evolution.

\subsection{Star Formation History and Evolution of the Blue and Red Sequence}
\label{sec:sfhist}
\begin{figure}
\plottwo{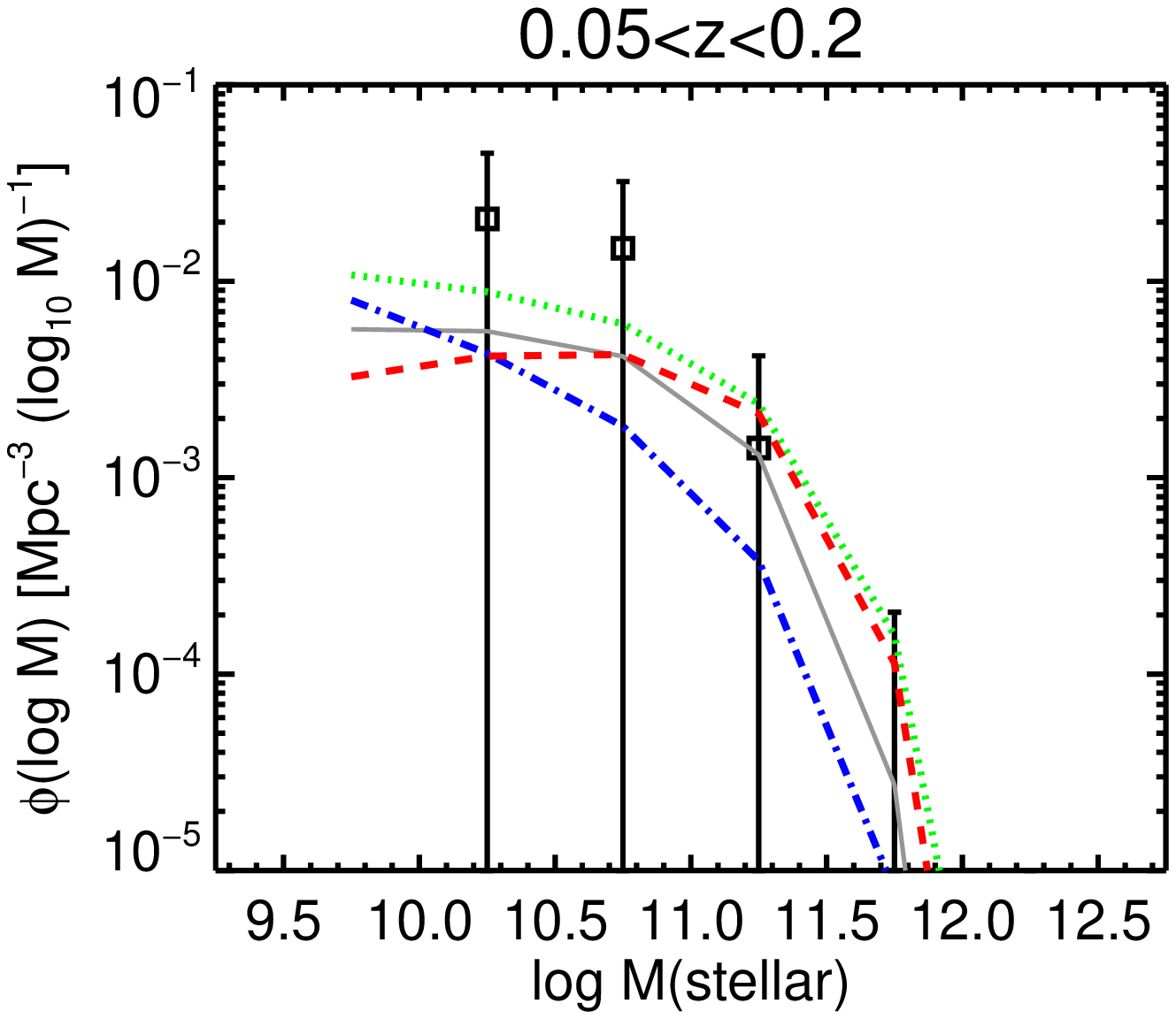}{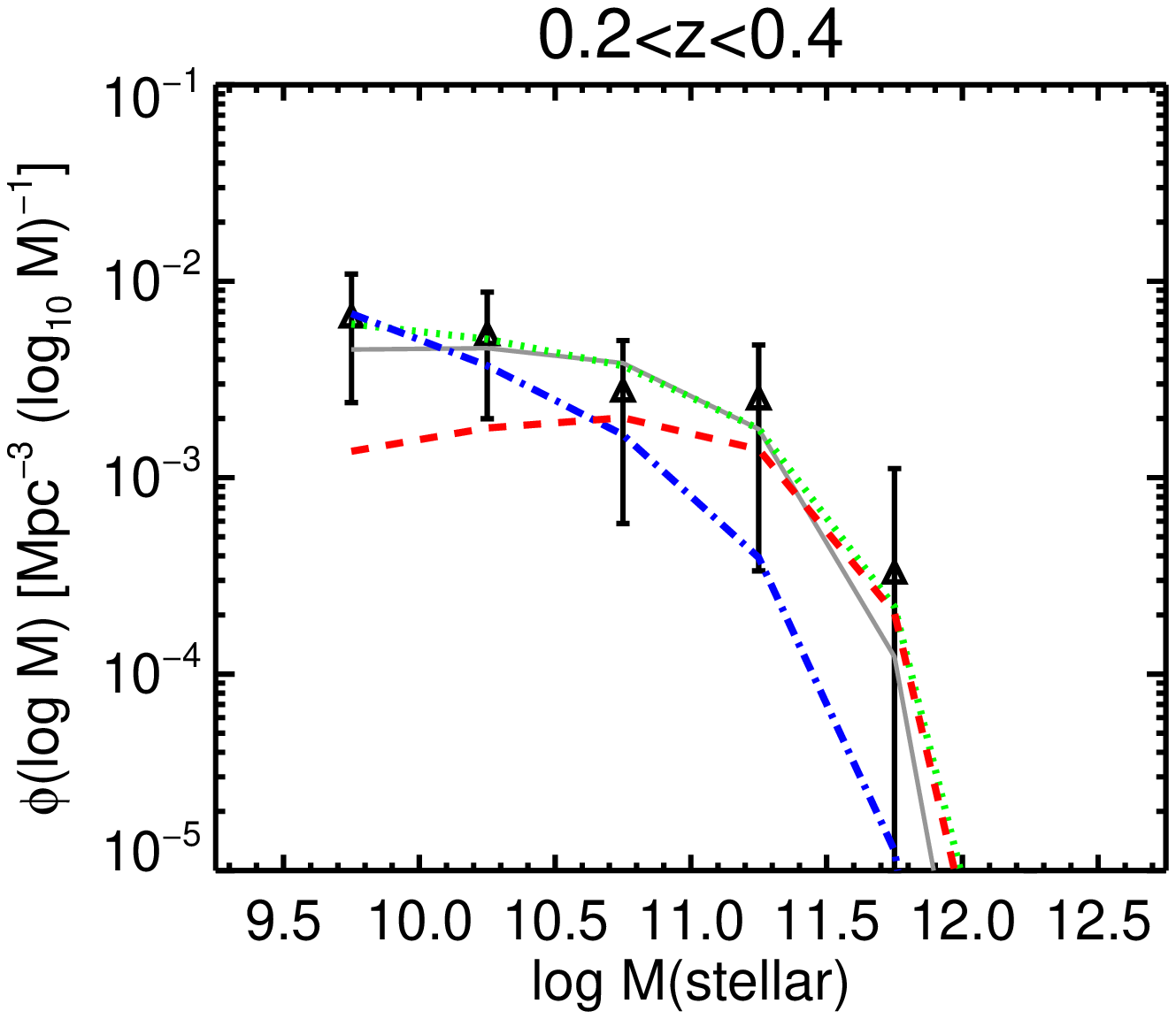}
\plottwo{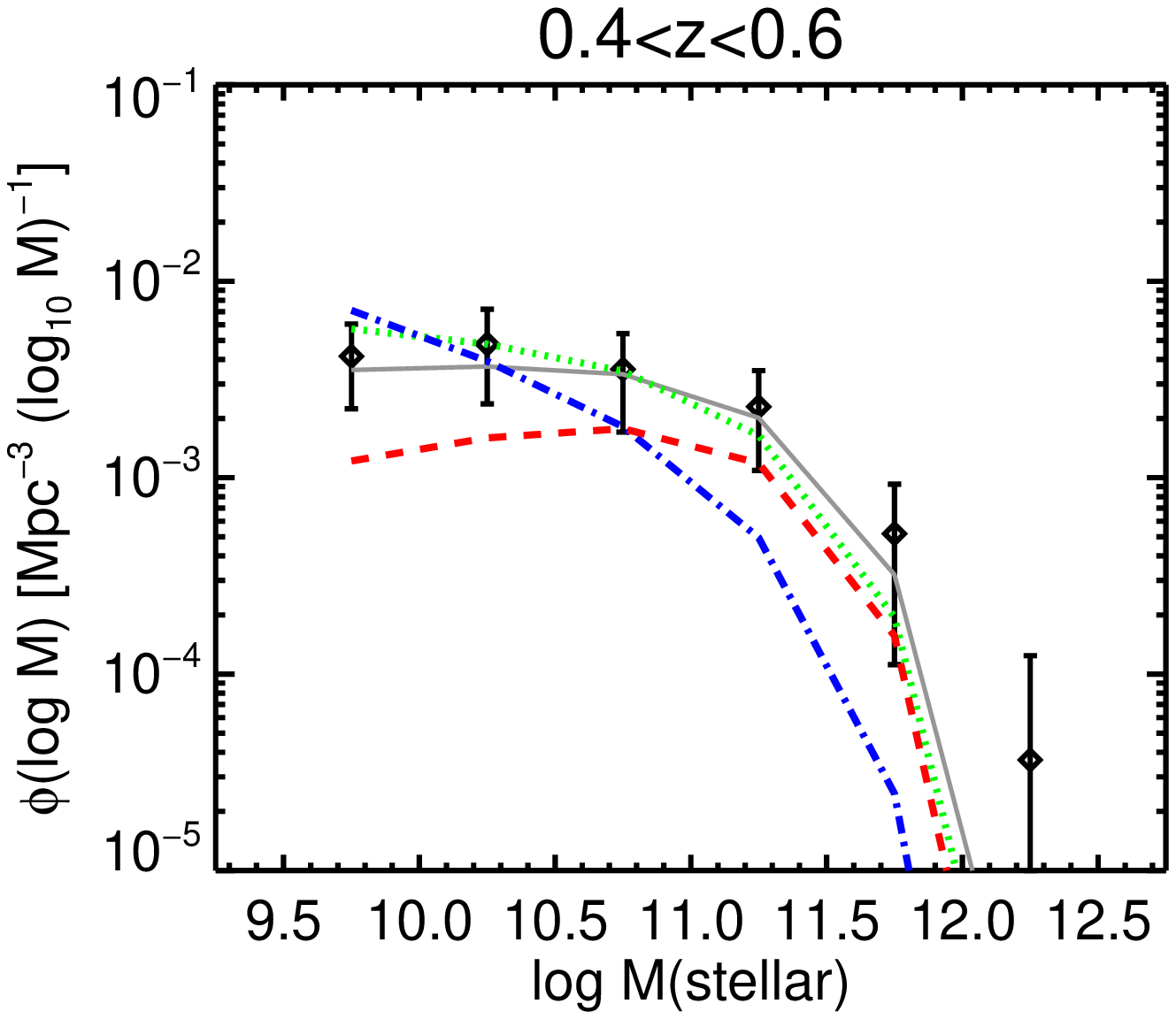}{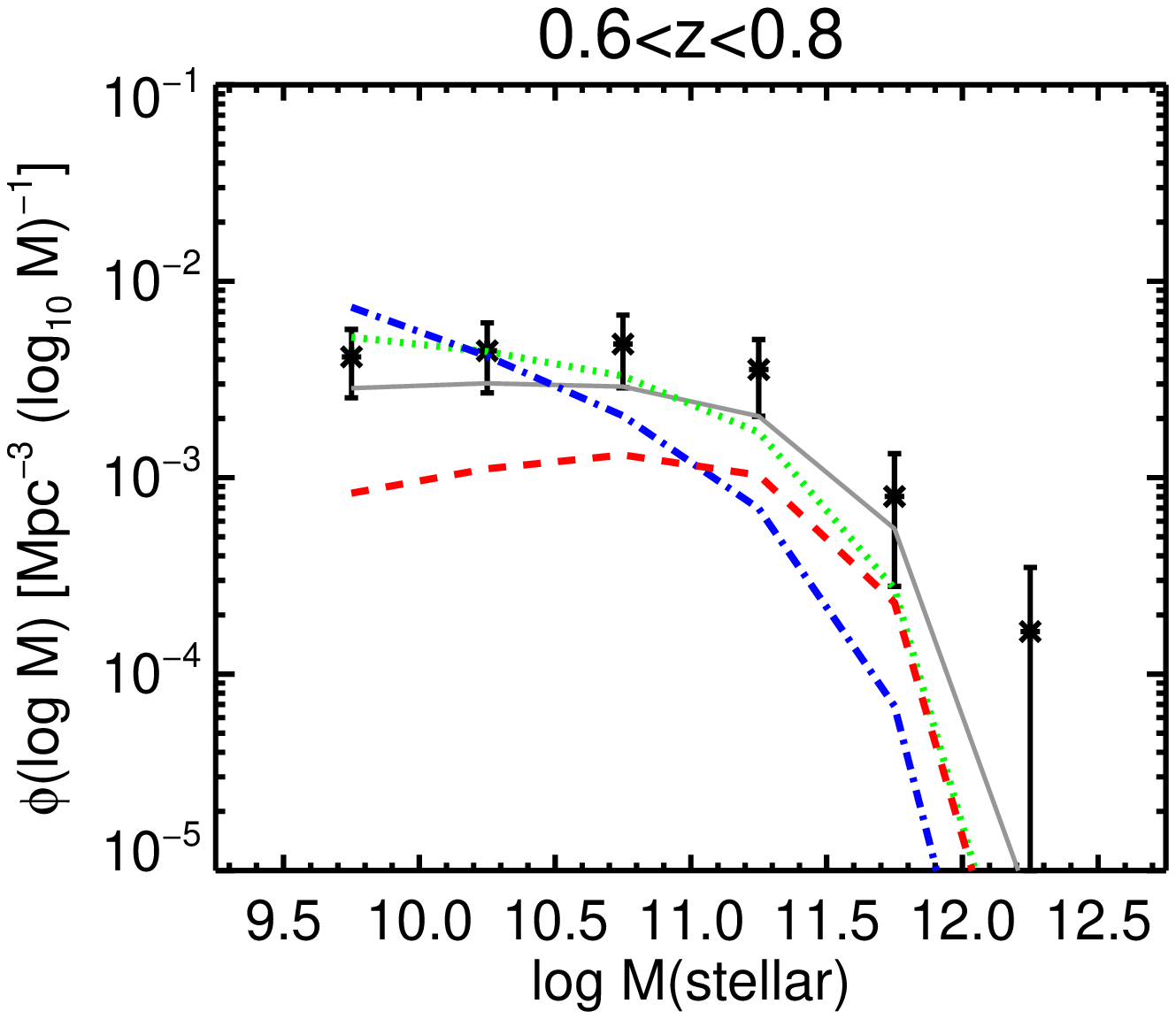}
\plottwo{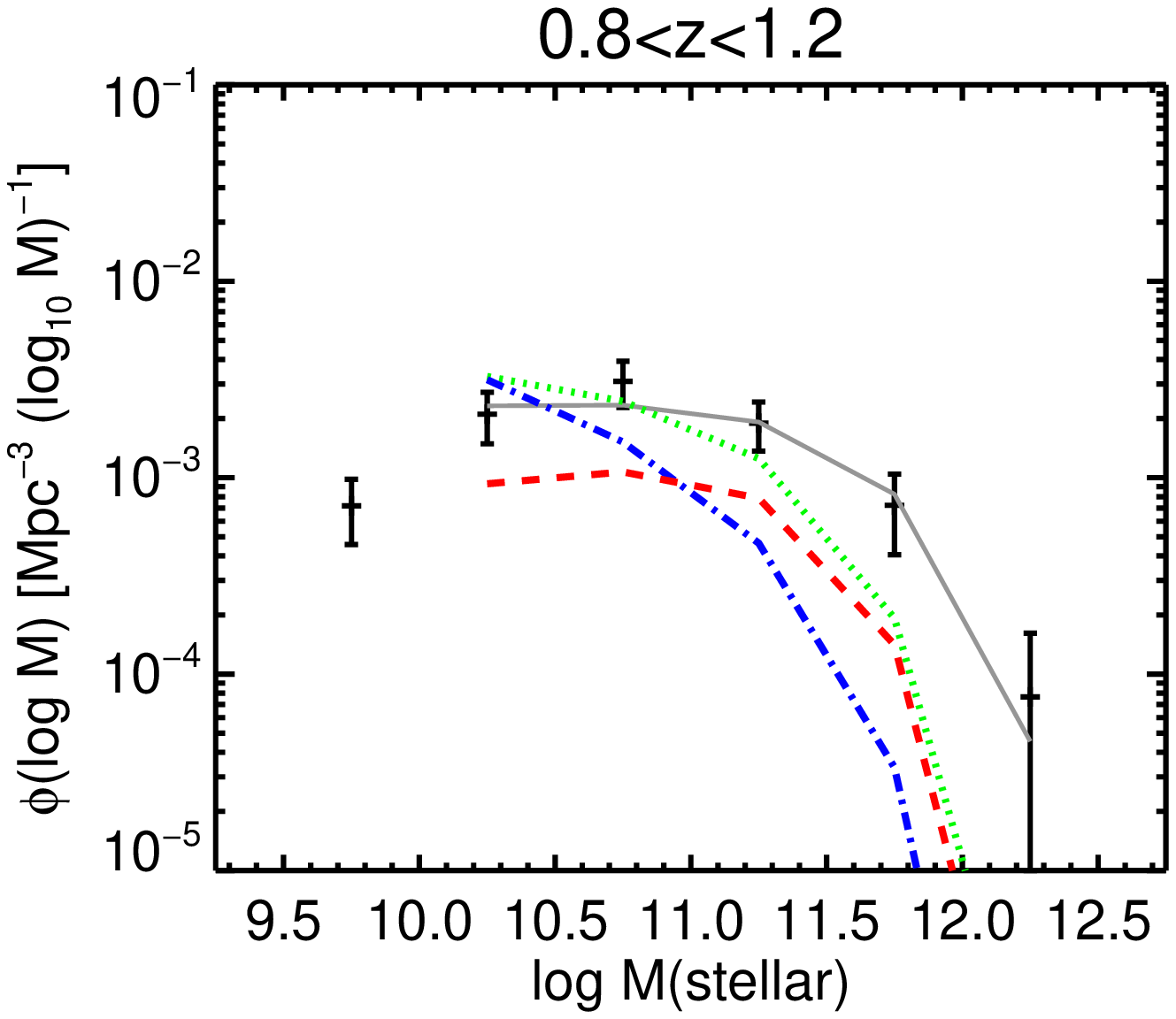}{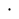}
\caption{Mass  distribution function $\phi(\log{\mass})$ vs. $\log{\mass}$ in each redshift bin.
Points and error bars give mass function derived in this paper. Solid line gives result
of fit to evolving Schechter function. Colored lines give Schechter functions derived
by \cite{borch06}. Green (dotted) is for entire galaxy sample, red for red-sequence galaxies,
and blue for blue-cloud galaxies. Note that the z=0 point is from \cite{bell03}, and our
z$\sim$1.0 bin is compared to the \cite{borch06} z=0.9 bin.}
\label{fig_phi_mass}
\end{figure}

We begin by determining the mass function by summing over SSFR in the
bivariate mass-SSFR function. The distribution in each redshift bin is displayed
in Figure \ref{fig_phi_mass}. The error bars include Poisson and cosmic variance, the
latter from \cite{somerville04}. Note that these combined errors will be highly
correlated between mass bins at a given redshift. This correlation is
not well represented by the plotted error bars, since there is
large covariance between mass bins in a given redshift bin. Note also that there is evidence for incompleteness
in the lowest mass bin ($\log{\mass}=9.75$) for z=1.0. Mass bins $\log{\mass}\leq 9.25$
were not used in the modelling of the previous section, because of obvious
incompleteness.

We model the mass function as an evolving Schechter function:
\begin{equation}
\phi(\mathcal{M},z) = {\ln{10} ~ \phi_*(z) }  \left( {\mathcal{M} \over {\mathcal{M}_*(z)}} \right) ^\alpha \exp{ \left( - {\mathcal{M} \over {\mathcal{M}_*(z)}} \right) }
\end{equation}
where
$\mathcal{M}_*(z) = \mathcal{M}_*(0) (1+z)^\beta$
and $\phi_*(z) = \phi_*(0) (1+z)^\gamma$.
We use a version of chi-square minimization
that accounts for the cosmic variance and
covariance between mass bins in a single
redshift bin (e.g., \cite{newman02})
The best fit parameters and one sigma errors (obtained
by marginalizing over other parameters) are
\begin{eqnarray}
\label{eqn:phi_fit}
\phi_*(0) & = & 0.0040 (-0.0023,+0.0036) Mpc^{-3} \log_{10} \mathcal{M}^{-1}   \\
\log\mathcal{M}_*(0) & = & 10.86 (-0.43, +0.33)   \\
\alpha & = & -0.93 (-0.10, +0.11)  \\
\beta & = & 2.4 (-1.0, +1.2)  \\
\gamma & = & -1.5 (-1.1, +0.5) 
\end{eqnarray}
The non-zero values of $\beta$ and $\gamma$ indicate evolution
in the characteristic mass and the number density,
although cosmic variance prevents tight
constraints on the evolutionary indices.
Note we have not parameterized an evolving
low mass slope.

We can compare the results of this mass function analysis with
the results of \cite{bell03} using a much larger cosmic volume at $z\sim0.15$.
Converted to our cosmology and corrected upward to a non-diet Salpeter IMF
(simply dividing their result by 0.7), they obtain 
$\phi_* = 0.0035 (0.0020) $ Mpc$^{-3}$  $\log_{10} \mathcal{M}^{-1}$, and
$\log\mathcal{M}_* = 11.17  (11.06) $ for all (late-type) galaxies vs. our values at z=0.125
$\phi_* =0.0040 (-0.0023 ,+0.0036) $ Mpc$^{-3}$ $\log_{10} \mathcal{M}^{-1}$, and
$\log\mathcal{M}_* = 10.86$. The mass cutoff is in fair agreement, but
our density is a factor of 2  larger if we compare to just their late-type (morphologically selected) sample.
The mass cutoff, slope and density parameters are highly correlated, and indeed
if we fix our mass cutoff (z=0) at 11.1 (which is within
our errors) we find $\phi_* =0.0025 $ Mpc$^{-3}$ $\log_{10} \mathcal{M}^{-1}$,
close to the \cite{bell03} value. If we fix the slope to $\alpha=-1.1$ we
find $\phi_* =0.0020 $ Mpc$^{-3}$ $\log_{10} \mathcal{M}^{-1}$ and
$\log\mathcal{M}_* = 11.1$.
Thus our results at $z\sim 0$ are consistent with \cite{bell03} within
the errors quoted above (dominated by cosmic variance).

We can also compare to the  evolving \cite{borch06} mass function which
uses all three COMBO-17 fields with a total of 25,000 galaxies.
We correct their masses to our standard Salpeter IMF by multiplying
by 1.8, as they suggest. At z=0.9, they obtain 
$\phi_*(z=0.9) = 0.0012 (0.0005) $ Mpc$^{-3}$ $\log_{10} \mathcal{M}^{-1}$
$\log\mathcal{M}_* = 11.08  (11.00) $ and , again for all (blue, color-selected) galaxies.
Using our evolutionary parameters, we find 
$\phi_*(z=0.9) = 0.0015 $ Mpc$^{-3}$ $\log_{10} \mathcal{M}^{-1}$ and
$\log\mathcal{M}_*(z=0.9) = 11.54$. 
Our results are compared to \cite{borch06} in Figure \ref{fig_phi_mass}.
Surprisingly, our results agree well with theirs for the entire galaxy sample
in all but the highest redshift bin. Our bin extends over (0.8<z<1.2), while
theirs is (0.8<z<1.0). Our mass function shows a distinctly higher mass cutoff
in the two highest redshift bins, which leads to the
stronger evolution in the characteristic mass. 
 
Searching for an explanation of this difference, 
we note that a significant fraction of our sample has red colors.
In the lowest redshift bin 
this is true even after the extinction correction.  For example, if
we eliminate galaxies with \ssfr$< 10^{-10.5}$, corresponding
to corrected $(NUV-H)_{0,AB}>4.25$, which from Figure \ref{fig_nuvh_dist}
can be seen to exclude the tail of the distribution, the mass function density parameter at z=0 falls
by a factor of two. This
suggests that a significant fraction of the mass (and even the star formation rate)
at low redshift is locked in high mass galaxies with low specific star formation rates and
red intrinsic (unextincted) colors. Some of these galaxies could be classified
as early-type in \cite{bell03} because their observed colors are even redder (as they are massive
and exhibit high IRX) while their morphologies could
be dominated by an evolved, bulge-like component.  The steepness
of the slope parameter obtained for color-selected blue samples \citep{bell03}
could also be a reflection of excluding extincted, reddened higher mass
galaxies and transition galaxies that still show some star formation.

The characteristic mass increases with redshift,
with $\Delta\log\mathcal{M}=0.7\pm0.4$, which can be compared
to the model change of $\log\mathcal{M}_t$ from 10.5 to 11.24. 
Thus the blue sequence mass function shows directly the
effects of downsizing.  The total mass of the blue sequence
obtained by integrating the mass function is a declining function of redshift
(since $\beta>-\gamma$),
$\rho_{\mathcal{M}} \simeq 4.8 \times 10^8 \rightarrow 2.8 \times 10^8 $M$_\odot$ Mpc$^{-3}$
(from z=0.8 to z=0).
Others \citep{borch06,blanton06} have found that the blue-sequence mass is constant with time,
and giving the uncertainty our results are not inconsistent with this
conclusion. In our model this occurs because the mass of the blue sequence
is moving from fewer massive galaxies to larger numbers of
lower mass galaxies. The decrease in blue luminosity density (e.g., \cite{bell04,faber05})
is a result of an increase in mass-to-light ratio due in turn to the falloff of the specific
star formation rate. The total stellar mass appears to remain constant or decline in spite of the
formation of new stars. Before exploring this point, we estimate the
star formation rate density and its evolution.

\begin{figure}
\plottwo{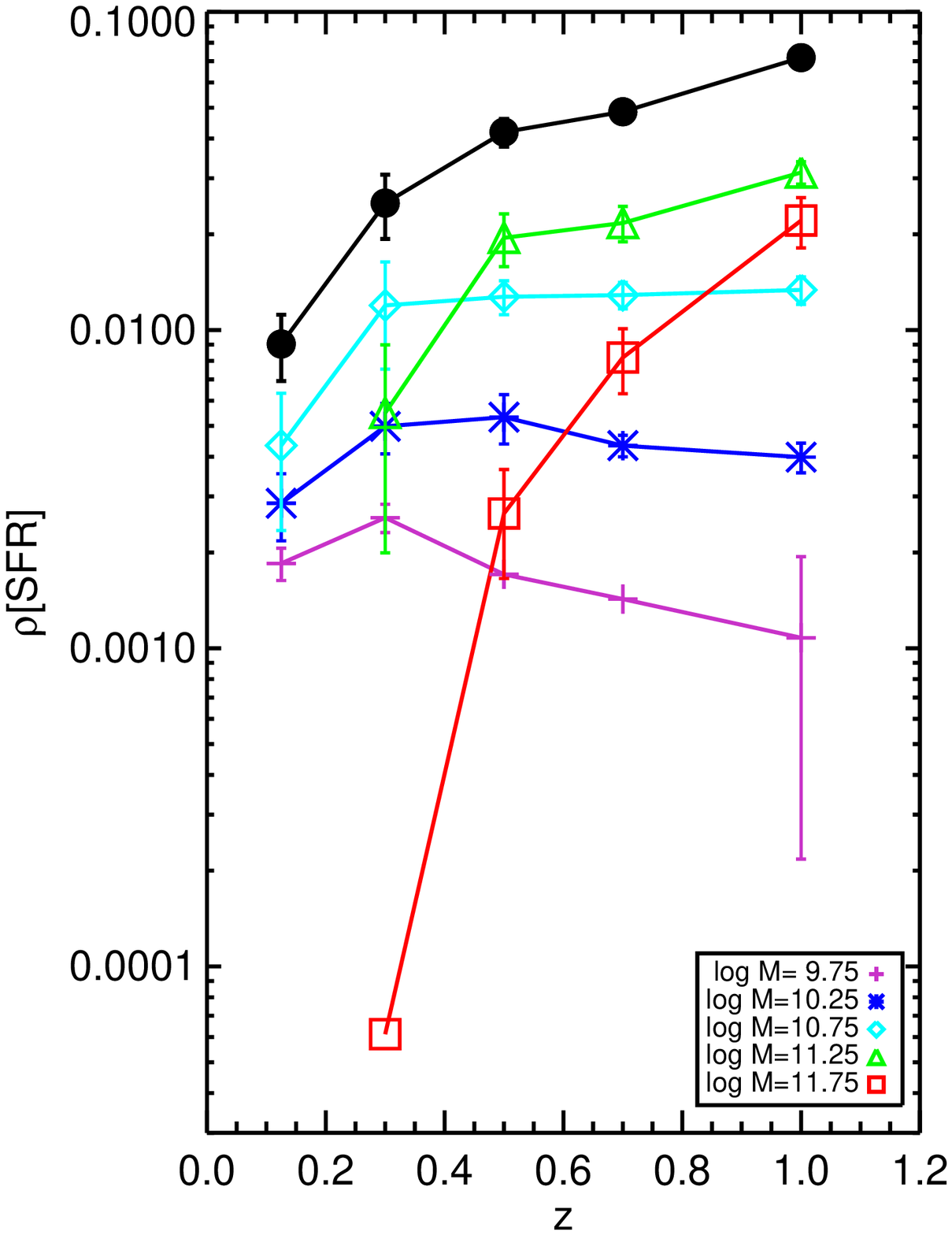}{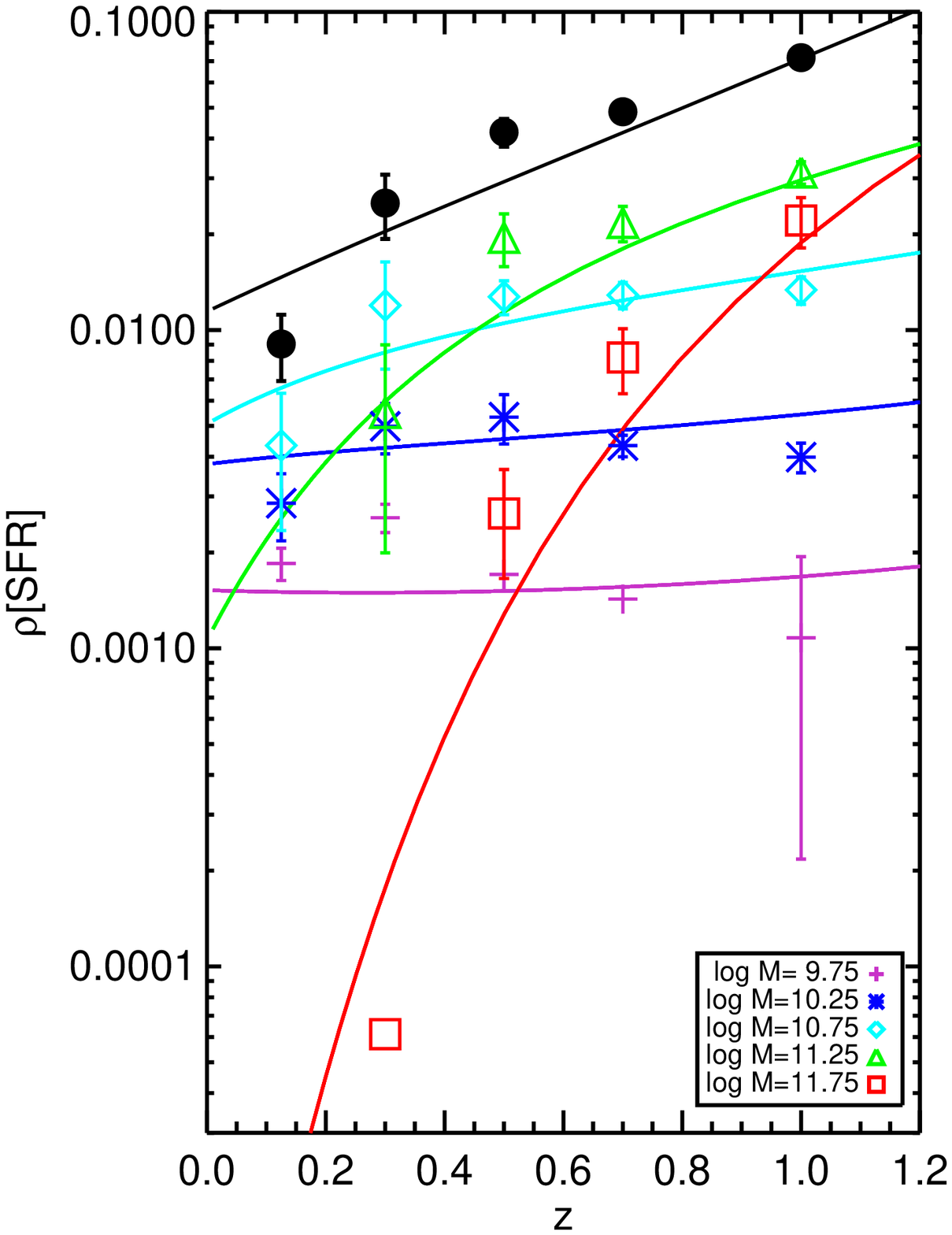}
\caption{Star formation rate density [\mass$_\odot$ yr$^{-1}$ Mpc$^{-3}$] vs. redshift derived
assuming a non-evolving mass function.
Error bars are based on bootstrap but do not
include cosmic variance, which adds a relative error of 0.37
to all the measurements. LEFT: density in each mass bin, with $9.5<\log{M}<10.0$ (purple), 
$10.0<\log{M}<10.5$ (blue), 
$10.5<\log{M}<11.0$ (cyan), 
$11.0<\log{M}<11.5$ (green), 
$11.5<\log{M}<12.0$ (red).
and total (black).
RIGHT: Compared with simple IRX-SSFR model, plotted in each mass bin (colors as in left panel)
and total (black).
 }
\label{fig_sfhist}
\end{figure}

In order to calculate the star formation rate history while minimizing the effects of
cosmic variance, we renormalize the observed mass function to the fit given in Figure \ref{fig_phi_mass}
and Equation \ref{eqn:phi_fit}
for all redshift bins. The results in each mass bin and the total
are shown in Figure \ref{fig_sfhist}. We also calculate the star formation rate
density evolution for the simple exponential model of the previous
section. This is shown in the right-hand panel of Figure \ref{fig_sfhist}.
Apparently the dominant galaxy population responsible for the
fall in SFR density from z=1 to z=0 is $10.5<\log{\mass}<11.5$ (cyan and green points).
This is consistent with the conclusion that the characteristic
mass $M_c$ derived in the previous section evolves from $10^{10.5}$
to $10^{11.25}$. \cite{zheng07}, also using COMBO-17 and Spitzer data, and
the stacking technique of \cite{zheng06}, have reached conclusions that
in many ways are similar to ours, but differ in some important details.
In particular, they find that the SFR in the highest mass bin
(log M$<$11.25 when converted to our IMF) does not fall more steeply
than that in the lower mass bins.  

Stars formed over $0<z<1$ will increase the total mass of the blue sequence
unless these galaxies transition to the red sequence. In order
to maintain the constant or declining blue-sequence mass, we calculate that the
average mass flux over $0<z<1$ must be $\rho_{B\rightarrow R}\simeq 0.03-0.05 $ \mass$_\odot$ yr$^{-1}$ Mpc$^{-3}$.
This is in agreement with the value we obtained by examining
transition galaxies at $z\sim0.1$ \citep{martin07a}. 

\section{Summary}

We have used COMBO-17, Spitzer, and GALEX data to study the co-evolution of the IRX and star formation
for galaxies over the mass range of $9.5<\log{M}<12.0$ and the redshift
range $0<z<1.2$. We have reached a number of interesting conclusions:

\begin{enumerate}

\item IRX grows with stellar mass in a way that mirrors the mass-metallicity relationship.
The rise of IRX with mass saturates at a characteristic mass, above which it appears to fall.

\item The SSFR is is roughly constant up to the same characteristic mass, above which it falls steeply.

\item The characteristic mass grows with redshift. 

\item At a given mass below the characteristic mass, the IRX grows with redshift.

\item The mass and evolutionary trends of the IRX and SSFR are 
reasonably fit by a simple gas-exhaustion model in which
IRX is determined by gas surface density and metallicity,
metallicity grows with time following a closed box model,
and SFR is determined by the exponentially falling gas density. The
SFR time constant scales with the mass as $\tau \sim M^{-0.7}$. 

\item The characteristic mass is a ``turnoff'' mass indicating galaxies that are
starting to move off the blue sequence.

\item The mass-metallicity relationship is understood to be determined largely by he
characteristic age of the galaxies. The mass-IRX relationship is also
influenced by gas exhaustion above the turnoff mass.

\item The factor of 6-8 rise in SFR density to z=1 is predominantly due
to galaxies in the mass range $10.5<\log{M}<11.5$, the turnoff mass
over the $0<z<1$ redshift range. 

\end{enumerate}

\begin{figure}
\plotone{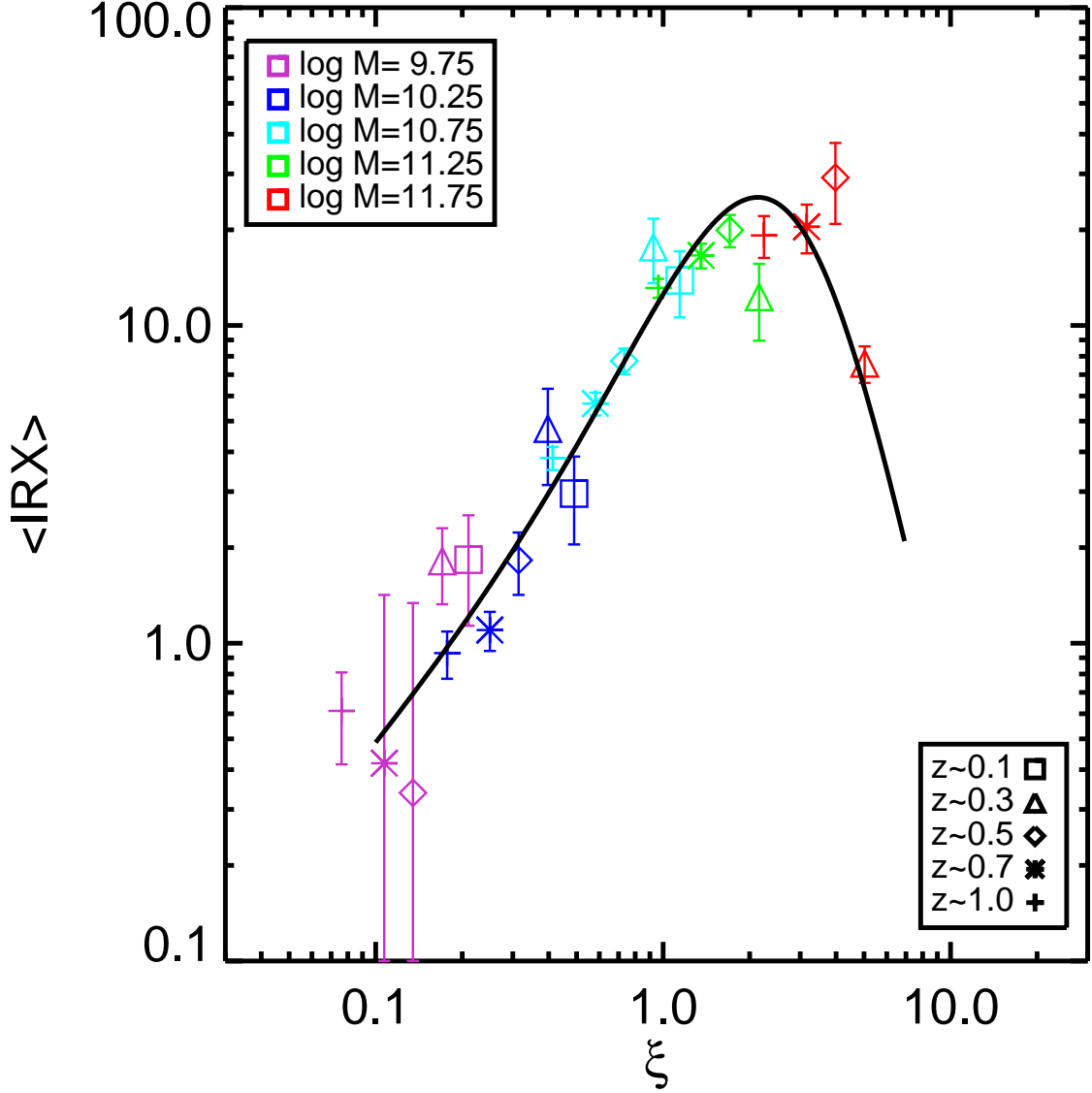}
\caption{IRX (linear version plotted on a logarithmic scale) vs. galaxy age parameter $\xi$ for all masses $9.5<\log{\mass}<12.0$ and all redshifts.
Color gives mass, with $9.5<\log{M}<10.0$ (purple), 
$10.0<\log{M}<10.5$ (blue), 
$10.5<\log{M}<11.0$ (cyan), 
$11.0<\log{M}<11.5$ (green), 
$11.5<\log{M}<12.0$ (red).
Symbol type gives redshift: 
 $0.05<z<0.2$ (plus), $0.2<z<0.4$ (star), $0.4<z<0.6$ (diamond), $0.6<z<0.8$ (triangle), and $0.8<z<1.2$ (square).
Solid line is from equation \ref{eqn:irx} and \ref{eqn:afuv}.}
\label{fig_irx_xi}
\end{figure}

These observations show directly
the steady build-up of heavy elements in the interstellar media of evolving galaxies,
and that the infrared excess IRX represents an excellent tool for selecting similar mass galaxies
at various stages of evolution. In particular, galaxies at early stages in their
evolution can be selected by their low IRX (Figure \ref{fig_irx_xi}). 

It is important to stress that these trends were uncovered by considering the
average properties, notably the IRX, of large numbers of galaxies.
A more sophisticated treatment would study the detailed distribution
of properties, for example the spread in SSFR in a given mass bin.
The simple scaling model of course predicts no spread at all for
a coeval population. This distribution may yield information
about the burst and formation history of galaxies. For example
it will be very interesting to study galaxies with an unusually low IRX
at a given epoch and mass to determine whether they are more
recently formed. It will also be interesting to compare this very simple picture
with the results of semianalytic models combined with the latest
numerical simulations. Finally, it is critical to improve the observational
basis of this work, most notably with a better understanding
of the FIR bolometric correction and its evolution, with a larger and deeper sample
of galaxies, and by extending the redshift range of this approach to determine
whether this simple picture continues to apply during the major
epoch of star formation.


\acknowledgments

GALEX (Galaxy Evolution Explorer) is a NASA Small Explorer, launched in April 2003.
We gratefully acknowledge NASA's support for construction, operation,
and science analysis for the GALEX mission,
developed in cooperation with the Centre National d'Etudes Spatiales
of France and the Korean Ministry of 
Science and Technology.

{\it Facilities:} \facility{GALEX}, \facility{SDSS}


\clearpage


\clearpage
\begin{deluxetable}{ll}
\tabletypesize{\scriptsize}
\tablecaption{Source Detection and Matching Statistics \label{tab_matchstats}}
\tablewidth{0pt}
\tablehead{
\colhead{Bands} & 
\colhead{Number} 
}
\startdata
 R$<$24 & 15882 \\
 IRAC1 $>$ 0.0005 mJy & 13754 \\
 MIPS24 $>$ 0.02 mJy & 3098 \\
 R$<$24, NUV$<$26.0 & 10298 \\
 R$<$24, NUV$<$26.0, FUV$<$26.0 & 4356 \\
 R$<$24, IRAC1$>$0.0005 & 8196 \\
 R$<$24, MIPS24$>$0.02 & 2090 \\
  R$<$24, NUV$<$26.0, MIPS24$>$0.02 & 1481 \\
 R$<$24, NUV$<$26.0, IRAC1$>$0.0005, MIPS24$>$0.02 & 1274 \\
 \enddata
\end{deluxetable}


\clearpage
\begin{deluxetable}{lll}
\label{tab_bcfit}
\tabletypesize{\scriptsize}
\tablecaption{Bolometric Correction Coefficients}
\tablecaption{$\log L_{FIR} = a_\lambda + b_\lambda \log (L[\lambda]/10^{10}L_\odot)$}
\tablewidth{0pt}
\tablehead{
\colhead{$\lambda$} &
\colhead{a$_\lambda$} & 
\colhead{b$_\lambda$} 
}
\startdata
12 & 0.710 & 0.037 \\
13 & 0.705 & 0.036 \\
14 & 0.778 & 0.033 \\
15 & 0.860 & 0.030 \\
16 & 0.884 & 0.026 \\
17 & 0.878 & 0.023 \\
18 & 0.874 & 0.020 \\
19 & 0.880 & 0.016 \\
20 & 0.874 & 0.011 \\
21 & 0.873 & 0.007 \\
22 & 0.863 & 0.003\\
23 & 0.854 & 0.000\\
24 & 0.824 & -0.004\\
 \enddata
\end{deluxetable}

\end{document}